\documentclass[aps,amssymb,amsmath,prx,twocolumn,showpacs,superscriptaddress,longbibliography]{revtex4-2}

\usepackage{graphicx}
\usepackage{dcolumn}
\usepackage{bm}
\usepackage{subfigure}
\usepackage{color}

\usepackage{amssymb,amsmath,amsfonts,latexsym,graphicx,verbatim}

\usepackage[margin=0.55in]{geometry}

\usepackage{ulem}

\usepackage{tikz}

\usepackage[english]{babel}
\usepackage{times}
\usepackage{dsfont}
\usepackage{latexsym}
\usepackage{fancyhdr}
\usepackage{float}
\usepackage{afterpage}
\usepackage{enumitem}
\usepackage{mleftright}

\definecolor{lR}{rgb}{1, 0.8, 0.79}

\usepackage{eso-pic, graphicx}

\usepackage{listings}
\usepackage{multirow}
\usepackage{xcolor,colortbl}

\usepackage{bbm}
\usepackage{upgreek}
\usepackage{amsmath}
\usepackage{newtxtext,newtxmath}
\usepackage{appendix}

\definecolor{Ablue}{rgb}{0.96,0.24,0.00}

\definecolor{Abluetitle}{rgb}{0.,0.24,0.51}

\definecolor{orange}{rgb}{0.96,0.24,0.00}

\definecolor{darkred}{rgb}{0.55, 0.0, 0.0}

\definecolor{darksalmon}{rgb}{0.91, 0.59, 0.48}
\definecolor{maroon}{cmyk}{0,0.87,0.68,0.32}

\setcitestyle{numbers,square,citesep={,\kern-.24em}}

\definecolor{mustard}{rgb}{1.0, 0.86, 0.35}

\addto\captionsenglish{\renewcommand{\figurename}{Fig.}}
\addto\captionsenglish{\renewcommand{\appendixname}{App.}}

\definecolor{Gray}{gray}{0.85}
\definecolor{LightCyan}{rgb}{0.88,1,1}
\newcolumntype{a}{$>${\columncolor{Gray}}c}
\newcolumntype{b}{$>${\columncolor{white}}c}

\usepackage{array}
\newcolumntype{L}[1]{$>${\raggedright\let\newline\\\arraybackslash\hspace{0pt}}m{#1}}
\newcolumntype{C}[1]{$>${\centering\let\newline\\\arraybackslash\hspace{0pt}}m{#1}}
\newcolumntype{R}[1]{$>${\raggedleft\let\newline\\\arraybackslash\hspace{0pt}}m{#1}}

\newcolumntype{P}[1]{>{\centering\arraybackslash}p{#1}}
\newcolumntype{M}[1]{>{\centering\arraybackslash}m{#1}}

\usepackage[colorlinks=true , citecolor=black,urlcolor=blue]{hyperref}

\newcommand{\xa}{\alpha}

\newcommand{\xd}{\delta}

\newcommand{\tm}{{\text -}}

\newcommand{\tacq}{t_{\R{acq}}}
\newcommand{\xg}{\gamma}
\newcommand{\xt}{\vartheta}

\newcommand{\xr}{\rho}
\newcommand{\xo}{\omega}
\newcommand{\xph}{\phi}

\newcommand{\app}{\approx}

\newcommand{\Bp}{B_{\R{pol}}}

\newcommand{\Cs}{{}^{13}\R{C}}

\newcommand{\fhet}{f_{\R{het}}}
\newcommand{\Bac}{B_{\R{AC}}}

\newcommand{\degree}{^{\circ}}

\newcommand{\mI}[0]{\mathcal{I}}

\newcommand{\mN}[0]{\mathcal{N}}

\newcommand{\mHdd}[0]{\mH_{\R{dd}}}
\newcommand{\mHac}[0]{\mH_{\R{AC}}}
\newcommand{\mHc}[0]{\mH_{\R{c}}}

\newcommand{\mHff}[0]{\mH_{\R{ff}}}

\newcommand{\ac}[0]{\R{AC}}

\newcommand{\xy}[0]{\xhat\tm\yhat}

\newcommand{\yz}[0]{\yhat\tm\zhat}

\newcommand{\xD}{\Delta}

\newcommand{\xT}{\Theta}

\newcommand{\xO}{\Omega}
\newcommand{\xPh}{\Phi}

\newcommand{\fr}[2]{\frac{#1}{#2}}

\newcommand{\res}{\R{res}}

\newcommand{\mH}[0]{\mathcal{H}}

\newcommand{\rt}{\rightarrow}

\newcommand{\beq}{\begin{equation}}
\newcommand{\eeq}{\end{equation}}
                  
\newcommand{\benum}{\begin{enumerate}}
\newcommand{\eenum}{\end{enumerate}}
                    
\newcommand{\bit}{\begin{itemize}}
\newcommand{\eit}{\end{itemize}}
\newcommand{\xhat}{\hat{\T{x}}}
\newcommand{\yhat}{\hat{\T{y}}}
\newcommand{\zhat}{\hat{\T{z}}}

\newcommand{\bea}{\begin{eqnarray}}
\newcommand{\eea}{\end{eqnarray}}

\newcommand{\bev}{\begin{verbatim}}
\newcommand{\eev}{\end{verbatim}}

\newcommand{\zt}{\times}

\newcommand{\qt}{\tau}

\newcommand{\lb}{\left(}
\newcommand{\rb}{\right)}
\newcommand{\lsb}{\left[}
\newcommand{\rsb}{\right]}

\newcommand{\pqty}[1]{\left( #1 \right)}
\newcommand{\bqty}[1]{\left[ #1 \right]}

\newcommand{\T}[1]{\textbf{#1}}
\newcommand{\I}[1]{\textit{#1}}
\newcommand{\R}[1]{\textrm{#1}}

\newcommand{\zl}[1]{\label{eqn:#1}}
\newcommand{\zr}[1]{Eq.\,(\ref{eqn:#1})}
\newcommand{\zfl}[1]{\protect\label{fig:#1}}
\newcommand{\zfr}[1]{\figurename\,\ref{fig:#1}}
\newcommand{\zar}[1]{\appendixname\,\ref{app:#1}}
\newcommand{\ztl}[1]{\label{table:#1}}
\newcommand{\ztr}[1]{Table \ref{table:#1}}
\newcommand{\zsl}[1]{\label{sec:#1}}
\newcommand{\zsr}[1]{Sec. \!\!\!\ref{sec:#1}}
\newcommand{\zal}[1]{\label{app:#1}}

\newcommand{\expec}[1]{\left\langle #1\right\rangle}

\newcommand{\ba}{\left\{ \begin{array}{lr}}
\newcommand{\ea}{\end{array}\right.}

\newcommand{\Tr}[1]{\textrm{Tr}\left\{{#1}\right\}}

\newcommand{\blist}[1]{
 \begin{list}{#1}
 \begin{align}
	 arrow
 \end{align}
 $\checkmark\star
  { \setlength{\itemsep}{3pt}
     \setlength{\parsep}{2pt}
     \setlength{\topsep}{3pt}
     \setlength{\partopsep}{0pt}
     \setlength{\leftmargin}{1em}
     \setlength{\labelwidth}{1em}
     \setlength{\labelsep}{0.5em} } }
\newcommand{\elist}{
  \end{list}  }

\DeclareMathSymbol{\vartheta}{\mathalpha}{letters}{"12}
\DeclareMathSymbol{\theta}{\mathalpha}{letters}{"23}
\DeclareMathSymbol{\phi}{\mathalpha}{letters}{"27}
\DeclareMathSymbol{\varphi}{\mathalpha}{letters}{"1E}

\newcommand{\bef}
{
\begin{figure}[htbp]
\centering
}

\newcommand{\eef}{\end{figure}}

\newcommand*\circled[1]{\tikz[baseline=(char.base)]{
            \node[shape=circle,draw,inner sep=2pt] (char) {#1};}}

\mleftright
\makeatletter
\@addtoreset{section}{part}
\makeatother

\newcommand{\beginsupplement}{%
        \setcounter{table}{0}
        \renewcommand{\thetable}{S\arabic{table}}%
        \setcounter{figure}{0}
        \renewcommand{\thefigure}{S\arabic{figure}}%
				
     }

\newcommand{\affA}{Department of Chemistry, University of California, Berkeley, Berkeley, CA 94720, USA.}
\newcommand{\affB}{Department of Physics, KTH Royal Institute of Technology, SE-106 91 Stockholm, Sweden.}
\newcommand{\affC}{Department of Physics, St. Kliment Ohridski University of Sofia, 5 James Bourchier Blvd, 1164 Sofia, Bulgaria.}
\newcommand{\affD}{Max Planck Institute for the Physics of Complex Systems, N\"othnitzer Str.~38, 01187 Dresden, Germany.}
\newcommand{\affE}{Chemical Sciences Division,  Lawrence Berkeley National Laboratory,  Berkeley, CA 94720, USA.}
\newcommand{\affF}{Tabor Electronics, Inc., Hatasia 9, Nesher 3660301, Israel.}
\newcommand{\affG}{Element Six Innovation, Fermi Avenue, Harwell Oxford, Didcot, Oxfordshire OX11 0QR, UK.}

\begin{document}
\title{Continuously tracked, stable, large excursion trajectories of dipolar coupled nuclear spins}
\author{Ozgur Sahin}\thanks{Equal contribution}\affiliation{\affA}
\author{Hawraa Al Asadi}\thanks{Equal contribution}\affiliation{\affA}
\author{Paul M. Schindler}\thanks{Equal contribution}\affiliation{\affD}
\author{Arjun Pillai}\affiliation{\affA}
\author{Erica Sanchez}\affiliation{\affA}
\author{Matthew Markham}\affiliation{\affG}
\author{Mark Elo}\affiliation{\affF}
\author{Maxwell McAllister}\affiliation{\affA}
\author{Emanuel Druga}\affiliation{\affA}
\author{Christoph Fleckenstein}\affiliation{\affB}
\author{Marin Bukov}\affiliation{\affD}\affiliation{\affC}
\author{Ashok Ajoy}\affiliation{\affA}\affiliation{\affE}

\begin{abstract}
We report an experimental approach to excite, stabilize, and continuously track Bloch sphere orbits of dipolar-coupled nuclear spins in a solid. We demonstrate these results on a model system of hyperpolarized $\Cs$ nuclear spins in diamond. Without quantum control, inter-spin coupling leads to rapid spin decay in $T_2^{\ast}{\approx}1.5$ms. We elucidate a method to preserve trajectories for over $T_2^\prime{>}27$s at excursion solid angles up to $16^{\circ}$, even in the presence of strong inter-spin coupling. This exploits a novel spin driving strategy that thermalizes the spins to a long-lived dipolar many-body state, while driving them in highly stable orbits. We show that motion of the spins can be quasi-continuously tracked for over 35s in three dimensions on the Bloch sphere. In this time the spins complete ${>}$68,000 closed precession orbits, demonstrating high stability and robustness against error. We experimentally probe the transient approach to such rigid motion, and thereby show the ability to engineer highly stable “designer” spin trajectories. Our results suggest new ways to stabilize and interrogate strongly-coupled quantum systems through periodic driving and portend powerful applications of rigid spin orbits in quantum sensing.
\end{abstract}

\maketitle

\begin{figure}[t]
 \centering
 {\includegraphics[width=0.475\textwidth]{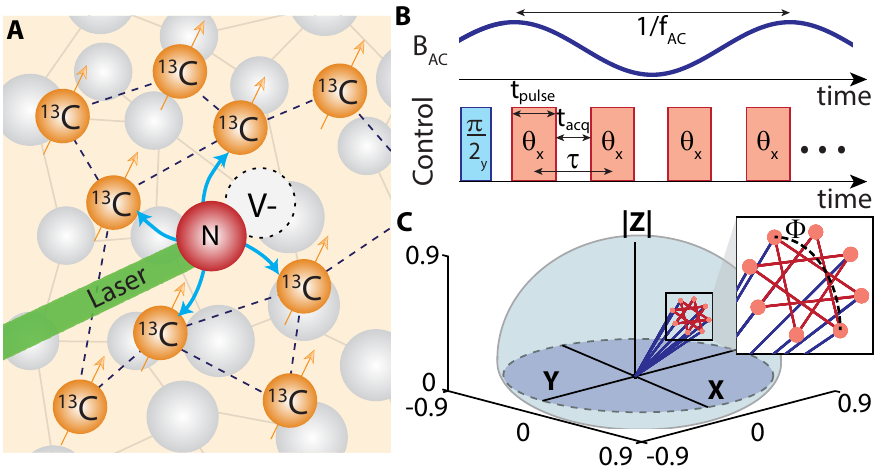}}
 \caption{
 \label{fig:schematic}
\T{Concept.} (A) \I{System:} ensemble of dipolar-coupled $\Cs$ nuclear spins hyperpolarized by NV centers at 38mT. (B) \I{Protocol:} $\xt$-pulses spin-lock the $\Cs$ nuclei at 7T, which are simultaneously subject to a time-varying field with amplitude $\Bac$ and angular frequency $\xo_{\R{AC}} {=} (2\pi) f_{\R{AC}}$ respectively. Pulse spacing is $\qt$, pulse width is $t_{\R{pulse}}$, and spin precession amplitude and phase are interrogated in $\tacq$ windows between pulses. (C) \I{Experimentally tracked} “8-point star” spin trajectory on a Bloch sphere during a $\xD t{=}4$ms window (zoomed in inset). Trajectory remains rigid (stable) for $T_2^\prime{\approx}26.4$s, and is continuously tracked for the entire period (see \zfr{polygons}). Extent of the spin excursion on the Bloch sphere is denoted by angle $\xPh$ (here $\xPh{\app}11.5^{\circ})$.}
	\zfl{concept}
\end{figure}
    
\section{\label{sec:intro}Introduction} 
Periodically driven quantum systems have attracted enormous interest for their ability to host novel far-from-equilibrium phases of matter ~\cite{goldman2014periodically,Bukov2015,eckardt2017colloquium, khemani2019brief,Else2020}, and for sustaining long-lived, highly stable states wherein absorption of energy can be controllably suppressed~\cite{singh2019quantifying,abadal2020floquet,peng2021floquet}. Consider, for instance, a network of dipolar coupled solid-state spins; their interaction drives rapid free induction decay of prepared transverse states in a very short time $T_2^{\ast}$. However, via Floquet prethermalization~\cite{Abanin15,mori_15,Beatrez21,peng2021floquet}, periodically driving the spins can greatly extend these state lifetimes from $T_2^{\ast}$ to $T_2^\prime$ in a manner that parametrically increases with the driving frequency.

However, this approach remains restricted to specific initial states, typically those aligned parallel to the driving field. More generally applicable schemes for stabilization, especially along \I{arbitrary} axes, remain elusive. In this paper, we propose and demonstrate a strategy to excite and stabilize \I{closed spin orbits} on the Bloch sphere, including those that span large angular excursions away from the driving axis. Our approach exploits novel applications of the eigenstate thermalization hypothesis (ETH) and quantum thermalization \cite{DAlessio2016,Deutsch2018}, combined with the engineering of a family of effective Hamiltonians~\cite{Beatrez22} under the simultaneous action of two orthogonal and frequency-separated driving fields. Stable orbits are then excited within the \I{micromotion} dynamics between these Hamiltonians~\cite{Desbuquois17}. In a model system of strongly interacting $\Cs$ nuclear spins in diamond with an intrinsic $T_2^{\ast}{\app}1.5$ms~\cite{Ajoy19relax}, we demonstrate the ability to stabilize highly tunable orbits with excursion angles ${>}16^{\circ}$ for a lifetime $T_2^\prime{>}27$s. This corresponds to an increase in the spin lifetimes of over $18,000$-fold, even in the presence of strong inter-spin dipolar couplings.

This same method allows us to quasi-continuously track the resulting spin motion in three-dimensions on a Bloch sphere over long periods; here we demonstrate the quasi-continuous tracking of spins as they traverse ${>}$68,000 cycles of the engineered stabilized orbits. While such state tracking is difficult in many experimental scenarios~\cite{Murch13,Zeiher21}, our method makes it easily attainable by (1) exploiting weak coupling of the nuclear spins to a readout cavity in a manner that imposes no back-action upon them~\cite{Schuster05}, (2) employing “hyperpolarization”, yielding considerable gains in measured signal to noise~\cite{Ajoy17, Ajoy18}, and (3) arranging a hierarchical, ${>}10^{5}$-fold separation of time scales between the rates of cavity-induced spin interrogation (signal sampling) and that of the spin orbiting. Ultimately, this rapid spin tracking unravels the emergence of stable spin orbits, including insights into the underlying dynamical thermalization processes that are key to their rigidity. Leveraging these special features, we demonstrate the ability to excite and track \I{“designer”} spin trajectories created via dynamically controlled Hamiltonians~\cite{Freeman98}. 

Overall, this work greatly expands the toolbox for engineering and stabilizing long-range strongly interacting quantum systems, and via their continuous tracking, portends diverse applications in quantum sensing~\cite{Degen17} and simulation~\cite{Weitenberg21}. In particular, the rigidity and symmetry of motion induced on the spins suggests new ways to detect weak external fields in the spin environment.

This paper is organized as follows. \zsr{model} presents the basis of our strategy for exciting and stabilizing spin orbits, \zsr{spin_tracking} describes the spin tracking protocol, \zsr{tracked_trajs} summarizes our experimental results in producing long-time rigid polyhedral spin orbits, and \zsr{emerging_traj} studies their emergent stability. \zsr{designer_trajs} extends these results to more complex "designer" trajectories. Technical details underlying these sections are presented in the Appendices (\zar{track_principle}--\zar{theory_quenches}).

\begin{figure*}[t]
  \centering
  {\includegraphics[width=0.95\textwidth]{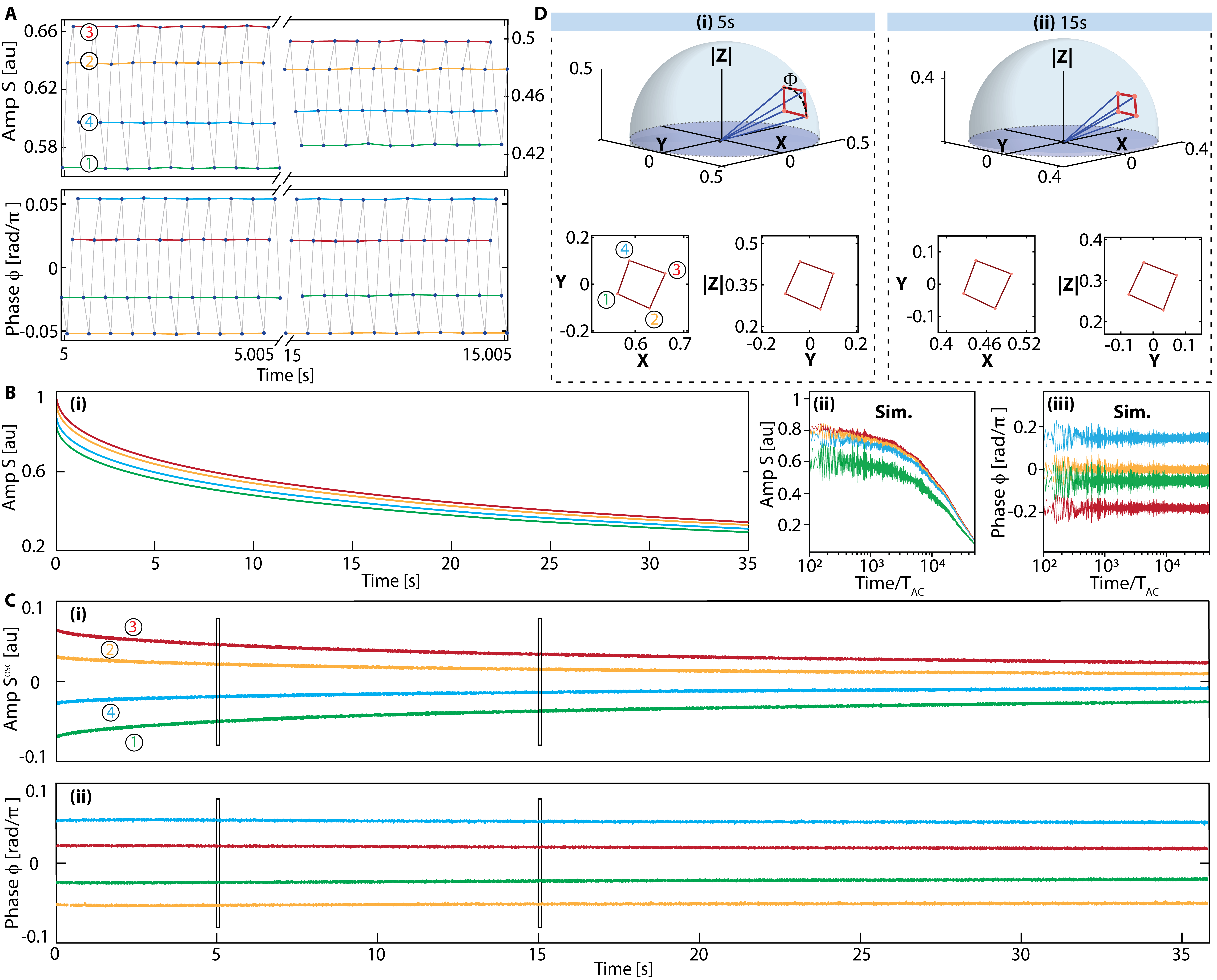}}
\caption{\T{Long-time continuous tracking of spin evolution.} (A) \I{Amplitude $S$ and phase $\varphi$} of spin precession (see \zar{track_principle}) for two representative $\xD t{=}$5ms time windows starting at 5s and 15s. Here $\xt{\app}\pi/2$, $\xo_{\R{AC}} {=} \xo_{\R{res}} {=} (2\pi) 1953.125$ Hz, matched in period with four pulses (\zfr{concept}B), and shown is an average of 30 experiments. Grey lines connect consecutive oscillating points, and colored lines (manifolds), labeled circled 1-4, connect every fourth point. Individual manifolds display a flat response in each $\xD t$ window.  (B) (i) \I{Full data for (A) over 35s}, showing signal $S$ with four manifolds separately colored as in (A). Decay lifetime is $T_2^{\prime}{\app} 32.5$ s. (ii-iii) Numerical simulation of signal $S$ and phase $\varphi$ using a system of $L{=}14$ spins (see \zar{theory_model} and \cite{SOM} for details of the numerical implementation; simulation parameters are given in     \zfr{longtime_simulation}). Simulation shows formation of four stable plateaus that ultimately decay to a featureless infinite temperature state (see \zar{prethermal_props}).
(C) \I{Oscillatory signal} (i) amplitude $S^{\R{osc}}$ and (ii) phase $\varphi$ extracted from (B) after each of the 273,250 pulses applied for the 35s period. Slow signal decay is evident, and the four parallel lines in $\varphi$ show that the oscillations persist regularly for many seconds. (D) \I{Bloch sphere representation} of spin trajectory in three dimensions (see \zar{track_method}), and as 2D projections for two $\xD t{=}5$ms windows starting at \I{(i)} $t{=}$5s and \I{(ii)} 15s (black narrow rectangles in (C)). Points corresponding to manifolds are marked. Lines join the centroid position of every fourth point in the $\xD t$ window. Quadrilateral shaped orbital trajectories are evident; shrinking with time reflects thermalization to an infinite temperature state. Excursion angle at $t{=}$5s is $\xPh{\app} 16.34\degree$. Evolution for 35s in (C) corresponds to 68,312 such precessions.}
	\zfl{longtime}
\end{figure*}

\begin{figure*}[t]
  \centering
  {\includegraphics[width=0.95\textwidth]{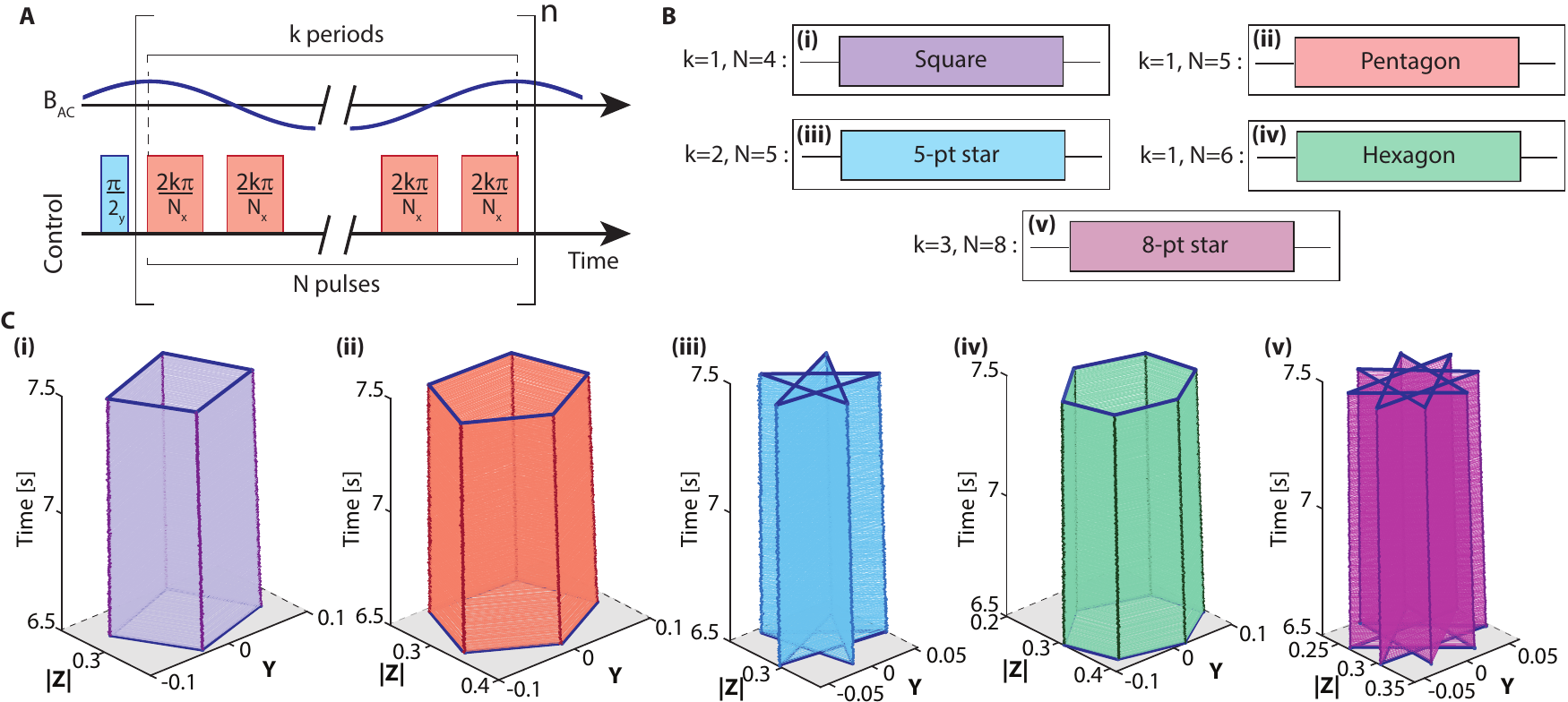}}
  \caption{\T{Long-time engineering and tracking of polyhedral spin trajectories.} (A) \I{Control sequence} consists of a train of pulses of angle $\xt{=}2k\pi/N$ simultaneous with $\zhat$-axis AC field with frequency such that $k$ periods are matched to $N$ total pulses. Bracketed portion is repeated $n({\gg}1)$ times. (B) \I{Block diagram} of sequences for different polygon spin trajectories \I{(i)-(v)}, showing corresponding $k$ and $N$ parameter values. (C) \I{Polyhedral spin trajectories} can be stabilized for multiple second long periods. Shown here are $\xD t{=}$1s sections of the tracked trajectories in the $\yz$ plane of the Bloch sphere; time is on the vertical axis. Top and bottom faces are highlighted in blue for clarity. For the different polyhedra (\I{(i)} square, \I{(ii)} pentagon, \I{(iii)} 5-point star, \I{(iv)} hexagon, and \I{(v)} 8-point star), $\xo_\R{AC}{=}$\I{(i)} $(2\pi)$ 1953.125Hz, \I{(ii)} $(2\pi)$ 1801.8018 Hz, \I{(iii)} $(2\pi)$ 1117.3184 Hz, \I{(iv)} $(2\pi)$ 1673.3601 Hz, and \I{(v)} $(2\pi)$ 733.1378 Hz, respectively (see \zar{movie} and Ref.~\cite{Bloch_movie} for movies of motion). Measured lifetimes $T_2'{=}$\I{(i)} 28.0${\pm}$0.3s, \I{(ii)} 27.8${\pm}$0.2s, \I{(iii)} 28.1${\pm}$0.2s, \I{(iv)}  25.7${\pm}$0.3s, \I{(v)} 26.4${\pm}$0.1s. We average over every 2 slices along the time axis for clarity; plots shown correspond to an average over 30 experiments.}
	\zfl{polygons}
\end{figure*}

\begin{figure*}[t]
  \centering
  {\includegraphics[width=0.95\textwidth]{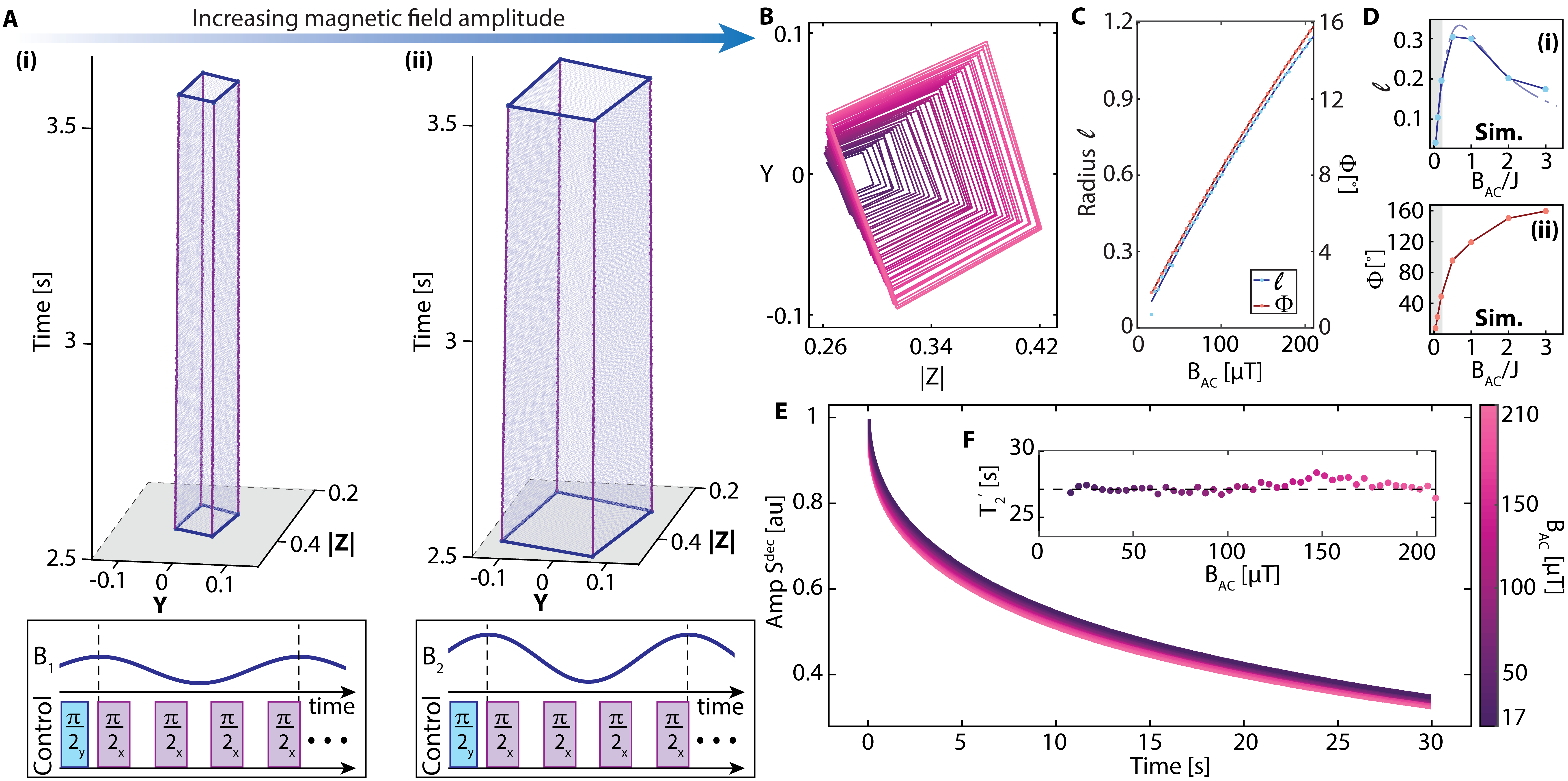}}
  \caption{\T{Stabilized spin trajectories at high amplitudes.}  (A) \I{Tracked spin trajectory} for a square-like trajectory (see \zfr{polygons}) shown for a $\xD t{=}$1s section under two representative fields: (i) $B_1 {\approx} 42\mu$T and (ii) $B_2 {\approx} 210\mu$T  (lower panels show schematic control sequences). Excursion angles are $\xPh{\app}4.05^{\circ}$ and $\xPh{\app}15.89^{\circ}$ respectively. (B) \I{Top-view} (in $\yz$ plane) of the spin trajectories with varying $\Bac$ (see colorbar in (E)). Curves here are an average over 50 slices. (C) Corresponding increase in excursion angle $\xPh$ (red points) and radius $\ell$ (blue points) with $\Bac$. Solid lines are spline fit guides to the eye. (D) Numerically simulated scaling of (i) radius $\ell$ and (ii) excursion angle $\xPh$ as a function of $\Bac/J$ (see \zfr{amplitude_simulation} for full data). In the former, the dashed line follows exact analytical predictions (see \zar{prethermal_props}). The shaded regions here correspond to the experimentally relevant regime of $B_{\R{AC}}$ values in (C). (E) \I{Spin trajectory decay profiles} $S^{dec}$ under different amplitudes $\Bac$ (see colorbar). Shown are smoothed raw amplitude $S$ curves (see SI~\cite{SOM} for discussion). (F) \I{$T_2'$ lifetimes} extracted from stretched exponential fits to profiles in (E). Data reveals corresponding lifetimes $T_2'{\approx} 27.0 \pm 0.2$s (dashed line), approximately independent of $\Bac$. Panel therefore demonstrates ability to stabilize spin trajectories with relatively large deviations from the $\xhat$ axis (see SI~\cite{SOM}).}
\zfl{amplitude}
\end{figure*}

\begin{figure}[t]
 \centering
 {\includegraphics[width=0.5\textwidth]{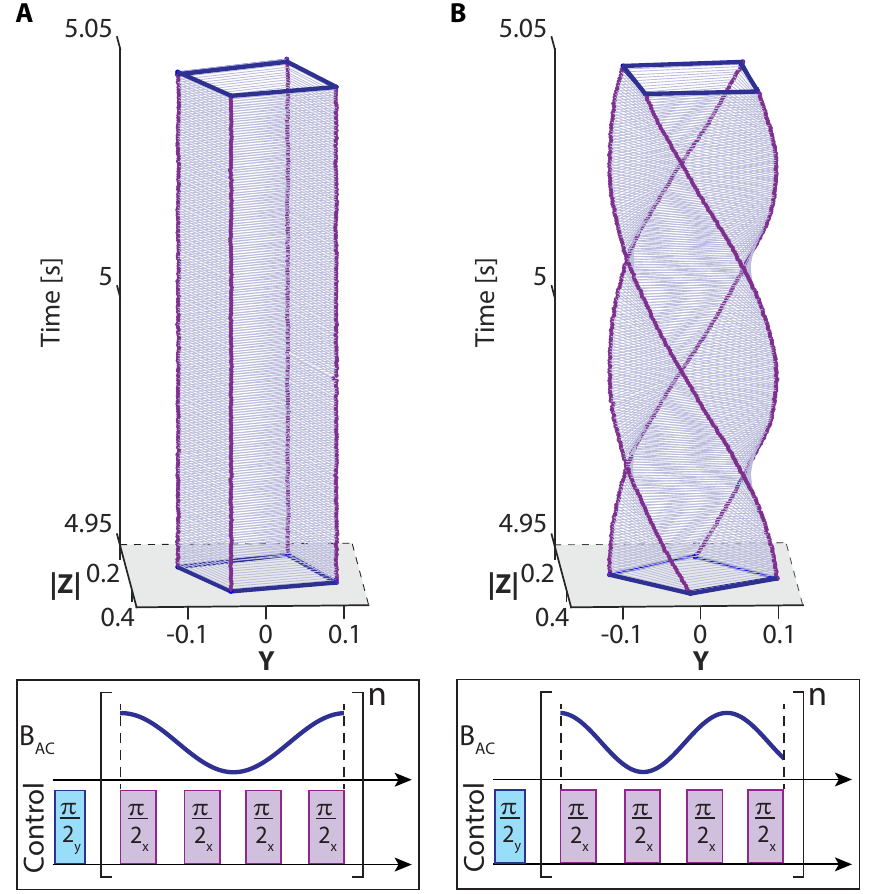}}
  \caption{\T{Spin trajectory response to frequency offset.} Tracked spin trajectories for $\xt{=}\pi/2$ in \zfr{concept}B, in the case of (A) a resonant AC field ($\xo_\R{AC}{=}\xo_\R{res}{=}(2\pi)$1953.125 Hz) and (B) an off-resonant field ($\xo_\R{AC}{=}\xo_\R{res}{+}(2\pi)$10 Hz). Data for a $\xD t {=} 0.1$s window is shown here. \I{Lower panels:} schematic of the experiments. Degree of off-resonance is exaggerated here for clarity. Frequency offset in (B) results in a stable “screw-like” trajectory with a square cross section. Screw pitch is given by the alias frequency ($|\xo_{\R{AC}} - \xo_{\R{res}}|$) with handedness determined by the sign of off-resonance deviation.}
	\zfl{twist}
\end{figure}

\begin{figure*}[t]
  \centering
  {\includegraphics[width=0.95\textwidth]{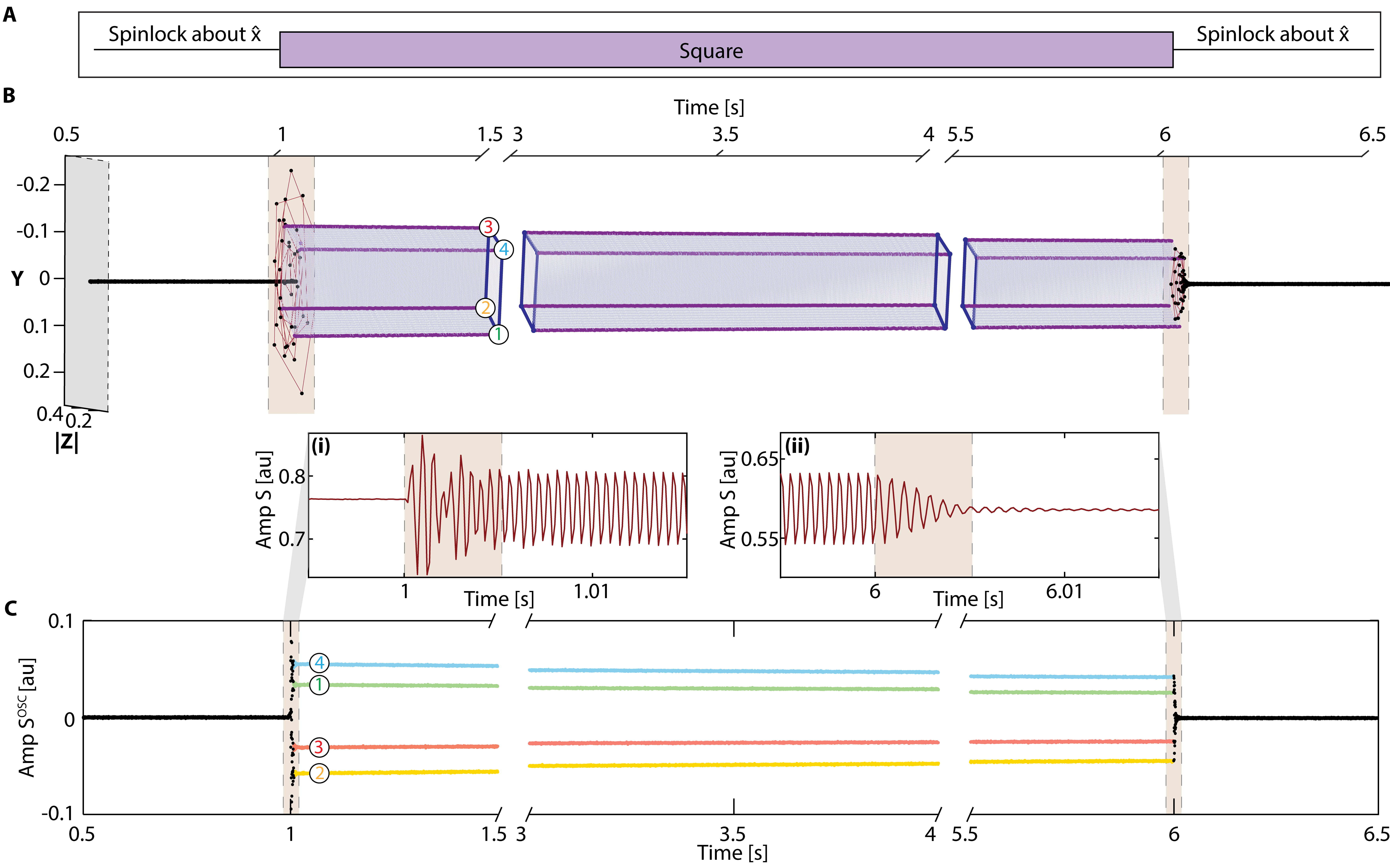}}
  \caption{\T{Emergence of a stabilized square-like spin trajectory} probed via a pulsed AC field. (A) \I{Schematic}. Spins are subject to pulsed spin-locking along $\xhat$ until $t{=}1$s, and then are exposed to an AC field causing a square-like trajectory for $\xD t{=}5$s. Subsequently, the field is turned off and the spins re-thermalize along $\xhat$. Interpulse spacing $\qt{=}128\mu$s and $\xo_{\R{AC}}{=}(2\pi)$1953.125 Hz (see \zfr{polygons}). (B) \I{Continuously tracked trajectory} shown in the $\yz$ plane, with a horizontal dimension representing time. Data is averaged over 5 slices everywhere, except in the beige shaded regions (denoted by dashed lines) to discern the transient dynamics (red traces). Broken time axis highlights square cross-sections of the spin trajectory (with points labeled 1-4 according to the manifolds in (C)). Slow signal decay is reflected in the decreasing size of the square cross-sections over time. (C) \I{Oscillatory amplitude} $S^{\R{osc}}$ as in \zfr{longtime}C(i). Splitting of the $\xhat$-locked spins into the 4 manifolds (circled 1-4) of the square trajectory, their collapse back to $\xhat$, and intermittent transient responses (shaded beige regions) are apparent here. \I{Insets (i)-(ii):} Zoom into raw amplitude response $S$ at the two transient regions (shaded beige). Transient lifetime is $\sim$4ms, comparable to native $T_{2}^{\ast}$. }	
  \zfl{emergence}
\end{figure*}

\begin{figure*}[t]
  \centering
 {\includegraphics[width=0.95\textwidth]{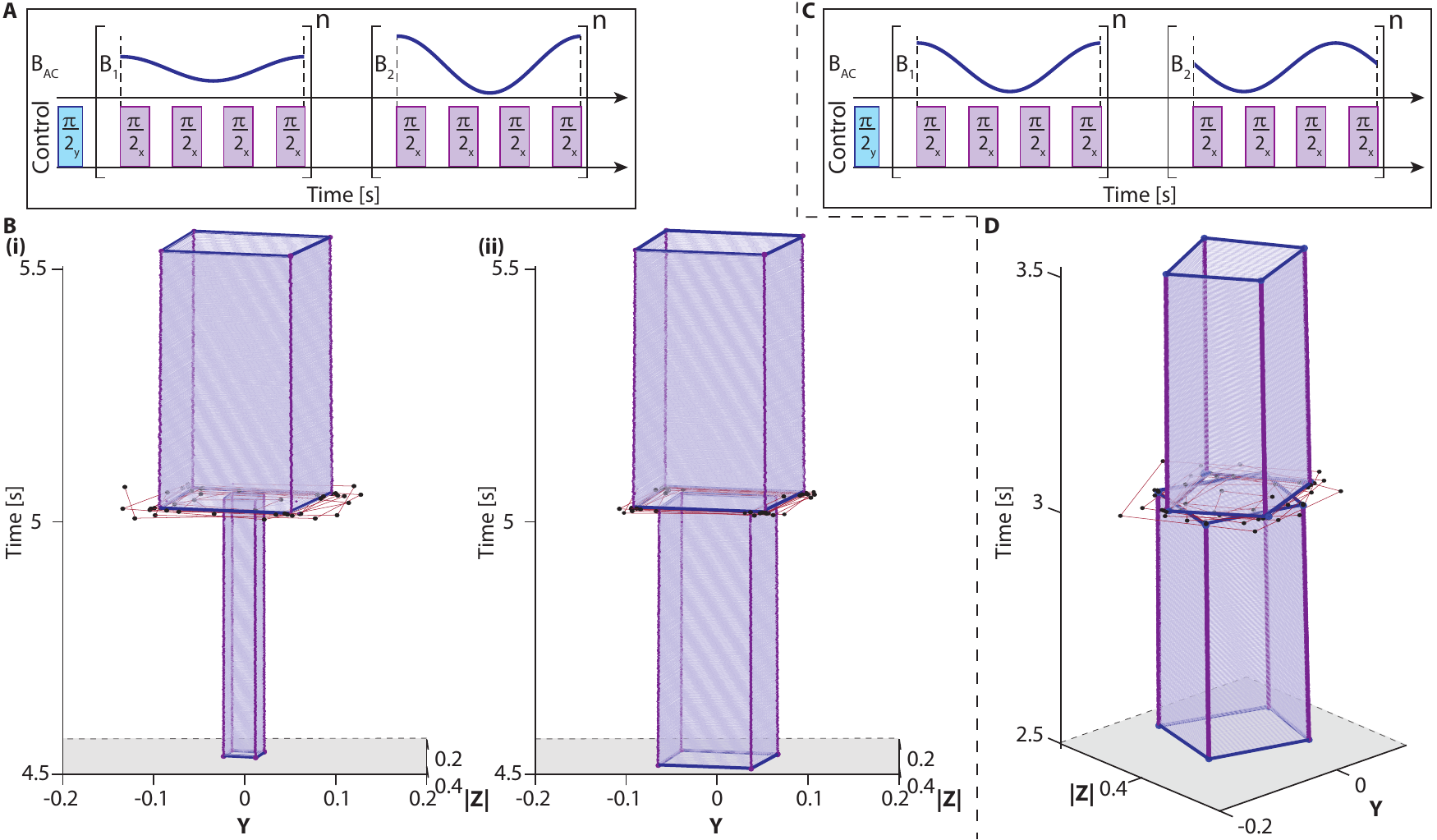}}
  \caption{\T{Transient dynamics upon amplitude and phase quench.} (A) \I{Schematic for amplitude quench} showing control sequence creating a square-like trajectory under an AC field $B_{1}$ for $t{=}5$s (first bracketed portion), which is then switched to an AC field $B_{2}$ with higher amplitude for $t{=}5$s (second bracketed portion). (B) (i)-(ii) \I{Tracked spin trajectories} shown near the switching region for two experiments corresponding to $B_{1}{\approx}{\{42, 126\}}\mu$T  and $B_{2}{=}185\mu$T  in (A) respectively. Emergence of a new stable trajectory is associated with a transient (red traces). (C) \I{Schematic for phase quench} similar to (A), but with AC field instead phase-shifted by $45^{\circ}$ at $t{=}3$ s. (D) \I{Tracked spin trajectories} with $B_{1} {=} B_{2} {=} 185\mu$T). Intermittent transient region (red traces) is again visible. }
	\zfl{amp_phase_hop}
\end{figure*}

\begin{figure*}[t]
  \centering
  {\includegraphics[width=0.95\textwidth]{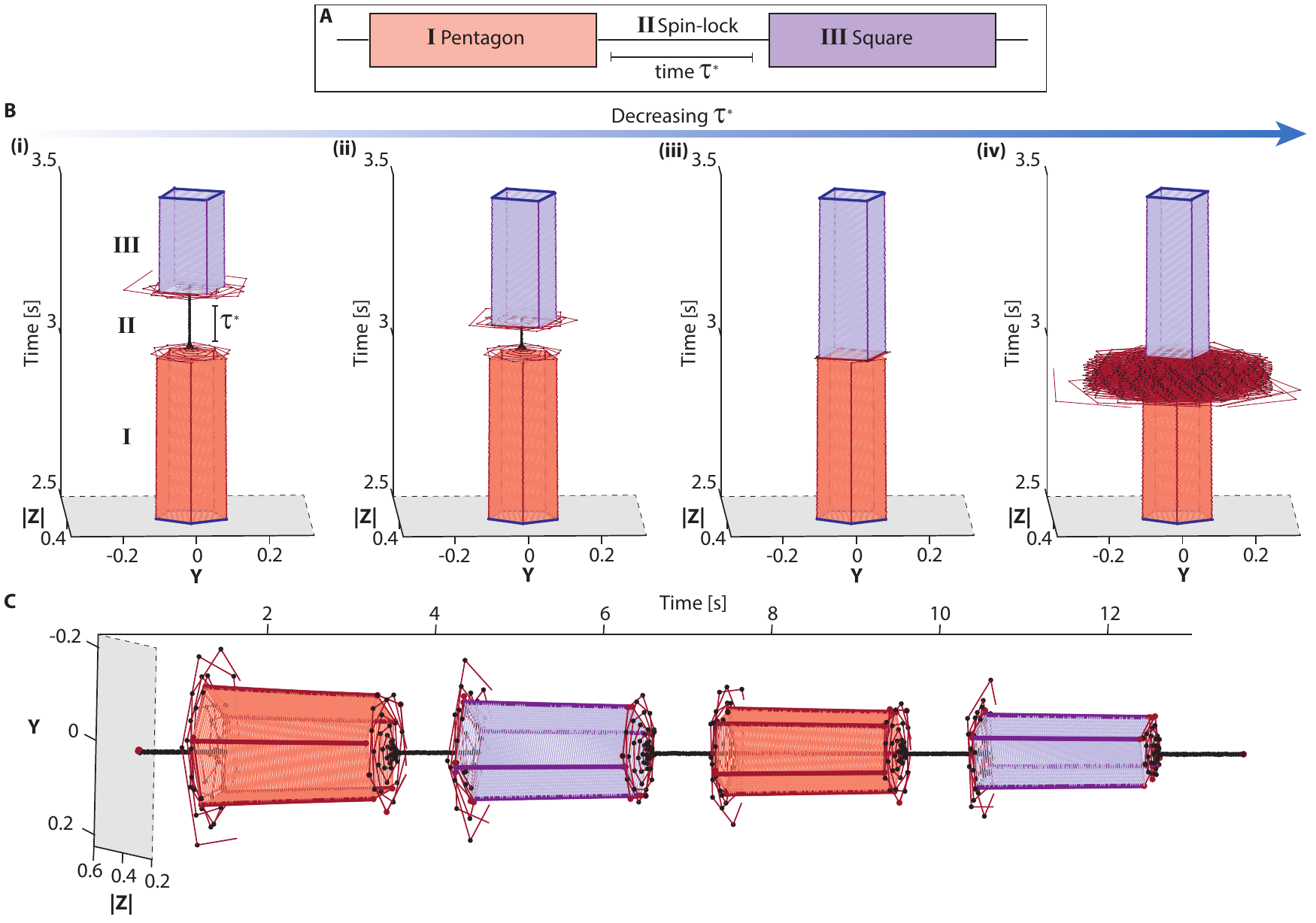}}
  \caption{\T{Designer spin trajectory evolution.} (A) \I{Schematic:} creation and stabilization of square and pentagon shaped trajectories (regions \textbf{I},\textbf{III}, see \zfr{polygons}), separated by a period $\qt_{\ast}$ where the spins are subject to only spin-locking (region \textbf{II}). (B) \I{Tracked spin trajectories} for varying $\qt_{\ast} {=} \{0.4, 0.2, 0, -0.2\}$s for panels \I{(i)-(iv)} respectively. Regions \textbf{I}-\textbf{III} are marked in panel \I{(i)} for clarity. In the  intersecting region in panel \I(iv), the pulse sequence is fixed to that of a square trajectory, but the spins are subject to two AC fields simultaneously (resonant with $N{=}4$ and 5 pulses respectively). (C) \I{Multiple-polyhedron spin trajectory} showing four periods of 2s long square and pentagon trajectories (separated by $\qt_{\ast}{=}1$s). See \zar{designer_decay} (\zfr{decay_profile}) for comparison of decay profile with a single square trajectory over the entire period.}
  \zfl{collisions}
\end{figure*}

\begin{figure*}[t]
  \centering
  {\includegraphics[width=0.8\textwidth]{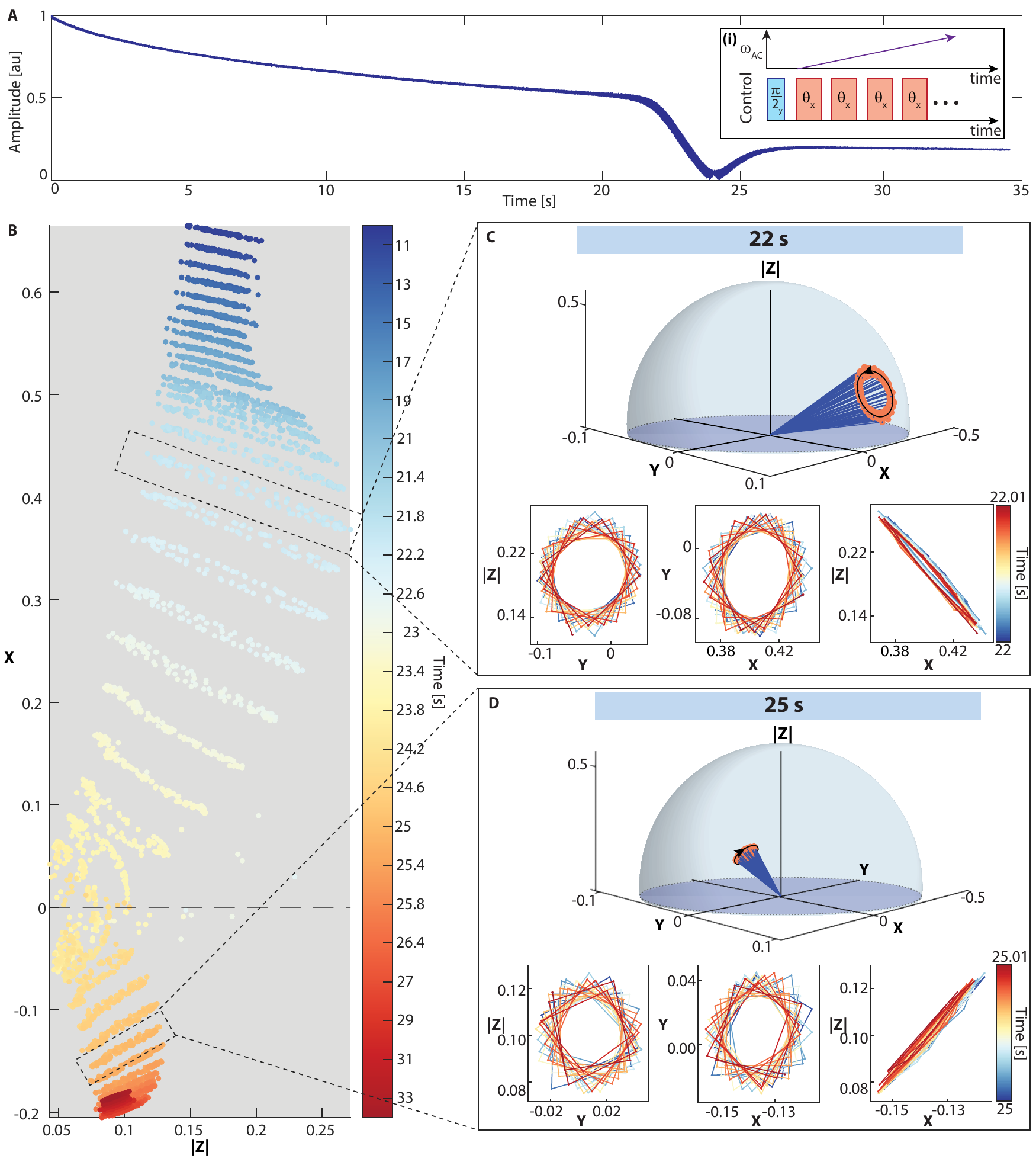}}
  \caption{\T{Continuous spin tracking under rapid adiabatic passage}. (A) \I{Signal amplitude} $S$ under a slow $\xo_{\R{AC}}$ drive swept in frequency, crossing resonance $\xo_{\res}$ at $t{\app}22$s. (i) \I{Pulse sequence} schmeatic showing simultaneous control sequence (here $\xt{=}\pi/2$) and AC field with linearly swept frequency $\xo_{\R{AC}}$. (B-D) Tracked spin evolution in three dimensions. (B) Spin evolution in the $\xhat\tm\zhat$ plane for various $t{=}10$ms windows, wherein motion resembles closed arcs. Here the individual time slices are non-linearly chosen for clarity (see colorbar).  (C-D) Panels display two representative slices at $t{=}22$s and $t{=}25$s, and show full motion for a $\xD t{=}10$ms interval (colorbar) in a 3D Bloch sphere representation and three 2D projections. Arrows denote the direction of spiral-like motion. Spin flip $\xhat{\rt}\tm\xhat$ is evident as a result of the adiabatic rapid passage.
  }
	\zfl{adiabatic}
\end{figure*}

\section{Exciting and Stabilizing Spin trajectories}
\zsl{model}
Our experimental system consists of hyperpolarized $\Cs$ nuclear spins dipolar-coupled with the interaction Hamiltonian~\cite{Duer04},
$\mHdd  {=} \sum_{n<m} b_{nm} (3 I_{nz}I_{mz} - \T{I}_n\cdot \T{I}_m)$, where $I_{n\xa}$ refer to spin-1/2 Pauli matrices of nuclear spin $n$, with $\xa{=}\{x,y,z\}$. The coupling strength $b_{nm} {=} c_\mathrm{exp}[3 \cos^2(\beta_{nm}) - 1]/r_{nm}^3$, with the constant $c_\mathrm{exp}{=}\mu_0 \hbar \gamma_n^2/4\pi $, and angle $\beta_{nm}{=}\cos^{-1}(\boldsymbol{r}_{nm}\cdot \boldsymbol{B_0}/r_{nm}B_0)$ of the internuclear vector $\boldsymbol{r}_{nm}$ in a magnetic field $B_0$. At natural abundance $\Cs$ concentration, $J{=}\expec{b_{nm}}{\app}0.66$kHz~\cite{Ajoy19relax, Beatrez21}, where $\langle\cdot\rangle$ denotes the median over internuclear vector positions. When prepared in a transverse state $\xr_I{=}I_{\nu}{=}\sum_j I_{j,\nu}$, where $\nu{=}\{x,y\}$, evolution under $\mHdd$ causes the spins to exhibit rapid decay in  $T_2^{\ast}{\approx} 1.5$ms~\cite{Beatrez21, SOM}. In addition to $\mHdd$, the spins are also subject to time fluctuating local fields, $\mH_z {=} \sum_{j} c_j(t) I_{jz}$, arising primarily from lattice paramagnetic impurities (P1 centers)~\cite{ajoy2020}.

One strategy to preserve spin polarization in such transverse states to long times is by engineering an effective Hamiltonian that conserves the magnetization along a $\hat{\boldsymbol\nu}$-axis to leading order~\cite{ajoy2020}. Such state protection can be understood within the framework of Floquet theory where the spin state is described by a prethermal ensemble under this effective Hamiltonian. In Ref.~\cite{Beatrez21}, we demonstrated that the resulting transverse lifetimes can exceed $T_2^\prime{=}90.9$s (a >60,000-fold extension).

This approach, however, only preserves spins along a \I{single} transverse axis, henceforth labeled $\xhat$ for convenience. Instead, as shown in \zfr{concept}C, our goal here is to preserve spins in a closed trajectory of {arbitrary} size, quantified by excursion angle $\xPh$ on the Bloch sphere away from $\xhat$. The protocol we employ is described in \zfr{concept}B. Spins are subject to a $\tau$-periodic train of spin-locking (SL) $\xt$ pulses~\cite{Rhim73,Rhim76} (of length $t_\mathrm{pulse}$ and interpulse spacing $\tau$), while simultaneously being driven by a lower-frequency oscillating (AC) magnetic field along $\zhat$. Larmor precession of the spins is continuously monitored in windows ($\tacq{\sim}\qt/2$) between the pulses; the induction signal reports on the precession amplitude and phase in each window. Hyperpolarization, carried out by transferring spin polarization from optically pumped Nitrogen Vacancy (NV) centers~\cite{Fischer2013,Ajoy17,Ajoy20}, allows high signal-to-noise readout (SNR${>} 10^3$ per window).

The resulting spin dynamics are most conveniently described by the rotating (dressed) frame Hamiltonian [cf.~\zar{theory_model}],
\begin{equation}
   \mH_{\R{rot}}(t) =  \mH_z+ \mHdd +  \delta I_z + \mH_{\R{SL}}(t)+\mHac(t) \,
   \zl{rot_hamiltonian},
\end{equation}
where the Hamiltonians corresponding to the orthogonal spin-locking (SL) and the continuous (AC) periodic drives are given by,
\begin{eqnarray}
    \mH_{\R{SL}}(t) &=& \Omega\; \Theta(t) I_x,\nonumber\\
    \mHac (t) &=& \Bac \cos( \xo_{\R{AC}} t + \xph_{\ac}) I_z \, . 
    \label{eq:rot_subhamiltonians}
\end{eqnarray}
Here, $\xO$ is the Rabi frequency. Time dependence of SL sequence is modeled by a $\tau$-periodic step-function $\xT(t)$; $B_{\ac}$ and $\xph_{\ac}$ refer to the amplitude and phase of the AC field, respectively. 

On-site random fields $\mH_z$ play an insignificant role in dictating the dynamics, and will be ignored for simplicity~\cite{Ajoy20DD}. Moreover, $\xO{\gg}\{\Bac, J\}$ and hence, during the SL pulses, we can also ignore the AC-field-driven evolution and the dipolar dynamics induced by $\mHdd$ to a good approximation (see \zar{theory_model}). These assumptions allow us to capture the nonequilibrium dynamics in the simplified Hamiltonian,
\begin{equation}
  \mH_\mathrm{rot}(t) \approx \begin{cases}
                \delta I_z + \Omega I_x,                             & t_n \leq t < t_n + t_\mathrm{pulse} \\
                \delta I_z +  \mHac(t) + \mHdd,  & t_n + t_\mathrm{pulse}\leq t < t_{n+1}
               \end{cases}\, , \label{eq:Hamiltonian}
\end{equation}
where $t_n{=}n\tau$,  $n\in \mathbb{N}$ (see \zfr{schematic}B).

In the following, we assume both drives to be commensurate, i.e., such that $N$-SL pulses are applied in the time the AC drive completes one period (see \zfr{polygons}A). This results in a Hamiltonian $\mH_\mathrm{rot}(t)$, periodic with frequency $\xo_\mathrm{AC}$, and allows the use of Floquet theory~\cite{Shirley65} to analyze the evolution of the system over one AC period $T_\mathrm{AC}{=}2\pi/\xo_\mathrm{AC}$. In particular, we define the unitaries:
\begin{equation}
   U_{F;k} = \exp\left(-i T_\mathrm{AC} \mH_{F;k}\right)
   \label{eq:unitary},
\end{equation}
where $\mH_{F;k}$ is a discrete family of stroboscopic Floquet Hamiltonians ($k{=}1,\cdots,N$), obtained by setting the beginning of the stroboscopic frame to be given by the $k{\R{th}}$ SL pulse~\cite{Bukov2015}.

While theoretical analysis is made complex due to the simultaneous application of continuous and kicked drives, it is analytically feasible when $\xt$ is close to $2\pi/N$, \I{i.e.} $\xt{=}2\pi/N+\delta \xt$ with $\delta \xt{\ll} 1$ (here $\xd$ refers to the detuning of the SL pulses from resonance). We employ a Floquet-Magnus expansion~\cite{Magnus1954} in the SL toggling frame over one AC drive period, to expand $\mH_{F;k}$ in the inverse frequency $\xo_\mathrm{AC}^{-1}$. In the high-frequency regime $\xo_\mathrm{AC}{\gg} \{J, \Omega, B_\mathrm{AC}, \delta\}$, the dynamics are governed by the time-independent effective Hamiltonian (see \zar{theory_floquet})
\begin{equation}
\label{eq:H_F}
   \frac{t_\mathrm{acq}}{\tau}\tilde\mH_{F;k}^{(0)} = \tilde{\boldsymbol{w}}_k(\alpha,\phi_\mathrm{AC}) \cdot \boldsymbol{I} + (1-3\sin^2\alpha)\sum_{n<m} b_{nm}\lb \frac{3}{2} \tilde{\mHff} -  \boldsymbol{I}_n\cdot \boldsymbol{I}_m\rb,
\end{equation}
with the flip-flop term $\tilde{\mHff} {=}I_{jz}I_{kz} + I_{jy}I_{ky}$, and $\alpha(\vartheta){=}\arctan(\delta/\Omega)$ (here the tildes denote operators transformed under the change of basis mapping $\lb \cos(\alpha) I_x + \sin(\alpha) I_z\rb \to I_x$, see App.~\ref{app:theory_floquet}). There is an emergence of a new effective static field
\begin{eqnarray}
\label{eq:w_k}
\tilde{\boldsymbol{w}}_k(\alpha,\phi_\mathrm{AC}) &=& 
\pqty{ \delta\sin\alpha + \delta \xt / \tau }\; \xhat  \\
&& + B_\mathrm{AC}\cos\alpha\sum_{n=1}^N \left(
-\sin n\vartheta\; \yhat
+\cos n\vartheta\; \zhat
\right)
f_{n+k}(\phi_\mathrm{AC}), \nonumber
\end{eqnarray}
which changes direction conditioned on $k$, since $f_{n+k}(\phi_\mathrm{AC}){=}\int_{t_n + t_\mathrm{pulse}}^{t_{n+1}} \sin\left(\frac{2\pi t}{T_\mathrm{AC}}+\phi_\mathrm{AC}\right)  \frac{\mathrm{d}t}{\tacq}$. Importantly, however, the interaction term in Eq.~\eqref{eq:H_F} remains independent of the stroboscopic frame index $k$.

Eq. \eqref{eq:H_F} is a key result of this work. $\tilde{\boldsymbol{w}}_k$ elucidates \I{micromotion} of the spins as they are successively kicked by the SL pulses. Within the duration of the parametrically long-lived prethermal plateau, the spins thermalize under the non-integrable Hamiltonian $\tilde\mH_{F;k}^{(0)}$. 
Hence, we anticipate formation of a prethermal state effectively captured by density matrix $\tilde{\xr}_F{\propto} \exp\left(-\beta\tilde\mH_{F;k}^{(0)}\right)$ with the inverse temperature $\beta$ set by the energy density of the initial state, in a similar manner to ETH~\cite{DAlessio2016,Deutsch2018}. It is straightforward to see that whenever the detuning $\delta{\neq}0$ ($\delta \xt {\neq} 0 $), the initial state $\tilde{\rho}_I{\propto} I_x$ has a small but non-vanishing energy density, $\mathrm{Tr}\left( \tilde{\rho}_I \tilde\mH_{F;k}^{(0)}\right){\propto} \pqty{\delta + \delta \xt/\tau}$, which leads to a finite temperature $\beta{>}0$ in the prethermal plateau. Hence, for ${\beta J{\ll} 1}$, we have $\tilde{\xr}_F{\propto} \tilde\mH_{F;k}^{(0)}$. 

From Eq.~\eqref{eq:H_F}, the prethermal state carries both single-body (magnetization) terms and “dipolar order”~\cite{Slichter61,Jeener65,Goldman74,Beckmann88}, where our realization of the latter stands in contrast to previous methods using the Jeener-Broekaert sequence~\cite{Jeener67,Jeener65,Cho06}. Let us denote by $\mI_{\nu}$ the magnetization components in the rotating frame. Then, in the prethermal plateau, the expectation value of the magnetization $\boldsymbol{M}{=}(\mI_x,\mI_y,\mI_z)$ is given by $\boldsymbol{M}{=} \Tr{\boldsymbol{I} \tilde{\rho}_F} {\propto}\tilde{\boldsymbol{w}}_k$. 
It points in different directions for every $k$ value within the drive period $T_\mathrm{AC}$. Since the micromotion dynamics are continuous and cyclic with the drive period, the magnetization vector will follow periodic excursions on the Bloch sphere. The extent of these excursions can be quantified by the angle $\Phi$ (see \zfr{longtime}D). Importantly, since the Floquet Hamiltonians $\tilde\mH_{F;k}^{(0)}$ are related to one another by a static change of basis, and the $\xt$-pulses just serve to cause micromotion between them, there is \I{no} transient in the dynamics of the thermalized spins between successive pulses. 

As a consequence, when observed at times $t_n+t_\R{pulse}$, the magnetization features $N$ metastable plateaus ($N{=}4$ in \zfr{longtime}). From the perspective of Floquet theory, the observed plateaus can be interpreted as a single plateau, periodically displaced in time by the micromotion dynamics. The lifetime of any of the plateaus is therefore determined by the intrinsic $T_2^\prime$ lifetime set by SL prethermalization; in our long-range 3D system, the plateau lifetimes depend parametrically as $T_2^\prime{\sim} (J\tau)^{-2}$~\cite{Beatrez22}. This forms the basis for exciting and preserving large closed orbits on the Bloch sphere in our experiment. To our knowledge, this is the first time that the observation of micromotion, combined with quantum thermalization, has been proposed as a stabilization mechanism for spin orbits.

\section{Continuous spin tracking on the Bloch sphere}
\zsl{spin_tracking}
An important practical goal of this paper is to be able to continuously \I{track} the rotating-frame spin trajectories in three dimensions on the Bloch sphere. The principle we employ is elucidated in \zar{track_principle}. Typically, spin trajectory tracking is challenging because it requires: (1) the ability to interrogate the spins without perturbing them, and (2) that state interrogation can occur simultaneously with the control field $\mHac$. To make such tracking viable, we employ a series of special features in our system:
\I{(i)} weak spin-cavity coupling that permits non-destructive measurements via RF induction,
\I{(ii)} a hierarchical ${>}10^5$-fold time-scale separation between the rate at which the induction signal is sampled and cyclic orbital motion (see \ztr{table}), and \I{(iii)} hyperpolarization, which yields high single-shot readout SNR.

\zar{track_method} elucidates the methodology employed. Quadrature information from each acquisition window following a SL pulse is employed to extract transverse spin projections in the rotating (dressed) frame, $\mI_{x}(t_j){=}S(t_j)\cos\varphi(t_j)$, and $\mI_{y}(t_j){=}S(t_j)\sin\varphi(t_j)$, where $S$ and $\varphi$ denote the amplitude and phase in this frame,  and variable $t_j$ captures quasi-continuous sampling at rate $\xo^s{=} (2\pi) \qt^{-1}$. In practice, obtaining $\varphi(t)$ requires removing the trivial phase accrued due to Larmor precession during each SL pulse, equivalent to a transformation from lab frame to rotating frame (see \zfr{fig2}). 

Since the spins undergo very little decay during successive cavity interrogation windows, $(\xo^s T_{2}^{\prime})^{-1}{\ll} 1$, it is possible to estimate the magnitude of the $I_{z}(t_j)$ component through a unitarity constraint, $|\mI_{z}(t_j)| {=} [\mN^2(t_j) - \mI_{x}^2(t_j)- \mI_{y}^2(t_j)]^{1/2}$; here $\mN(t)$ refers to the instantaneous norm of the magnetization vector (see \zar{track_method}). This allows us to track and plot the total magnetization $M(t){=}\{\mI_{x}(t),\mI_{y}(t),|\mI_{z}(t)|\}$ (sampled at $\xo^s$) on a Bloch hemisphere for long times, yielding a quasi real-time 3D map of their orbital trajectories (\zfr{concept}C).

\section{\label{sec:tracked_trajs}Long-time continuously tracked spin trajectories} 
\zfr{longtime} demonstrates the ability to excite, stabilize, and track spin trajectories on the Bloch sphere for long periods (here 35s). \zfr{longtime}A shows the tracked amplitude $S$ and phase $\varphi$ employing the protocol in \zfr{concept}B with $\xt{=}\pi/2$ and $B_{\R{AC}}{=}$ 185$\mu$T, shown for two $\xD t{=}$5ms windows starting at $t{=}5$s and $t{=}15$s, respectively. Evidently, the amplitude decay during the ${\approx}10$s intervening period is relatively small (cf.~left and right y-axes in \zfr{longtime}A). More striking, perhaps, is the regularity of the four-point periodic oscillations (gray lines), reflecting the micromotion dynamics. To reveal this more clearly and highlight its stability, in \zfr{longtime}A we join every fourth point by lines of different colors and label these manifolds as $\circled{1}$-$\circled{4}$. 

The full signal $S$ up to $t{=}35$s is shown in \zfr{longtime}B(i). We use the same color convention as in \zfr{longtime}A, but ignore the gray line joining successive points. Spin prethermalization causes each signal manifold to decay with a stretched exponential profile $\exp[-(t/T_2^\prime)^{1/2}]$ with $T_2^\prime{\approx}33$s, cf. \zfr{longtime}B(i) (see also Ref.~\cite{schultzen2022glassy}). \zfr{longtime}B(ii-iii) shows full numerical simulations of $S$ and $\varphi$ following Eq.~\eqref{eq:Hamiltonian}, employing a system of $L{=}14$ spins and averaging over multiple disorder manifestations (see SI~\cite{SOM}). It is possible to discern four stable plateaus (colored) that undergo slow prethermal decays to infinite temperature. The small size of the simulations results in larger fluctuations compared to the experimental data in \zfr{longtime}B(i). 

To highlight the micromotion dynamics, it is helpful to separate the effect of the background decay. Indeed, the prethermal decay profile, $S^{\R{dec}}$, can be extracted by smoothing $S$ with a moving average filter. Then the micromotion is captured by the oscillatory part $S^{\R{osc}}(t){=}S(t) - S^{\R{dec}}(t)$, and is shown in \zfr{longtime}C(i) for $t{=}35$s. Corresponding phase $\varphi(t)$ signal is shown in \zfr{longtime}C(ii).  Data here comprises 273,250 spin-lock pulses and the completion of 68,312 four-point periodic orbits. High stability of the micromotion is reflected in the observation that the four amplitude and phase manifolds remain separate and almost parallel over the entire 35s period. We note that the stability persists in spite of ever present RF inhomogeneity in the SL pulses; see \zar{ampl_distortion} and \zar{flip_distortion} for a full discussion of the origin of this enhanced robustness. While finite computer memory limitations impose restrictions to the total observation period in current experiments, from the data in \zfr{longtime}C, we estimate that the stable dynamics should persist up to $t{>}200$s.

The high stability makes the spin tracking method described in \zar{track_method} viable and carry a low degree of error. This is reflected in \zfr{longtime}D where we reconstruct the full 3D dynamics for two $\xD t{=}$5ms windows starting at (i) $t{=} 5$s and (ii) $t{=} 15$s (narrow rectangular windows in \zfr{longtime}C) on a half Bloch sphere. Spins trace a four-point orbit on the Bloch sphere that  resemble quadrilaterals; we will refer to the trajectories as being \I{“square-like”} for convenience. Comparison of the two panels in \zfr{longtime}D reflects the apparent shrinking of the orbits due to prethermal decay during the intervening period between these $\xD t$ windows. Lower panels in \zfr{longtime}D show corresponding 2D projections in the $\yz$ and $\xy$ planes. Plotted are ${\app}$10 points per manifold (labeled as in \zfr{longtime}C) that correspond to the extracted spin positions in the $\xD t{=}$5ms window; due to high SNR and stability, the points here are essentially overlapping. Red solid lines join the points in both the 2D and 3D representations. The excursion angle in \zfr{longtime}D(i) is $\Phi{\app}16.34^{\degree}$, and we ascribe the overall tilt of the trajectory away from $\xhat$ to detuning of the SL pulses (see SI~\cite{SOM}). This is also supported by the fact that the $\xhat\tm\zhat$ projections trace arcs (see \zar{movie} and \zfr{movie}).

Panels in \zfr{longtime}D, corresponding to small $\xD t{=}5$ms windows, only represent a tiny sliver ($\sim10^{-4}$ times) of the full dynamics tracked in \zfr{longtime}C. Indeed, it is possible to plot the resulting trajectory for the \I{entire} 35s period, but the motion is most conveniently captured by focusing on a single projection. \zfr{polygons}C shows the results of such a visualization, where we cast the $\yz$ projection on a horizontal plane and time on the vertical axis.

In particular, \zfr{polygons}C(i) shows the tracked motion in \zfr{longtime}C for a $\xD t{=}1$s window centered at $t{=}7$s. The spin evolution traces an approximately rhombohedral prism (here the top and bottom faces are highlighted in blue for clarity). Stability of the motion is particularly clear in the straightness of the prism sidewalls. The rotational orientation of the prism depends directly on the AC phase $\xph_{\ac}$ in Eq.~\eqref{eq:Hamiltonian} (for \zfr{polygons}C(i), $\xph_{\ac}{=}0$). Data plotted here comprises 7,813 pulses (separated by $\tau{=}128\mu$s), and for clarity we average over 3 consecutive points in each manifold.

Spin stabilization and tracking can be extended to other shapes via the generalized protocol described in \zfr{polygons}A. It comprises $\xt{=}2\pi k/N$ pulses and imposes that $\xo_{\R{AC}}$ is such that the AC field completes $k$ full periods every $N$ pulses, i.e., $\xo_{\R{AC}}{=}2\pi k/(N\tau)$. We will refer to the AC field in this case as satisfying a \I{resonance condition}, with $\xo_{\R{AC}}$ here being the "resonant frequency" $\xo_{\R{AC}}{=}\xo_{\R{res}}$. In reality, as detailed in \zar{flip_distortion}, this condition can be even more relaxed and made robust against error in $\xt$. 

Sequences in \zfr{polygons}B then produce different “polyhedral” spin trajectories, the tracked results of which are shown in \zfr{polygons}C for a (i) square, (ii) pentagon, (iii) 5-point star, (iv) hexagon, and (v) 8-point star. The latter is also shown on a Bloch hemisphere in \zfr{concept}C. Measured $T_2'$ lifetime values of the corresponding trajectories are listed in caption of \zfr{polygons}C. The 5-point and 8-point star (\zfr{polygons}B(iii,v)) represent examples of a family of non-convex shapes created by employing $\xt{>}\pi/2$ in \zfr{concept}B. We note that in \zfr{polygons}B(v), a single Floquet period occupies $N\qt{=}1.364$ms, a significant proportion of the coupling induced $T_2^{*}{\app}1.5$ms. It is interesting that stabilization of the spin trajectory can be retained even in this limit.

Remarkably, the protocol in \zfr{polygons}A will robustly produce polyhedral trajectories even if the AC field applied is significantly distorted from a pure sinusoid as long as the resonance condition is maintained  (see \zar{ampl_distortion} for a detailed discussion). This is because the micromotion imposes that each manifold in $S^{\R{osc}}$ (corresponding to each edge of the prism) depends, via Eq.~\eqref{eq:w_k}, on the \I{full} AC field over one period rather than its instantaneous value after every pulse (see \zar{ampl_distortion}). 

To probe the relative stability of the trajectories with the excursion angle, we excite square-like trajectories (as in \zfr{polygons}C(i)) of varying sizes by changing the AC amplitude $|B_{\R{AC}}|$. \zfr{amplitude}A shows two representative tracked trajectories with $\Bac {=}$42 and 210$\mu$T respectively. \zfr{amplitude}B shows the $\yz$ projection averaged over a $\xD t{=}15$ms window centered at $t{=}3$s for a range of $\Bac$ values (see colorbar in \zfr{amplitude}E).  The upper limit of $\Bac$ values employed here is due to heating in the AC field coil. In \zfr{amplitude}C, we extract the corresponding (average) radius $\ell$ of the "square” trajectories (blue points) and corresponding excursion angles $\xPh$ (red points). Ultimately, at $\Bac{=}$210$\mu$T we estimate an excursion angle of $\xPh {=} 15.89 \degree$.
Analytical expressions and numerical simulation (see App.~\ref{app:prethermal_props} and SI~\cite{SOM}) yields good qualitative agreement; these results are shown in \zfr{amplitude}D(i-ii) respectively. 

\zfr{amplitude}E shows the corresponding $S^{\R{dec}}$ decay profiles obtained by smoothing raw amplitude $S$ curves (see colorbar for $\Bac$ values). Overlapping curves in \zfr{amplitude}E indicate that the respective decay rates are approximately independent of the AC field amplitude $\Bac$. This is clearer in \zfr{amplitude}F, where we extract corresponding stretched exponential decay constants $T_2^\prime$. Data reveals $T_2^\prime {\approx} 27.0\pm 0.2$s (horizontal dashed line) is independent of $\Bac$ to a good approximation. This is supported by the numerical simulation in \zfr{amplitude_simulation} where we extract $T_2^\prime$ as a function of $\Bac/J$ for an $L{=}14$ system. These results suggest the ability to stabilize, via inter-spin interactions, large spin trajectories on the Bloch sphere while also continuously tracking the resulting motion — a key result of this paper.

Rigidity of the trajectories is also retained when the frequency $\xo_{\R{AC}}$ deviates from the resonance condition. This is illustrated in \zfr{twist} comparing the trajectories on-resonance ($\xo_{\R{AC}}{=}\xo_{\R{res}}$) and off-resonance ($\xo_{\R{AC}}{=}\xo_{\R{res}}+(2\pi)10$Hz). A “screw-like” trajectory with a 10Hz pitch (corresponding to $|\xo_{\R{AC}} - \xo_{\R{res}}|$) is then traced in \zfr{twist}B, with a handedness determined by the sign of the off-resonance detuning.

\section{\label{sec:emerging_traj}Emergence of stabilized spin trajectories}
Let us now consider the \I{emergence} of these stable spin trajectories, starting first with the spins locked along $\xhat$. \zfr{emergence} shows the evolution of the tracked trajectory upon “turning on” the $\Bac$ field corresponding to a "square-like" trajectory at $t{=}1$s, and subsequently turning it off at $t{=}6$s (see schematic in \zfr{emergence}A). In a visualization focused on the $\yz$ projection (see \zfr{emergence}B), the spins start off tracing a line along the time dimension, subsequently settling into a stable "square-like" trajectory under $\Bac$ (stability evident from the cross sections), that finally collapses back into a line when $\Bac$ is turned off. Both switching events are associated with rapid transient dynamics, captured by the red traces in the dashed shaded regions in \zfr{emergence}B. In this visualization, averaging over $5$ frames is carried out everywhere except in the dashed shaded regions to preserve the transients. \zfr{emergence}C shows the full $S^{\R{osc}}$ data with the manifolds labeled following the same convention as in \zfr{longtime}C (i) (corresponding edges are marked in \zfr{emergence}B). 

The transients capture the re-thermalization of the spins upon the Hamiltonian quench~\cite{Polkovnikov11} under the switching action of $\Bac$. To quantify the timescale for this thermalization, insets \zfr{emergence}C(i)-(ii) show the raw amplitude signals $S$ in the dashed transient regions. We discern a transient lifetime of ${\approx}4$ms in both cases, and the close resemblance to $T_2^{\ast}$ suggests that  thermalization here is driven by inter-spin interactions. 
 
The transient dynamics are strongly model-dependent and do not exhibit universal features inherent to the subsequent prethermal plateaus. A closed-form analysis is out of reach for our nonintegrable dipolar system; in addition, numerical simulations on a classical computer are challenging. Nevertheless, via an understanding of the thermal properties of the pre- and post-quench equilibrium state, it is possible to make a few statements about the properties of these ``on'' and ``off'' processes. In particular, when the AC field is turned on, the system undergoes transient dynamics from the initial state $\tilde{\rho}{\propto} I_x$ to the prethermal state $\rho_F{\propto} \beta \mH^{(0)}_{F;\,k}$. In contrast, turning the AC field off yields the prethermal state $\rho_F{\propto} \beta \mH^{(0)}_{F;\,k}\pqty{B_\mathrm{AC}{=}0}$ at the \textit{same} (inverse) temperature $\beta$, but with regard to a different Hamiltonian. Note that $\mH^{(0)}_{F;\,k}\pqty{B_\mathrm{AC}{=}0}$ is independent of $k$ and differs from $\mH^{(0)}_{F;\,k}\pqty{B_\mathrm{AC}{\neq}0}$ by the emergent on-site field, i.e. $\tilde{\boldsymbol{w}}_k\pqty{B_\mathrm{AC}{=}0}$ has only a non-vanishing $\xhat$ component. While it is difficult to make rigorous statements about the transient, these observations suggest that switching the AC field on can lead to a rich transient behaviour whereas the ``off'' transients only correspond to a simple decay of the $N$ plateaus into a single one (see App.~\ref{app:theory_quenches}).

In fact, the insets in \zfr{emergence}C show clear differences between the two transients. While turning on the AC field appears to cause "spikes" in the signal (\zfr{emergence}C(i)) where $S$ momentarily goes higher or lower than the prethermal oscillations, the off-transient is free from large oscillations and only preserves terms in $\rho$ collinear with $\xhat$ which are subsequently stabilized by prethermalization. 

To experimentally quantify the size of the transients, in \zfr{amp_phase_hop} we consider traced trajectory evolutions under amplitude quenches corresponding to an abrupt change in AC field amplitude (\zfr{amp_phase_hop}A-B) or phase (\zfr{amp_phase_hop}C-D). In particular, we expose the spins to a large $\xD B_\mathrm{AC}$ from $B_{\mathrm{AC},1}{\approx}42\mu$T to $B_{\mathrm{AC},2}{\approx}185\mu$T (\zfr{amp_phase_hop}B(i)) and a smaller $\xD B_\mathrm{AC}$ from $B_{\mathrm{AC},1}{\approx}126\mu$T to $B_{\mathrm{AC},2}{\approx}185\mu$T (\zfr{amp_phase_hop}B(ii)); additionally, we conduct a phase step change ($\xph_{\ac,1}{=}0^{\degree}$ to $\xph_{\ac,2}{=}45^{\degree}$) under constant $\Bac$ amplitude (\zfr{amp_phase_hop}D). The corresponding control sequences are described in \zfr{amp_phase_hop}A,C. \zar{theory_quenches} provides a more detailed theoretical discussion of the quench dynamics and associated transients.

\section{\label{sec:designer_trajs}"Designer" spin trajectories} 

The flexibility demonstrated in \zfr{emergence}--\zfr{amp_phase_hop} directly lends itself to the ability to construct dynamically changing \I{“designer”} spin trajectories. As an example, \zfr{collisions} demonstrates a pentagonal-trajectory (region $\T{I}$) that is dynamically changed to a square (region $\T{III}$) after a variable delay $\qt_{\ast}$  (region $\T{II}$). This is schematically described in \zfr{collisions}A. \zfr{collisions}B plots the resulting trajectories with decreasing $\qt_{\ast}$, until $\qt_{\ast}{=}0$ (\zfr{collisions}B(iii)), and ultimately to where the two shapes seemingly "collide" (\zfr{collisions}B(iv)). This idea is extended in \zfr{collisions}C where multiple square and pentagonal trajectories are repeatedly "opened" and "closed". Comparison with a single square trajectory (see \zfr{decay_profile}) reveals that the decays within each "open" region are slow and quantitatively match the decay during a single square trajectory.
However, the closing and reopening of the shapes leads to loss in signal. This is in agreement with theoretical findings, which predict a loss in signal by a factor ${1-\sigma}$, where $\sigma$ is a (small) parameter depending on the microscopic details, due to re-prethermalization along different axes (see App.~\ref{app:theory_quenches}).

While so far we have considered designer trajectories under quenched fields, its scope can also be expanded to adiabatically varying fields. As an example, in \zfr{adiabatic} we consider experiments under a frequency chirped field $\mHac(t) {=} \Bac\cos(\xo_\mathrm{AC}(t)t)\zhat$, where $\xo_\mathrm{AC}(t) {=} \xo_\mathrm{AC}(0) + \xd\xo t/T$ is a linearly ramped frequency, with $\xo_\mathrm{AC}(0){=}1$kHz, $\xd\xo{=}2$kHz, and $T{=}35$s (\zfr{adiabatic}A(i)). Its effect is to drive a rapid adiabatic passage~\cite{Garwood01} (RAP) of the spins prepared along $\xhat$ towards $-\xhat$. This is borne out by the measurements in \zfr{adiabatic}A, where we display the net amplitude signal $S(t)$. The sharp dip in the signal corresponds to the spins precessing close to the $\yz$ plane, and occurs when the instantaneous frequency crosses the dressed frame resonance $\xo_{\ac}(t) {\app}(t_{\R{pulse}}/\qt)\xO$. \zfr{adiabatic}B shows the tracked motion on the $\xhat\tm\zhat$ projection for chosen instants along the motion (see colorbar). Phase unwrapping of $\varphi$ becomes more challenging around the resonance condition (see discussion in SI~\cite{SOM}). \zfr{adiabatic}C-D shows the extracted Bloch reconstructions of the spins at two instants ($t{=}22$s and $t{=}25$s), on either side of the RAP resonance condition, clearly demonstrating the adiabatic flipping of the spins.

\section{\label{sec:outro}Conclusions and Outlook}

This work has great potential to be extended in diverse directions, both fundamental and applied. The ability to stabilize and continuously track spin orbits in interacting spin ensembles opens avenues to use them in quantum sensing, such as in magnetometry~\cite{Boss17,Schmitt17}, gyroscopes~\cite{Ajoy2012,Ledbetter2012,Jaskula2019,Jarmola21}, relayed NMR detection~\cite{Sahin21}, and dark matter searches~\cite{Budker14}. Our method may make \I{dense} spin ensembles practical for these use cases by exploiting thermalization-controlled micromotion dynamics. This would substantially expand the scope of quantum sensing beyond the conventional dilute (single-spin) limit~\cite{Barry2020,Zhou2020}. Beyond high-field magnetometry with $\Cs$ nuclear spins~\cite{Sahin21}, we anticipate that this approach can be extended to NV center ensembles~\cite{Wolf2015} and polar molecules~\cite{Micheli06} where analogous driving schemes may be implemented.  Quantum sensing may benefit even further from this control scheme because the symmetry imposed on the spin orbits (\zfr{polygons}) will allow higher selectivity for specific $\mHac$ frequencies of interest over noise and other frequencies, which do not produce any symmetry of the spin motion. Finally, spin orbits under a gradient may serve as an encoding strategy for hyperpolarized MR imaging~\cite{Lv21}. 

While our experiments have focused on hyperpolarized $\Cs$ nuclear spins in diamond, organic molecules which host triplet photoexcitable electrons~\cite{Henstra2014,Tateishi2014,Negoro2018} might prove even more compelling. For instance, Ref.~\cite{Niketic15} measured the $T_1$ lifetimes of optically hyperpolarized ${}^{1}$H nuclear spins in pentacene to be ${>}100$h at 77K. When applied to such systems, stabilized nuclear orbits might serve akin to “magnetic compass needles” with ultralow damping~\cite{Kimball16}, enabling continuously tracked magnetometry as described in this work. 

More fundamentally, the experiments here provide a new way to interrogate thermalization dynamics in interacting quantum systems~\cite{zu2021emergent}, and thereby provide a concrete means to verify predictions from the eigenstate thermalization hypothesis. Moreover, the use of Floquet prethermal plateaus has remained largely unexplored in the context of state manipulation thus far. The robustness of the spin orbits to errors (\zar{ampl_distortion}-\zar{flip_distortion}) suggests practical applications for quantum gates without needing to decouple interactions between the spins.

Finally, the principle for long-time quasi-continuous spin tracking developed here is generally applicable to wide classes of control fields $\mHc(t)$. As such, this portends impactful applications in quantum feedback~\cite{Lloyd2000,Vijay12}, Hamiltonian estimation~\cite{Burgarth17} and in the design of optimal control protocols~\cite{Khaneja05,Sakellariou00,Goodwin18,Ajoy12b}.

\section*{Acknowledgments}

We thank  C. Meriles, C. Ramanathan, J. Reimer, and D. Suter for valuable discussions. A.A. acknowledges funding from ONR (N00014-20-1-2806) and DOE STTR (DE-SC0022441). C.F. acknowledges support from the European Research Council (ERC) under the European Union’s Horizon 2020 research and innovation program (grant agreements No. 679722 and No. 101001902). M.B. was supported by the Marie Sklodowska-Curie grant agreement No 890711. Computational work reported on in this paper was performed on the MPI PKS and W\"urzburg HPC cluster.


\begin{appendix}

\begin{figure}[t]
  \centering
  {\includegraphics[width=0.35\textwidth]{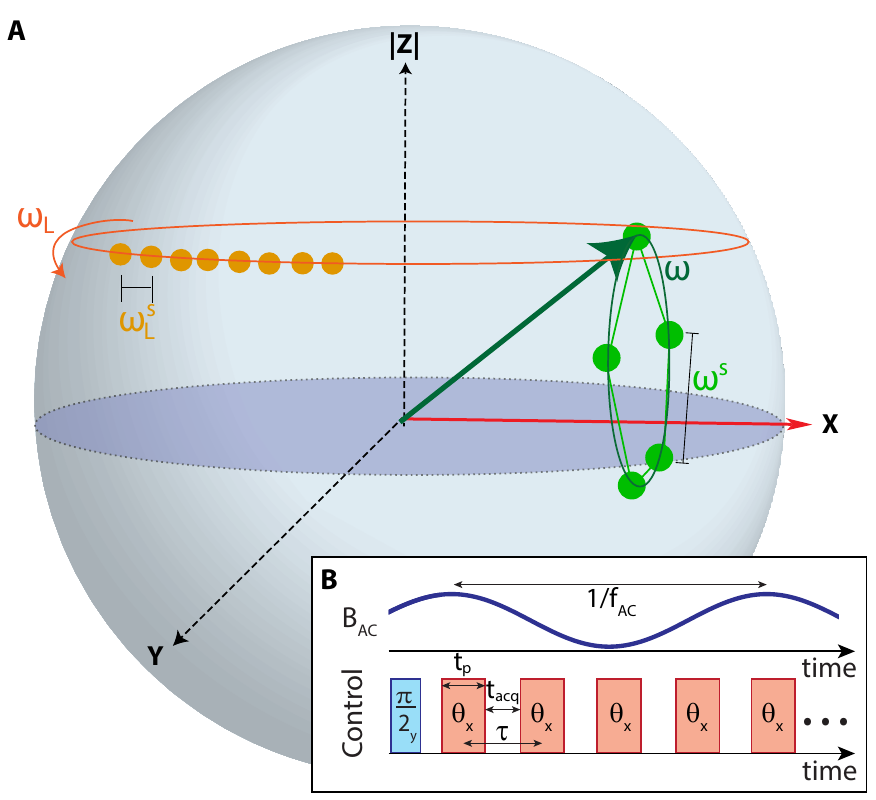}}
  \caption{\T{Spin tracking principle} shown for AC field $\mHac$ applied simultaneously with a spin-locking field along $\xhat$ (schematic in (B)). (A) In the rotating frame, spins (green arrow) undergo orbital motion (green circle) under the joint action of both fields; the AC field frequency $\xo_{\R{AC}}$ is matched to that of a full precession, which completes in time $(\xo/(2\pi))^{-1}{=}(t_{\R{pulse}}/\qt)\xO$. Trajectory is sampled at rate $\xo^s$ (green points) after each SL pulse; at resonance, $\xo^s {=} (2\pi/\xt)\xo$ leading to the polyhedrally sampled trajectories in \zfr{polygons} (here a pentagon is sampled). For each such sample, the spins are interrogated by an RF cavity via spin precession in the lab frame (orange circle), at rate $\xo_L$. This precession is sampled at rate $\omega_L^s$ (orange points). Hierarchical time-scale separation (see \ztr{table}) ensures the quasi-continuous tracking of the trajectory at $\omega^s$ with high SNR.}
	\zfl{tracking_concept}
\end{figure}

\begin{table}[t]
\begin{tabular}{|c|c|c|c|c|}\hline\hline
Parameter   & Particular & Value  & Norm. Value   \vspace{-3pt}\\
 &  &  &  \\\hline
 $(T_{2}')^{-1}$   & Prethermal decay rate & $\sim$10mHz  & $10^{-2}$\\
$\xg\Bac$   & Strength of AC field & 0.01-1kHz  &
1\\
$\xo/(2\pi)$ &  Dressed frame res. freq.& 0.1-20kHz  & 10-2000   \\ 
$\xo^s/(2\pi)$ & Trajectory sampling rate  & 1kHz-1MHz  & $10^2-10^5$ \\
$\xo_L/(2\pi)$& Cavity resonance frequency  &
75MHz& 7.5$\zt10^6$ \\
$\xo_L^s/(2\pi)$ & Sampling of cavity evolution  & 1GHz  & $10^8$ \\
\hline\hline
\end{tabular}
\caption{\T{Table elucidating time-scale separation} between parameters of spin evolution and cavity interrogation (see \zfr{tracking_concept}). Typical values employed in experiments are listed in third column, while the last column shows them normalized with respect to $\Bac$. There is a ${\sim}10^5$-fold separation between $\xo$ and $\xo_L^s$ that allows quasi-continuous sampling. }\ztl{table}
\end{table}

\begin{figure*}[t]
  \centering
 {\includegraphics[width=0.95\textwidth]{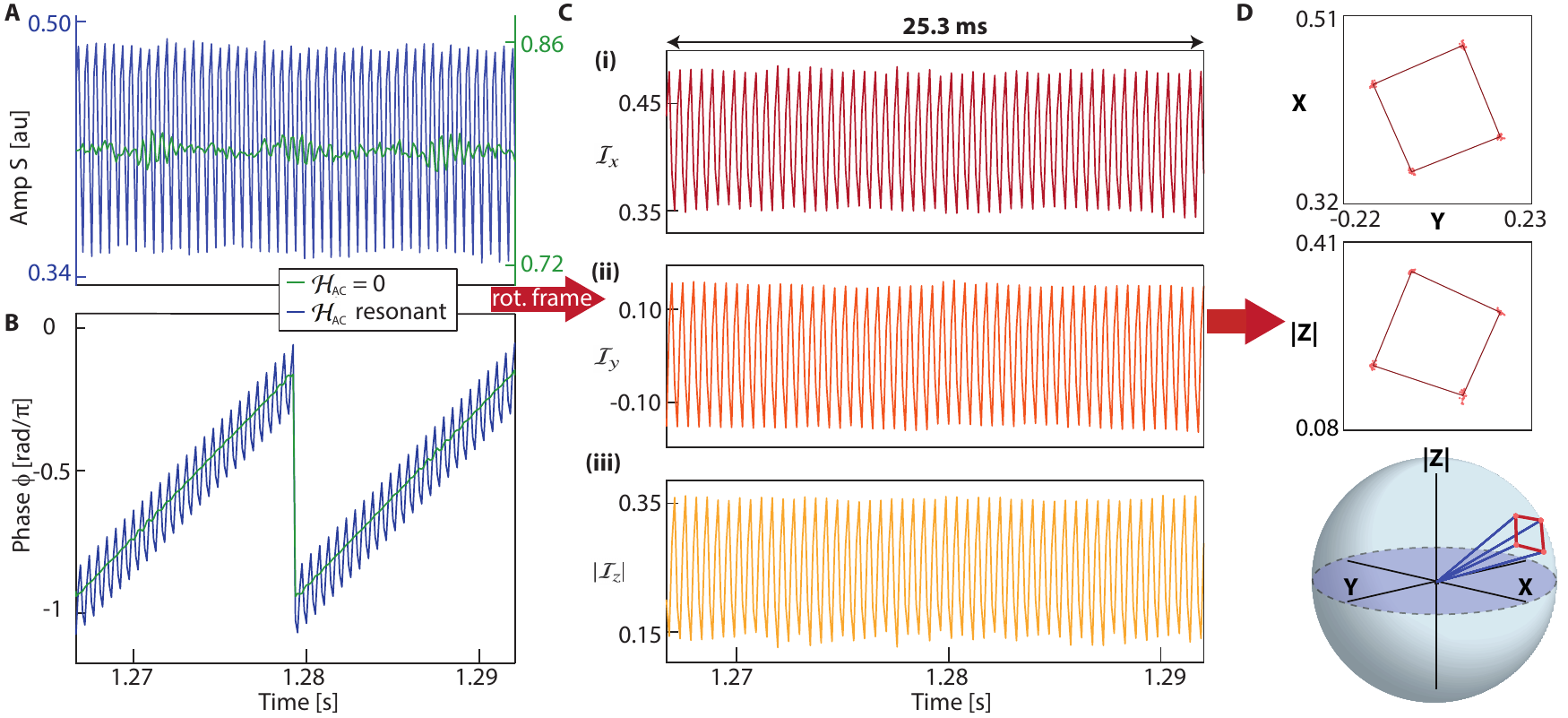}}
 \caption{\T{Principle.} (A-B) \I{Amplitude and phase response} showing signals $S$ and $\varphi$ respectively for a representative $\xD t{=}25.3$ms window under: (i) no applied control field $\mHac(t){=}0$ (green), and (ii) a resonant sinusoidal AC field at $\xo_{\R{AC}}{=}(2\pi) 1953.125$Hz (blue). Sequence here employs $\xt{=}\pi/2$, $\xo^s{=}(2\pi)$ 7.81kHz, and full data is collected over 35s (see \zfr{longtime}). In the former case ($\mHac(t){=}0$), spins are locked along $\xhat$. Small wiggles are due to 60 Hz noise pickup (see SI~\cite{SOM}). Ramp-like phase pattern indicates phase accrued during pulses. (C) \I{Dressed frame cartesian coordinate} trajectory components. Here (i)-(ii) $I_{x,y}$ components are unwrapped from data in (A-B), and (iii) $|I_z|$ is calculated via a unitary constraint. (D) \I{Bloch sphere representation} of data in (C) shown as 2D projections (upper panels), and in three dimensions (lower panel). Points display the sampled rotating frame trajectory over $\xD t$, which manifest in four distinct sets of points. Tracing the centroid of the four sets of points results in a trajectory resembling a quadrilateral.}
\zfl{fig2}
\end{figure*}

\begin{figure}[t]
  \centering
  {\includegraphics[width=0.5\textwidth]{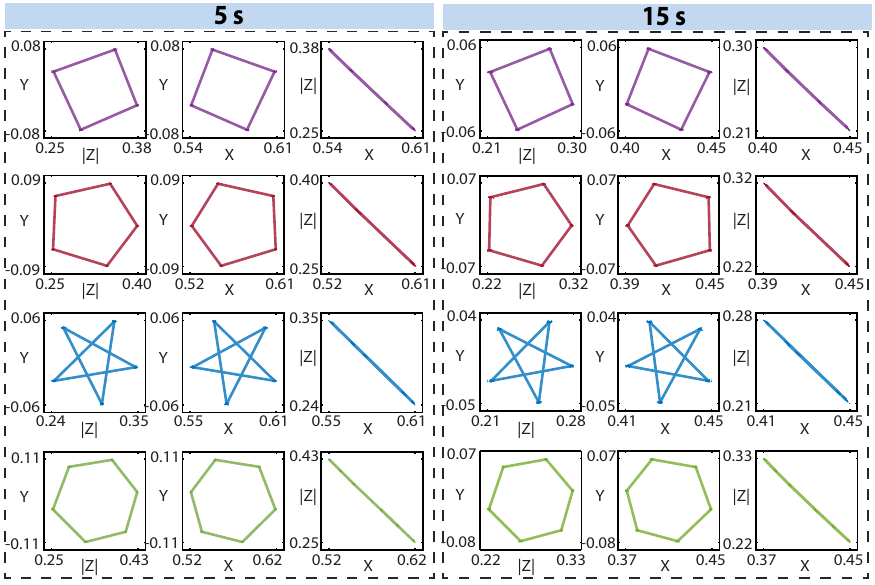}}
    \caption{\T{Bloch sphere trajectories} for representative tracked polyhedral spin orbits in \zfr{polygons}C shown as individual projections on the $\yz, \xy$, and $\xhat\tm\zhat$ planes. We focus on two $\xD t{=}$40ms windows of data starting at $t{=}5$s and $t{=}15$s respectively. Data here is shown after 15 averages. Points show the reconstructed spin positions on the Bloch sphere while solid lines join their respective centroid. Stability of the motion is reflected in the relatively small decay between the panels at different time points. Arc-like projections in the $\xhat\tm\zhat$ plane indicate that trajectories are akin to “tilted” 2D shapes plastered on the 3D Bloch sphere. Data here is shown as a movie at Youtube link~\cite{Bloch_movie}.}
    \zfl{movie}
\end{figure}

\section{Principle of Continuous Spin Trajectory Tracking}
\zal{track_principle}
In this section we describe the principle of spin tracking employed in this paper; schematically shown in \zfr{tracking_concept}A. Indeed, considering the protocol in \zfr{tracking_concept}B, the pulsed SL pulses \I{dress} the spins such that the energy gap $\xo{=} \eta\xO$, where $\xO$ is the Rabi frequency, and $\eta{=}t_{\R{pulse}}/\qt$ is the pulsing duty cycle. Micromotion under the action of the control field $\mHac(t)$ results in spin evolution in a rotating (dressed frame) trajectory. In our experiments, this rotating-frame spin trajectory is ``sampled” quasi-continuously at a rate $\xo^s/{2\pi}{=}\qt^{-1}$ (green points). Practically, this is accomplished by interrogating the spins via an RF cavity by allowing them to precess in the lab frame (orange circle) at Larmor frequency $\xo_L$ in $\tacq({\sim}\qt/2)$ windows between the SL pulses. Crucially, cavity interrogation is arranged to be very rapid: if the rotating-frame trajectory is sampled at rate $\xo^s$, then for each such sample the cavity readout is carried out multiple times, sampled at a rate $\xo_L^s$ (orange points) such that $\xo_L^s/\xo^s{>}10^4$. Our strategy therefore revolves around ensuring a hierarchal time-scale separation (see \ztr{table}) such that, $\xo_L^s{\gg} \xo_L{\gg} \xo^s{>} \xo {>} \Bac$. This leverages the relatively low $\xo_L$ Larmor frequencies of the nuclear spins, and their ability to be driven for long periods ($T_2^\prime$), but interrogated rapidly (at $\xo_L^s$). \ztr{table} shows typical values as employed in experiment; the orders of magnitude overall time-scale separation proves the key to \I{quasi-continuous} state tracking.

To see this more clearly, consider some comparisons in \ztr{table}. First, the wide separation $\xo_L/\xo{\gg}1$, allows the applied AC fields at $\xo_{\ac}({\app}\xo$) to be filtered from the cavity response ($\xo_L$). This permits RF spin interrogation to proceed simultaneously with the application of the AC field $\mHac(t)$. Second, the hierarchy $\xo^s{\ll}\xo_L$ means that cavity interrogation can be carried out for multiple Larmor cycles for each sampled point in the rotating-frame spin trajectory (\zfr{tracking_concept}A). This yields a high readout  signal-to-noise (SNR) per sampled point on the trajectory, and $\xo^s{\gg}(T_2^\prime)^{-1}$ means that the tracking is for most practical purposes, quasi-continuous. Finally, the large ratio $\xo_L^s/\xo_L$ means it is possible to carry out exquisite filtration of the cavity response, suppressing all components except those exactly at $\xo_L$, also yielding SNR gains. 

\section{\label{app:spin_reconstruction}3D Spin Reconstruction Strategy}
\zal{track_method}
Following the principle employed in \zar{track_principle}, let us now describe how the position of the spins is tracked in three-dimensions on a Bloch sphere. Methodologically, we rely on digital homodyne detection of the cavity response at frequency $\xo_L$ (see Methods ~\cite{SOM}). In particular, a Fourier transform (FT) of the cavity readout in each SL readout window (separated by $\qt$) is carried out, and the amplitude and phase at $\xo_L$ report on the components of the spin vector along the equatorial $\xy$ plane in \zfr{tracking_concept}A. For clarity, we will use $t_j$ to denote the quasi-continuous time variable discretized by $\qt$ in these measurements. \zfr{fig2}A-B then illustrates the data obtained in a 25.3ms time window with $\xt{=}\pi/2$ for two exemplary cases: \I{(i)} first, in the absence of the AC field, i.e., $\mHac{=}0$ (green line) i.e., and \I{(ii)} second, with an AC field applied on resonance, $\mHac{=}\Bac\cos(\xo_{\R{res}} t)$ (blue line), the case of the “square-trajectory” in \zfr{polygons}C(i). When $\mHac{=}0$, the constant amplitude signal $S(t_j)$ reflects the spins being locked along $\xhat$. However, considering the phase (\zfr{fig2}B), ramps arise due to a trivial phase accrual under Larmor precession during each $\xt$-pulse. Removing this phase via a linear fit (see SI~\cite{SOM}) allows us to extract the effective phase evolution in the dressed frame, $\varphi(t_j)$. Combining information from the two quadratures then allows direct access to the transverse projections (see \zfr{fig2}C), $\mI_{x}(t){=}S(t)\cos(\varphi(t))$, and $\mI_{y}(t){=}S(t)\sin(\varphi(t))$, where $\mI_{\nu}$ specifies magnetization in the rotating frame. $\mI_{x}$ and  $\mI_{y}$ for the square trajectory AC field case are displayed in \zfr{fig2}C (i-ii). The four-point periodic oscillations reflect micromotion driven orbital motion in the rotating frame.

In principle, $\mI_{x}$ and  $\mI_{y}$ is the maximal extent of information that can be extracted from these measurements.  However, as is evident in \zfr{fig2}C, the spins undergo very little decay during successive cavity interrogation windows, $(\xo^s T_{2}^{\prime})^{-1}{\ll} 1$. This allows the ability to extract the full 3D vector $\vec{\mI}$ information from only homodyne measurements of signal and phase, by reasonably estimating also the $I_{z}(t_j)$ component (sans its sign) using, $|I_{z}(t_j)| = [\mN^2(t_j) - I_{x}^2(t_j)- I_{y}^2(t_j)]^{1/2}$  (\zfr{fig2}C(iii)), where norm $\mN(t) {<} 1$ accounts for spin decay up until time $t$. Following \zfr{amplitude}E, the decay profile $\mN(t)$ reflects the prethermalization process. Combining this information to form $\vec{I}(t_j)$ allows us to plot (track) the dressed-frame dynamics on a Bloch hemisphere, shown in \zfr{fig2}D for the individual projections on the $\xy$ and $\yz$, and in three dimensions on a Bloch sphere. In the 25.3ms period shown, there are ${\sim}50$ precessions about $\zhat$ (see \zfr{fig2}D); the points here display the $\xo^s$ samples (\zfr{tracking_concept}A) and the lines join their centroids for clarity. This manifests as the quadrilateral spin projections observed in \zfr{fig2}D (also highlighted in \zfr{longtime}D). 

We now elucidate upon the assumptions employed for this reconstruction. The instantaneous norm of the spin vector $\mN(t)$ is hard to measure quantitatively because a drop in the signal $S$ can arise both due to relaxation and the tilt of the nuclear spin vector away from the $\xy$ plane, which are hard to distinguish. Although the prethermal lifetimes are almost identical in the absence and presence of $\mHac$, there is still slight variation with the magnitude of $B_{\R{AC}}$ employed (\zfr{amplitude}F). This makes an absolute normalization strategy employing the $\mHac{=}0$ case as a reference difficult. To circumvent this problem, we focus instead on only imposing that the \I{relative} normalization between two time instants in the spin trajectory are accurate. For this, in the long-time reconstructions in this paper (e.g. in \zfr{longtime}), we assume $\mN(t)$ can be evaluated by a small offset $S^{\R{dec}}(t)+\xd_0$, where $\xd_0{=}0.05S^{\R{dec}}(t{=}0)$. While this yields a (constant) error (${\propto} \xd_0^2$) in the absolute values of the projections on $\xhat$ axis, it does not affect the relative relationship of spin trajectory projections at different instants $t_j$ of the spin evolution, still yielding a semi-quantitative real-time 3D map of the spin trajectories. This is illustrated by comparison of the two panels in \zfr{longtime}D, where the relaxation-mediated shrinking of the Bloch sphere trajectories are evident at long times. 

\section{Movies and projections corresponding to polyhedral trajectories}
\zal{movie}
As a complement to \zfr{polygons}C of the main paper, \zfr{movie} presents all three projections of the excited spin trajectories, and compares the motion at $t{=}5$s and $t{=}15$s similar to \zfr{longtime}D. The panels once again emphasize that the orbital motion is highly stable. A special feature of the visualization here is that the $\xhat\tm\zhat$ projections of the motion can be seen to be arc-like. This illustrates that the motions in \zfr{polygons} are akin to 2D polygonal shapes pasted atop the 3D Bloch sphere. Movies of the tracked motion can be seen on Youtube~\cite{Bloch_movie}.

\begin{figure}[t]
 \centering
 {\includegraphics[width=0.5\textwidth]{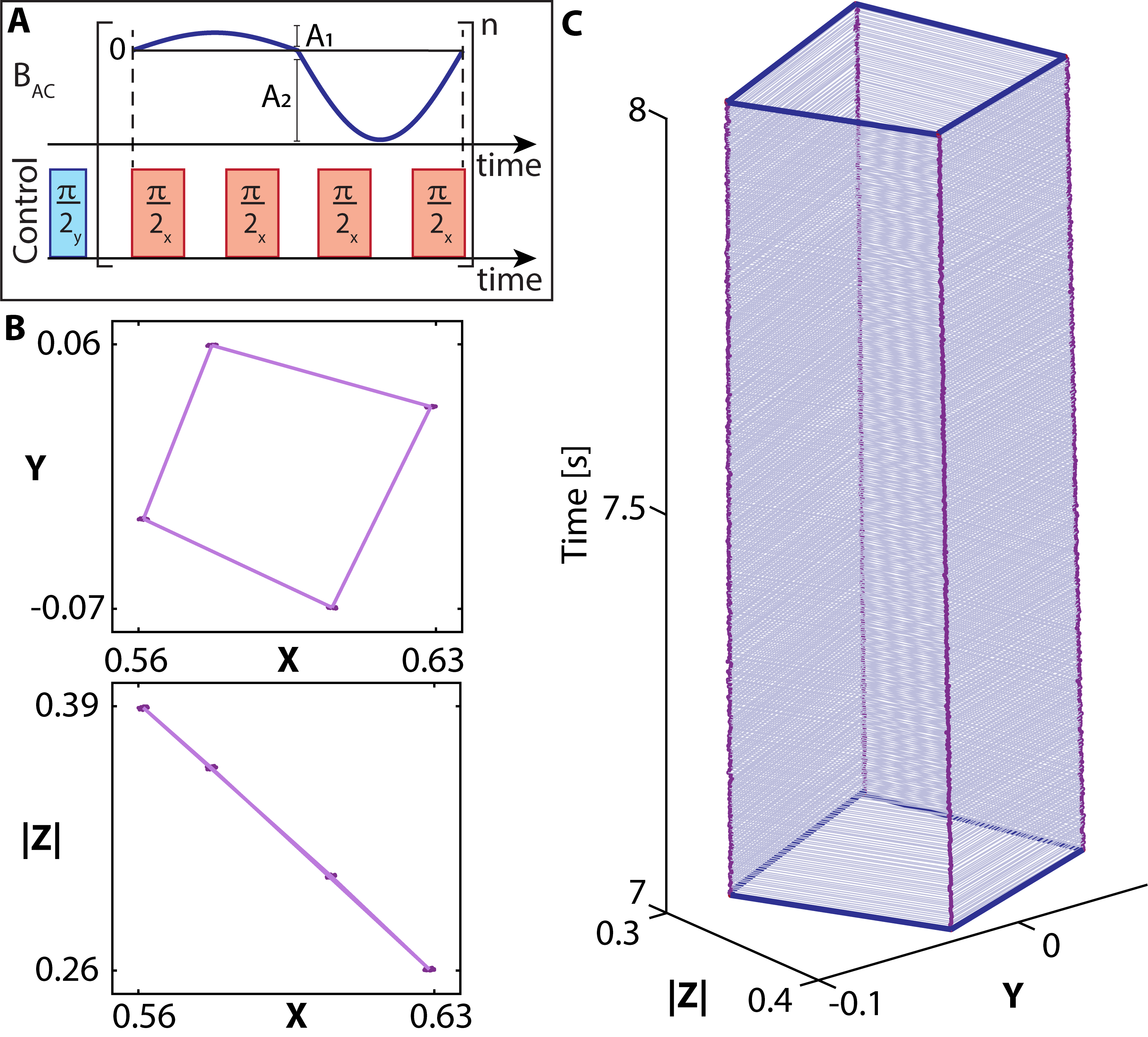}}
  \caption{\T{Robustness of spin trajectory to variations in AC field.} Tracked spin trajectory for a representative “distorted” sine-wave AC field. (A) Schematic of experiment; $\xt {=} \pi/2$ and the AC field is arranged to be resonant. Here, the second half-period has amplitude $A_2 {=} 5A_1$. (B-C) Tracked spin trajectory in a $\yz$ plane over time on the vertical axis (C), and as individual $\xy$ and $\xhat\tm\zhat$ projections (B) at $t {=}$5s. Trajectory traced remains “square-like” in spite of the introduced distortion.}
	\zfl{robust}
\end{figure}

\begin{figure}[t]
  \centering
  {\includegraphics[width=0.5\textwidth]{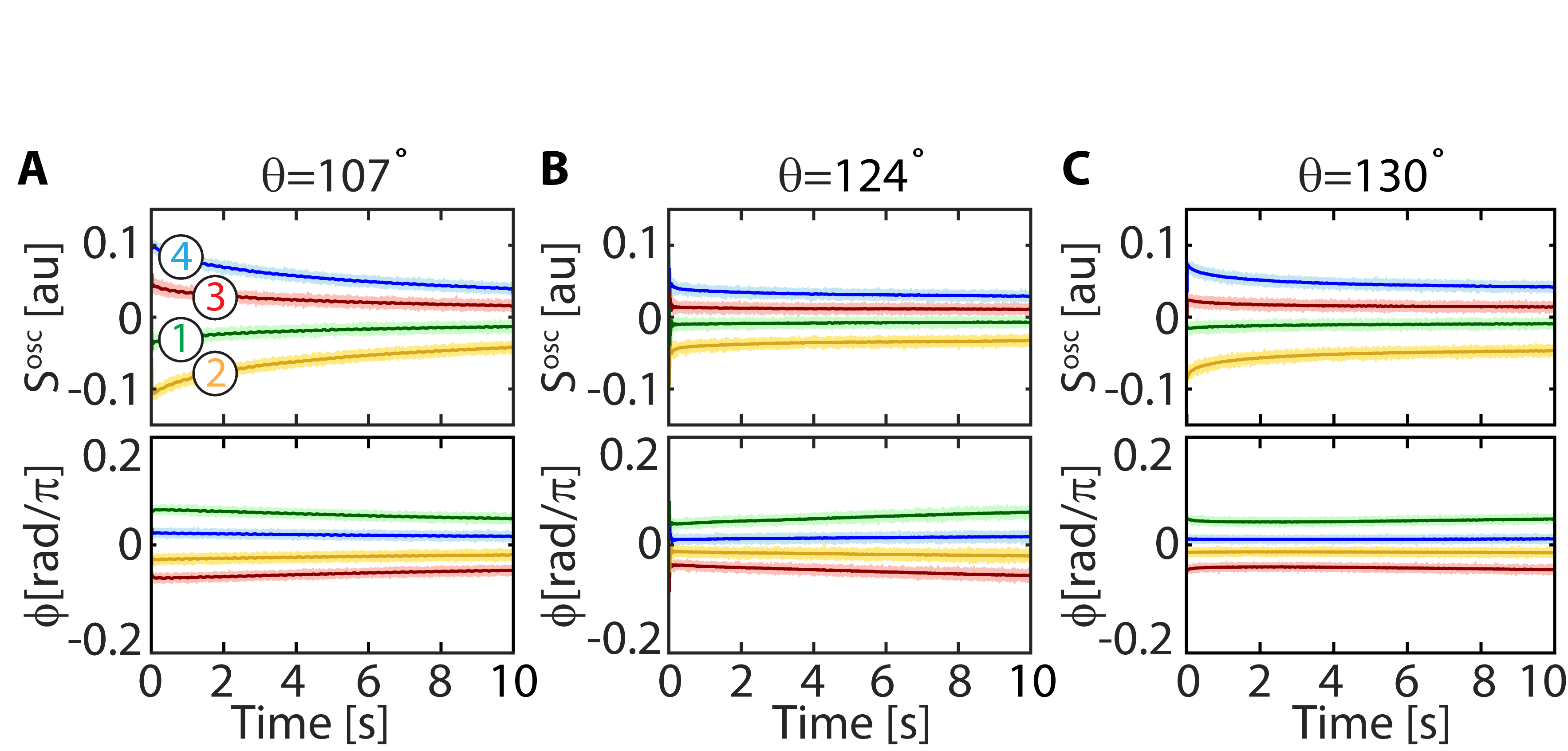}}
    \caption{\T{Robustness to deviations in SL flip angle.} Shown are characteristic plateaus shown for signal $S^{\R{osc}}$ and phase $\xph$ similar to \zfr{longtime}C of the main paper, but for SL pulse angles $\xt$ chosen to be $\xt{=}\{107^{\circ},124^{\circ},130^{\circ}\}$ respectively. AC field frequency is chosen to be resonant in each case, $\xo_{\ac}{=}\xo_{\res}$ such that one AC period completes in exactly $4\qt$. Four parallel plateaus (labeled circled 1-4 as in \zfr{longtime}) are visible indicating stable square-like spin orbits in spite of the $\xt$ deviating from $\pi/2$. }
    \zfl{flip_angle}
\end{figure}

\begin{figure}[t]
 \centering
 {\includegraphics[width=0.5\textwidth]{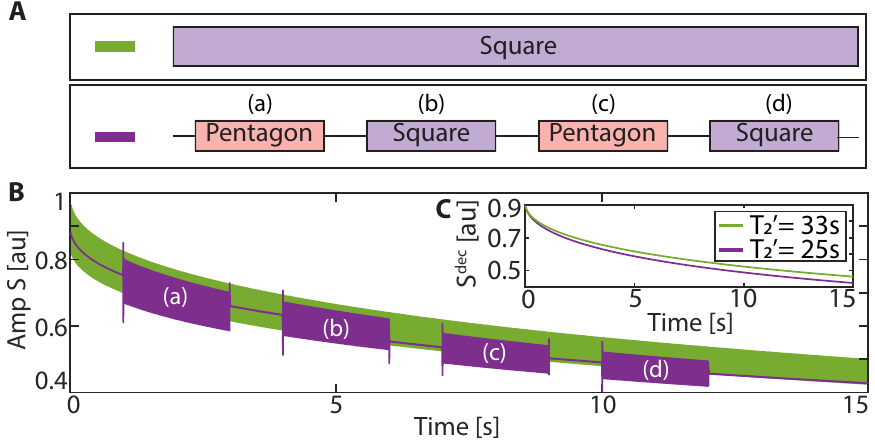}}
  \caption{\T{Decay of designer spin trajectory.} Comparison of decay of \zfr{collisions}C, consisting of multiple separated  pentagons and squares, to that of a single square of the same $\Bac$ amplitude. (A) Schematic of experiments being compared (cf. \zfr{polygons}C(i) and \zfr{collisions}C). (B) Raw signal $S$ plotted in both cases. Motion under $\Bac$ appear as bands because the oscillations (see \zfr{longtime}A) are rapid compared to the total 15s period under study. Envelope of oscillations allows a means to compare both cases, and demonstrate that the signal decay in each $\Bac$ window closely matches that of the original square trajectory. Transient upon each Hamiltonian quench event is clearly visible. (C) Comparison of decay profiles $S^{\R{dec}}$ in both cases, obtained by smoothing respective data in (B). We measure a relative stretched exponential decay constant of $T_2^\prime{=}32.6\pm0.3$s and $T_2^\prime{=}25.3\pm0.1$s for the trajectories, demonstrating that the decays remain relatively unaffected even after multiple $\Bac$ switching events. }
	\zfl{decay_profile}
\end{figure}

\section{Robustness of trajectories to amplitude distortion}
\zal{ampl_distortion}
A remarkable feature of the spin orbits is that they are robust to the exact functional form of the AC field.  Indeed, it is possible to obtain orbits similar to \zfr{polygons}C even when the AC field is significantly distorted from a sinusoid. An immediate consequence of this resilience is that amplitude fluctuations (noise) in the AC field have very little effect to the obtained spin orbits.

To illustrate this, in \zfr{robust} we carry out experiments with an exemplary AC field that is constructed by combining two half-period sinusoids with a five-fold amplitude ratio (schematically shown in \zfr{robust}A). $\xt {=}\pi/2$ SL pulses are employed here, and the overall period of the AC field is resonant: one full AC field period matches that of four SL pulses ($\xo_{\ac}{=}2\pi/(4\qt)$). We then carry out spin tracking on the resulting orbits; as the projections in \zfr{robust}B indicate, the obtained trajectories continue to closely resemble a “square”, as expected for a pure undistorted sinusoid as in \zfr{polygons}C(i). The obtained trajectory also remains equivalently rigid, as evident in tracked data in \zfr{robust}C.

The origin of this robustness can be seen from Eq.~\eqref{eq:w_k} where the vector family $\tilde{\boldsymbol{w}}_k$ is constructed out of functions $f_n(\xph_{\ac})$) that are evaluated as an integral over the \I{entire} period $T_{\R{period}}$ of the AC field, and not on the instantaneous value of the amplitude $\Bac(t)$ in-between the pulses (see also \zar{theory_floquet}). Hence, a distortion in $\Bac(t)$ only affects the area enclosed, but not the orbital shape itself (see also \zar{flip_distortion}).

\section{Robustness of trajectories to SL flip angle error}
\zal{flip_distortion}
In this Appendix, we demonstrate that the “resonance condition” as employed in the main paper can be very relaxed, offering a high degree of robustness with respect to the choice of the SL flip angle $\xt$. Indeed, the ability to apply ${>}250$k pulses in the presence of RF inhomogeneity means that the obtained stable orbits are highly robust to $\xt$. In this section, we consider experimentally the situation when $\xt{\neq}2\pi/N$, but deviates slightly from it, while remaining the same for all applied pulses. In the Supplementary Information ~\cite{SOM}, we consider the alternate situation where the pulses have small differences in flip-angle between them.

\zfr{flip_angle} shows experiments for three representative values of $\xt{=}\{107^{\circ},124^{\circ},130^{\circ}\}$. In each case we satisfy a \I{relaxed} version of the resonance condition, only requiring that one AC period completes in a total period $N\qt$, i.e. $\xo_{\ac}{=}\xo_{\res}{=}2\pi/(N\qt)$. Experimentally, this condition is easier to satisfy because it only requires a precise period matching between the AC field and the SL pulses.

In \zfr{flip_angle}, we demonstrate this for a representative case of $N{=}4$ and use the notation employed in \zfr{longtime}, showing signal $S^{osc}$ and phase $\xph$ over a 10s period. The four plateaus demonstrate that the stable square-like trajectory forms in this case, even though the pulses deviate from $\pi/2$. Overall, \zfr{robust} and \zfr{flip_angle} demonstrate that stable spin orbits are robust against error in both the AC field amplitude and SL pulses. As long as the SL pulses and the AC field are period-matched (mutually resonant), the spin orbits will remain identical and stable. We envision this feature will have important implications for quantum sensing with hyperpolarized nuclear spins.

\section{Stability of designer trajectories}
\zal{designer_decay}
We now demonstrate that the designer trajectories we engineer, for instance in \zfr{collisions}C with multiple “opening” and “closing” events of polyhedral spin excursions, are inherently robust. To see this, \zfr{decay_profile} shows a comparison between the decay of the trajectory in \zfr{collisions}C, tracked for $t{=}15$s, to that of a single square trajectory over the same period (see schematic in \zfr{decay_profile}A). The designer trajectory here consists of pentagonal and square excursions created with $\xo_{\R{AC}}{=}(2\pi)$1801.8018Hz and $\xo_{\R{AC}}{=}(2\pi)$1953.125Hz respectively, which are switched on for $\xD t{=}2$s in an alternating fashion. Plotted is the full raw signal $S$; the rapid AC field driven oscillations appear here as apparent bands in the signal as shown in \zfr{decay_profile}B.

It is evident that every switching event $\I{((a)-(d))}$ in \zfr{decay_profile}B yields a transient. The “closing” transient results in pre-thermalization of the spins so that only the component collinear with $\xhat$ is retained (see also \zfr{emergence}C(ii)). Since the switching instants are not exactly matched to a complete period to either $\xo_{\R{AC}}$, this leads to a step-like signal jump at every closure. Similarly, an opening transient results in prethermalization to the corresponding $\tilde{\boldsymbol{w}}_k$ trajectory in Eq.~\eqref{eq:w_k}. Importantly however, as \zfr{decay_profile}B illustrates, the overall decay profile during the excursion periods closely matches that of the original pentagonal one, as can be seen by comparing the envelope of the respective decays in \zfr{decay_profile}B ($S^{\R{dec}}$). In \zfr{decay_profile}C we plot the smoothed version of the signal in \zfr{decay_profile}B, where the individual decays follow stretched exponentials with $T_2\prime{=}32.6$s and $T_2^\prime{=}25.3$s respectively. Overall, spin trajectories are highly robust even when subject to multiple AC field switching events.


\begin{figure}[t]
 \centering
 {\includegraphics[width=0.49\textwidth]{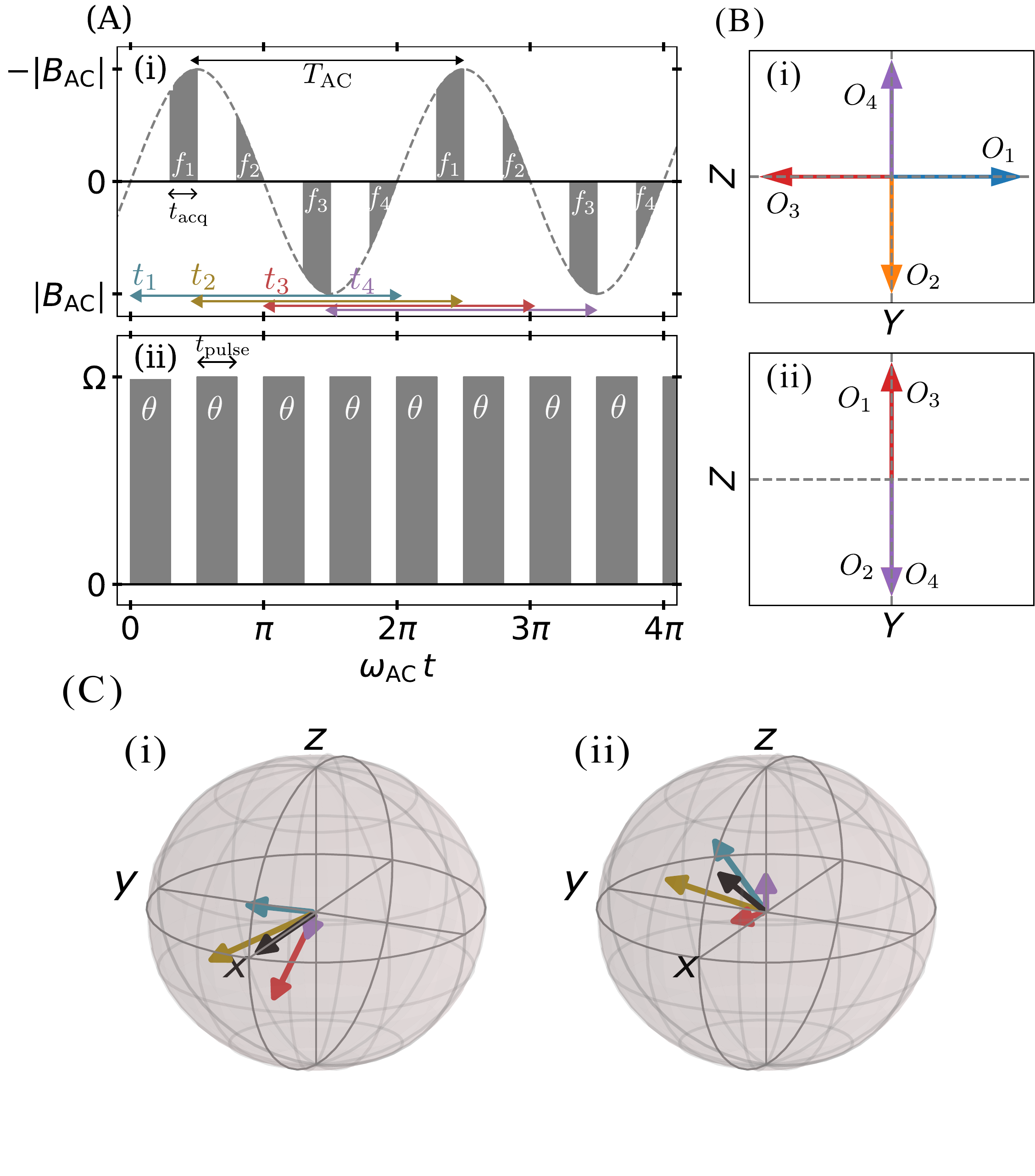}}
 \caption{
 \T{Theoretical evaluation of spin driving protocol}. (A) \I{Schematic.} Panels show the AC field and SL pulses (see \zfr{concept}B). For simplicity here $N{=}4$ SL pulses are matched to one AC period $T_\mathrm{AC}{=}2\pi/\omega_\mathrm{AC}$.   Shaded regions respectively denote time regions where AC and kicks dominate. Functions $f_n$, encapsulating the effect of the AC field during the $\tacq$ windows, are shown. Families $k$ correspond to evaluating the action of the sequence considering different starting points, $t_k{=}T_\mathrm{AC}k/N$, schematically represented by the colored arrows. (B) Schematic representation of operators $O_n{=}U_{\R{SL}}^{-n}I_z U_{\R{SL}}^{n}$ for pulses $U_{\R{SL}}$ throughout the sequence, shown for the case of (i) $\xt{=}\pi/2$ and (ii) $\xt{=}\pi$. (C) \I{Bloch sphere representation} of corresponding magnetizations $\boldsymbol{M}$, plotted for (i) vanishing detuning ($\xd{=}0$), and (ii) with finite detuning $\xd$. Colored vectors correspond to case $\xt{=}\pi/2$, whereas the black vector corresponds to the case $\xt{=}\pi$. In the latter case, all four plateaus have equal magnetizations. Detuning serves to tilt the orbits away from the $\xhat$ axis.}
\zfl{sketch_theory}
 \end{figure}

\section{\label{app:theory_model}Theoretical model and rotating frame Hamiltonian}

In this section, we present a derivation of the rotating frame Hamiltonian in \zr{rot_hamiltonian}. Consider first that the lab frame Hamiltonian of the interacting system of $\Cs$ nuclear spins is given by,
\begin{equation}
    \mathcal{H}_\mathrm{lab}\lb t\rb = \omega_L I_z + \mHdd + \mathcal{H}_z + \mathcal{H}_\mathrm{AC}\lb t \rb + \mathcal{H}_{\R{SL},\, \mathrm{lab}}\lb t \rb \, \zl{lab_hamiltonian}
\end{equation}
with the Larmor frequency $\omega_L{=}\xg_nB_0$, where $\xg_n$ is the magnetogyric ratio and $B_0{=}7$T is the bias magnetic field. \zr{lab_hamiltonian} also includes the dipole-dipole interaction $\mHdd$, the time-fluctuating on-site fields $\mathcal{H}_z$, and the AC drive $\mathcal{H}_\mathrm{AC}$, as defined in Eq.~\eqref{eq:rot_subhamiltonians}. Now, in the lab frame, the pulsed spin-locking~(SL) drive can be described by the Hamiltonian,
\begin{equation}
    \mathcal{H}_{\mathrm{SL},\, \mathrm{lab}}\lb t \rb = \cos(\omega_\mathrm{SL} t) \Omega \Theta\lb t \rb I_x \, ,
\end{equation}
where we assume the carrier frequency is applied close to the Lamor frequency resonance, $\omega_\mathrm{SL}=\omega_L+\delta$, with the small detuning given by $\xd$ (with $\delta/\omega_L{\ll}1$).

As is typical at high field, the Larmor frequency $\omega_L$ exceeds all other energy scales of the Hamiltonian~\zr{lab_hamiltonian} by at least three orders of magnitude (see \ztr{table}). We can remove this large energy scale by changing to a rotating frame with respect to $W\lb t \rb = \exp(-i (\omega_\mathrm{SL} t + \phi ) I_z )$.
In principle, a generic transformation to the rotating frame can also have a static (i.e., change-of-basis) contribution, which introduces an offset in the experimentally observed phase $\phi$ [for more details see SI~\ref{SOM:xy_phase}]; for simplicity, we will choose $\phi{=}0$ hereinafter.
Note that the transformation $W(t)$ also affects $\mH_{\mathrm{SL},\, \mathrm{lab}}$ which is transformed to $\mH_{\mathrm{SL}}=\Omega \Theta\lb t \rb I_x$ upon performing a rotating wave approximation. 
Taking all these considerations into account, it is straightforward to arrive at  the rotating frame Hamiltonian as specified in \zr{rot_hamiltonian},
\begin{equation*}
    \mH_{\R{rot}}(t) =  \mH_z + \mHdd +  \delta I_z + \mH_{\R{SL}}(t)+\mHac(t) \, .
\end{equation*}

We now specify the assumptions we will employ to aid the theoretical analysis. First, the on-site random fields $\mH_z$ play an insignificant role for the dynamics of the system, and as such we will ignore them. Second, while the detuning $\delta$ is small compared to the Larmor frequency $\omega_L$, we emphasize that it can be comparable to, or larger than, the remaining energy scales in the Hamiltonian, i.e., our experiments can be carried out even in the regime where $\xd{\gtrsim} J$. In fact, a finite detuning is crucial to theoretically capture the experimentally observed tilt of the spin orbits away from the $\xhat$ axis and towards the $\yz$ plane, see App.~\ref{app:theory_floquet} and Ref.~\cite{SOM}. Finally, an important hierarchy of energy scales in the rotating frame Hamiltonian \zr{rot_hamiltonian} is that the SL Rabi frequency $\Omega$ exceeds all other scales by at least an order of magnitude: $\Omega{\gg} J {\approx} \left| \mHac \right|$. Therefore, it is sufficient to keep only the contribution of the $\xhat$-kicks and detuning $\delta$ during the SL pulses. These approximations lead directly to the Hamiltonian in Eq.~\eqref{eq:Hamiltonian}.

\section{\label{app:theory_floquet}Derivation of the leading-order Floquet Hamiltonian}
In the following, we derive the family of Floquet Hamiltonians $\mH_{F;\, k}^{(0)}$, given in Eq.~\eqref{eq:H_F} of the main text, starting from the simplified Hamiltonian $\mH_{\R{rot}}(t)$ in Eq.~\eqref{eq:Hamiltonian} and employing the assumptions elucidated above (see \zar{theory_model}).

Our discussion will be based on the sequence in \zfr{sketch_theory}A for clarity. In general, there exist $N$ distinct unitaries  $U_{\mathrm{acq},\, k+j}$ (labelled by index $k$,  $k{=}1,\,{\cdots},\,N$) corresponding to the different instants $t_k{=}T_\mathrm{AC}k/N $ that we consider to represent the start of a period. The Floquet unitary describing the time evolution over one AC period $T_\mathrm{AC}{=}2\pi/\xo_\mathrm{AC}$ can be written as
\begin{equation}
	\begin{aligned}
		U_{F;\,k}  &= U_{\mathrm{acq},\, k+N} U_\mathrm{SL} \dots U_{\mathrm{acq},\, k+2} U_\mathrm{SL} U_{\mathrm{acq},\, k+1} U_\mathrm{SL} \\
		&\equiv \prod_{j=1}^N U_{\mathrm{acq},\, k+j} U_\mathrm{SL} \label{eq:floquet_unitary},
	\end{aligned}
\end{equation}
where $U_{\mathrm{acq},\, j }\!=\!\exp(-it_\mathrm{acq} \mH_\mathrm{acq})$ and $\mH_\mathrm{acq}\!=\!\mH_\mathrm{dd} + \delta\, I_z + f_j(\phi_\mathrm{AC}) I_z$; $f_{n}(\phi_\mathrm{AC}){=}\int_{t_n + t_\mathrm{pulse}}^{t_{n+1}} \sin\left(\frac{2\pi t}{T_\mathrm{AC}} +\phi_\mathrm{AC} \right)  \mathrm{d}t/{t_\mathrm{acq}}$ is the effective phase developed due to evolution between two successive SL pulses. This is highlighted in \zfr{sketch_theory}. The unitary operator modeling the action of the SL pulses can be written as,
\begin{equation}
	\begin{aligned}
		U_\mathrm{SL}   &= \exp(-i t_\mathrm{pulse}  \lsb \Omega I_x + \delta I_z \rsb )    \\
		&= \exp(-i \xt  \lsb \cos(\alpha) I_x + \sin(\alpha) I_z \rsb ) \, ,
	\end{aligned}\label{eq:sl_unitary}
\end{equation}
where we identify $\vartheta{=}t_\mathrm{pulse} \sqrt{\Omega^2+\delta^2}$ as the total angle acquired during a single pulse, and $\alpha{=}\arctan(\delta/\Omega)$ represents the deviation of the pulse direction from the $\xhat $ axis. 

In order to simplify the analysis, we consider a static change of basis which maps $\lb \cos(\alpha) I_x + \sin(\alpha) I_z\rb \to I_x$. In effect, this allows us to ignore the trivial tilt of the spin orbits away from the $\xhat$ axis due to the SL pulse detuning, and introduce it back later. The basis change here amounts to a global static rotation about the $\yhat$-axis by an angle $\alpha{=}\arctan(\delta/\Omega)$, and can be described by the unitary $V{=}\exp(-i\alpha I_y)$. In what follows, we will denote operators transformed in this basis by a tilde.

We will also consider the experimentally relevant scenario (see \zfr{polygons}) where each of the SL pulses accumulates a flip angle $\xt={}2\pi/N+\delta\xt$, with $\delta\xt \ll 1$~(see also Ref.~\cite{SOM}~\ref{SOM:detuning}), such that their combined rotation adds up to (almost) unity over one full AC period. To account for the deviation away from $2\pi/N$, we split up the action of the SL pulse in two parts, $\tilde{U}_\mathrm{SL}{=}\exp(-i\delta \xt I_x) \exp(-i 2 \pi/N I_x)$. We include the first part into $\tilde{U}_{\mathrm{acq},\, k}$, i.e., $\tilde{U}_{\mathrm{acq},\, k} \to \tilde{U}_{\mathrm{acq},\, k}^\prime {=} \tilde{U}_{\mathrm{acq},\, k}\exp(-i\delta \xt I_x)$, and isolate the commensurate kicks $\tilde{U}_\mathrm{SL}\to \tilde{U}_\mathrm{SL}^\prime {=} \exp(-i 2 \pi/N I_x)$.
For readability we will drop the prime superscript on the unitaries.

In order to describe the time evolution over stroboscopic times $nT_\mathrm{AC}+t_k$, where $t_k$ is fixed,  we make use of Floquet's theorem; it states that the stroboscopic evolution is generated by a time-independent Floquet Hamiltonian $\mH_{F;\, k}$ defined via,
\begin{equation}
	U_{F;\, k} 
	= \exp(-i T_\mathrm{AC} \mH_{F;\, k}),
	\label{eq:floquet_def}
\end{equation}
where the index $k$ explicitly indicates that the Floquet Hamiltonian depends on the time instant that defines the start of the sequence, $t_k {=} k/N T_\mathrm{AC}$.
However, as our model in Eq.~\eqref{eq:Hamiltonian} is both non-integrable and long-range interacting, solving for the exact Floquet Hamiltonian is intractable. Instead, we make use of the high-frequency expansion $\mH_{F;\, k}{=}\sum_n \omega_\mathrm{AC}^{-n} \mH_{F;\, k}^{(n)}$.
Using the Floquet-Magnus Expansion, or Baker-Campbell-Hausdorff formula \cite{Eckardt2015}, we can combine the unitaries in Eq.~\eqref{eq:floquet_unitary} into one unitary $\exp(-i  \sum_j t_\mathrm{acq} \tilde{\mH}_{\mathrm{acq},\, k+j} + 2\pi/N I_x)$, up to an error of order $\mathcal{O}(t_\mathrm{acq} J ,\, 2\pi/N)$, where $J$ is the energy scale of the Hamiltonian.

Since the kick angle $2\pi/N$ is large ($\sim \mathcal O(1)$), also terms higher order in $2\pi/N$ (i.e., $\mathcal{O}(t_{\mathrm{acq}J,\, j} |,\, (2\pi/N)^n)$), contribute to the leading order expression for $\tilde{\mH}_{F;\, k}^{(0)}$ in $T_\mathrm{AC}$.
Instead, we can account for contributions of the SL pulse to the Floquet Hamiltonian exactly, by writing Eq.~\eqref{eq:floquet_unitary} in the toggling frame as,
\begin{equation}
	\tilde{U}_{F;\,k}
	= \prod_{j=1}^N \tilde U_{\mathrm{acq},\, k+j} \tilde U_\mathrm{SL}
	=\tilde{U}_\mathrm{SL}^{N} \prod_{j=1}^N \tilde{U}_\mathrm{SL}^{-j} \tilde{U}_{\mathrm{acq},\, k+j} \tilde{U}_\mathrm{SL}^j\, , \label{eq:toggling_frame}
\end{equation}
which is obtained by introducing $N$ identities, $\mathbf{1}{=}U_\mathrm{SL}^{-j} U_\mathrm{SL}^{j}$.

This leads to an average Hamiltonian, which is valid up to $\mathcal O(T_\mathrm{AC})$-corrections, and reads as
\begin{equation}
	\begin{aligned}
		\frac{t_\mathrm{acq}}{\tau}\tilde{\mH}_{F;\, k}^{(0)}  &=\! \sum_{j=1}^N \tilde{U}_\mathrm{SL}^{-j} \tilde{\mH}_{\mathrm{acq},\, k+j} \tilde{U}_\mathrm{SL}^j \\
		&=\! \tilde{\boldsymbol{w}}_k(\alpha,\phi_\mathrm{AC}) \cdot \boldsymbol{I} + (1\! -\!3\sin^2\alpha)\!\sum_{n<m}\! b_{nm}\lb \frac{3}{2} \tilde{H}_\mathrm{ff} -  \boldsymbol{I}_n\cdot \boldsymbol{I}_m\rb\, , 
	\end{aligned}\label{eq:full_H_F}
\end{equation}
with the flip-flop term $\tilde{\mHff} {=}I_{jz}I_{kz} + I_{jy}I_{ky}$, and where,
\begin{equation}
	\begin{aligned}
		\tilde{\boldsymbol{w}}_k \pqty{\alpha, \phi_\mathrm{AC}} &= \pqty{\sin(\alpha) \delta + \frac{\delta \xt}{ \tau }} \xhat + \tilde{u}_k\pqty{\alpha,\, \phi_\mathrm{AC}}.
	\end{aligned}
\end{equation}
with,
\beq
\tilde{\boldsymbol{u}}_k(\alpha,\, \phi_\mathrm{AC}) = B_\mathrm{AC}\cos\alpha\sum_{n=1}^N \left(
		-\sin n\vartheta\; \yhat
		+\cos n\vartheta\; \zhat \right) f_{n+k}\pqty{\phi_\mathrm{AC}}. \label{eq:field_micromotion}
\eeq
A priori, it is unclear whether the sum in Eq.~\eqref{eq:field_micromotion} would yield a nonzero number, since both terms under the sum vanish individually, i.e. $\sum_n f_n {=}0$ and $\sum_{n=1}\left( -\sin n\vartheta\; \yhat+\cos n\vartheta\; \zhat \right) {=}0$.
However, for the special case $\xt\approx 2\pi/N$ they interfere constructively to yield a finite contribution.

To describe this more clearly, let us consider the exemplary case of $N{=}4$, sketched in \zfr{sketch_theory}, for $\xt{\app}\pi/2$ and $\xt{\app}\pi$. Both cases are sketched in \zfr{sketch_theory}C (i), with the colored and black vectors respectively. First, we note that by symmetry, and regardless of $\xt$, $f_1{=}-f_3$ and $f_2{=}-f_4$, due to the periodicity of the sine function (see \zfr{sketch_theory}A). Let us first consider the case of $\xt{\app}\pi/2$. Defining the vectors $\boldsymbol{\nu}_{n}\pqty{\vartheta} {=} \left( -\sin n\vartheta\; \yhat+\cos n\vartheta\; \zhat \right)$, we then have the quantities, 
\begin{equation*}
	\begin{aligned}
		\boldsymbol{a}_{1,\, 3} &= \boldsymbol{\nu}_{1,\,3}\pqty{\pi/2} = \pm \yhat & \boldsymbol{a}_{2,\, 4 }&= \boldsymbol{\nu}_{2,\,4}\pqty{\pi/2} = \mp \zhat\:,
	\end{aligned}
\end{equation*}
and following Eq.~\eqref{eq:field_micromotion}, the emergent field is then given by
$\tilde{\boldsymbol{u}}_k(\alpha,\, \phi_\mathrm{AC}) {=} B_\mathrm{AC}\cos\alpha\sum_n f_{n+k}(\xph_{\ac}) \boldsymbol{a}_n$. Now, the two pairs of weights cancel each other out, i.e., $f_1{=}-f_3$ and $f_2{=}-f_4$, and are associated with anti-parallel vectors, see \zfr{sketch_theory}B. This leads to an enhancement instead of a cancellation, and hence $\tilde{\boldsymbol{u}}_{k=0}(\xt=\pi/2)= 2 f_1 \yhat - 2 f_2 \zhat$ (see \zfr{sketch_theory}C (i)). The other vectors $\tilde{\boldsymbol{u}}_{k}$ ($k=1,2,3$) are simply related to the first one by micromotion-induced rotation about the $\xhat$-axis.

On the other hand however, for the case of $\xt{\app}\pi$, we have,
\begin{equation*}
	\begin{aligned}
		\boldsymbol{a}_{1,\, 3} &= \boldsymbol{\nu}_{1,\,3}\pqty{\pi} =   \zhat & 
		\boldsymbol{a}_{2,\, 4 }&= \boldsymbol{\nu}_{2,\,4}\pqty{\pi} = - \zhat,
	\end{aligned}
\end{equation*}
In this case, the interference is destructive  yielding $\tilde{\boldsymbol{u}}_{k=0}\pqty{\xt=\pi}{=} (f_1+f_3 - f_2 - f_4)\zhat=\boldsymbol{0}$ (see black vector in \zfr{sketch_theory}C (i)).

From Eq.~\eqref{eq:full_H_F} it follows directly that the $N$ different stroboscopic Floquet Hamiltonians are related by a rotation about the $\xhat$-axis: $\tilde{\mH}_{F;\, k^\prime}^{(0)}{=} \tilde{U}_\mathrm{SL}^{ k^\prime - k } \tilde{\mH}_{F;\, k}^{(0) } \tilde{U}_\mathrm{SL}^{k - k^\prime }$.
Notice that the leading order Floquet Hamiltonians $\tilde{\mH}_{F;\, k}^{(0)}$ only differ in their single particle terms $\tilde{\boldsymbol{w}}_k(\alpha,\phi_\mathrm{AC}) \cdot \boldsymbol{I}$. 
Moreover, in the absence of an AC drive, the Hamiltonians $\tilde{\mH}_{F;\, k}^{(0)}\pqty{B_\mathrm{AC}{=}0}$ become $k$-independent and are invariant under global rotations about the $x$-axis, i.e., $\lsb \tilde{\mH}_{F;\, k}^{(0)}\pqty{B_\mathrm{AC}{=}0},\, I_x \rsb{=}0$.
To sum up, introducing a finite AC drive leads to non-trivial emergent fields, but it does not affect the dipole-dipole interaction term, to lowest order in $T_\mathrm{AC}$.
We emphasize that these arguments still hold true after undoing the basis transformation $V$; however, the conserved axis (for $B_\mathrm{AC}{=}0$) is rotated into $\hat{\T{n}}{=}\cos(\alpha)\xhat {+} \sin(\alpha)\zhat$.

These expressions are derived in the simplified tilde basis that does not account for the tilt of the trajectories away from the $\xhat$ axis on account of SL pulse detuning. In order now to write down the average Hamiltonian in the original rotating frame, we carry out the transformation  $\mH_{F;\, k}^{(0)}= V^\dagger \tilde{\mH}_{F;\, k}^{(0)}  V$, resulting in,
\beq	
\frac{t_\mathrm{acq}}{\tau}{\mH}_{F;\, k}^{(0)}  
		= \boldsymbol{w}_k(\alpha,\phi_\mathrm{AC}) \cdot \boldsymbol{I} + (1-3\sin^2\alpha)\sum_{n<m} b_{nm}\lb \frac{3}{2} H_\mathrm{ff} -  \boldsymbol{I}_n\cdot \boldsymbol{I}_m\rb\, , \label{eq:H_F_rotframe}
\eeq
with the transformed flip-flop Hamiltonian
\begin{equation}
    \begin{aligned}
    H_\mathrm{ff}   &=  \cos^2(\alpha) I_{jz} I_{kz} + \sin^2(\alpha) I_{jx}I_{kx} +I_{jy}I_{ky}\\
                    &\quad - \cos(\alpha)\sin(\alpha) \pqty{I_{jx} I_{kz} + I_{jz} I_{kx}},
    \end{aligned}
\end{equation}
and the single particle contributions
\begin{equation}
	\begin{aligned}
		\boldsymbol{w}_k \pqty{\alpha, \phi_\mathrm{AC}} &= \boldsymbol{v}\pqty{\alpha} + \boldsymbol{u}_k\pqty{\alpha,\, \phi_\mathrm{AC}}\, \label{eq:fields_rotated_frame}
	\end{aligned}
\end{equation}
where,
\begin{equation}
	\begin{aligned}
		\boldsymbol{v}\pqty{\alpha}                          &= \pqty{\sin(\alpha) \delta + \frac{\delta \xt}{\tau} } \pqty{\cos(\alpha) \xhat + \sin(\alpha) \zhat}, \\
		\boldsymbol{u}_k\pqty{\alpha,\, \phi_\mathrm{AC}}    &= B_\mathrm{AC}\cos\alpha\sum_{n=1}^N \left[
		-\sin n\vartheta\; \yhat \right. \\
		&\quad+\left.\cos n\vartheta\; \pqty{\cos(\alpha) \zhat - \sin(\alpha) \xhat} \right] f_{n+k}\pqty{\phi_\mathrm{AC}}
	\end{aligned}
\end{equation}
to leading order in the AC period $T_\mathrm{AC}$. 

Since undoing the basis transformation $V$ hides the simplicity of the result and undoing the transformation is straightforward, we will hereinafter carry out most of the analysis in the transformed frame and only undo the transformation for the final expressions.

We emphasize that, due to the finite detuning~($\delta{\neq}0$), the vectors $\boldsymbol{w}_k$ are not centered around the $\xhat$-axis (see SI~\cite{SOM}).
Instead, the points measured on the Bloch sphere are centered around $\hat{\T{n}}{=}\cos(\alpha) \xhat + \sin(\alpha) \zhat$, and thus feature a finite $\zhat$-tilt that moves the midpoint of the polygon into the lower or upper half of the Bloch-sphere as sketched in \zfr{sketch_theory}C. This is important for obtaining closed shapes as we are only able to measure the modulus of the $\zhat$-magnetization in the experiment~(see \zfr{sketch_theory} and SI~\cite{SOM}).

\section{\label{app:prethermal_props}Properties of the state in the prethermal plateau}

The leading order Floquet Hamiltonian $\tilde{\mH}_{F;\, k}^{(0)}$ describes the steady-state of the system in the prethermal plateau \cite{Abanin2017}. 
When measured at times $n T_\mathrm{AC}+ k\tau$ for some integers $n,k<N$, we thus anticipate the system to prethermalize to a state that is well-described by a prethermal density matrix,
\begin{equation}
	\tilde{\rho}_{F;\, k} = \exp(-\beta \tilde{\mH}_{F;\, k}^{(0)})/Z \, , \label{eq:thermal_state}
\end{equation}
with $Z{=}\mathrm{Tr} \exp(-\beta \tilde{\mH}_{F;\, k}^{(0)})$.
Due to the unitarity of the time evolution, the temperature $\beta_k^{-1}$ is determined by the initial energy density $\varepsilon_{k}[\tilde{\rho}_0] {=}\Tr{\tilde{\rho}_0 \tilde{\mH}_{F;\, k}^{(0)})}/L {\overset{!}{=}}  \Tr{\tilde{\rho}_{F;\, k} \tilde{\mH}_{F;\, k}^{(0)}}/L$ which matches the energy density of the state evolved in the prethermal plateau.
For an initial state with magnetization $\mu$ along the $x$-direction $\tilde{\rho}_0 {\propto} (\boldsymbol{1} + \mu I_x)$, using $\sin(\alpha) {=}\delta/\sqrt{\delta^2 + \Omega^2}$, we find that the initial energy density 
\begin{equation}
	\varepsilon_k =\Tr{\tilde{\rho}_0 \tilde{\mH}_{F;\, k}^{(0)}}/L = \mu {\left| \pqty{ \delta/\sqrt{\delta^2 + \Omega^2}} + \pqty{\delta \xt/\tau} \right|} \, , \label{eq:energy_density}
\end{equation}
in general does not vanish, as long as either $\delta \xt$ or $\delta$ are finite. 
Equation~\eqref{eq:energy_density} is a direct result of the tracelessness of the spin operators $\mathrm{Tr}\pqty{I_a}{=}\mathrm{Tr}\pqty{I_{na}I_{mb}}{=}0$, for $a,\,b{=}x,\,y,\,z$ and $n{\neq} m$, and $\mathrm{Tr}\pqty{I_a I_a}/Z{=}L$.

Notice that the initial magnetization, and with it the inverse temperature, is relatively small; this justifies the high-temperature approximation $\tilde{\rho}_{F;\, k}{\approx}(\mathbf{1} -\beta \tilde{\mH}_{F;\, k}^{(0)})/Z$. 
Within the high-temperature approximation, using the relation 
$
|\boldsymbol{w}_k|^2{=}
|\tilde{\boldsymbol{w}}_k|^2 $,
for the inverse temperature we find 
\begin{equation}
	\beta_k = -\frac{ \varepsilon_k }{ h_\mathrm{dd}^2 + {|\boldsymbol{w}}_k|^2 } \label{eq:temperature} \, .
\end{equation}
Note that the (inverse) temperature at which the system prethermalizes is set by the relative phase $\phi_\mathrm{AC}$ between the AC and the SL pulse drives; e.g., for $\phi_\mathrm{AC}{=}0$, we have $\beta{=}\beta_{k=0}$, etc. We stress that, once the system has reached the prethermal plateau, its state has a single well-defined temperature throughout the entire micromotion dynamics. 

In Eq.~\eqref{eq:temperature},
\begin{equation}
	h_\mathrm{dd}^2 = 2 \pqty{1 - 3 \sin^2 \alpha} \sum_{n<m} b_{nm}^2 /L \propto J^2 \label{eq:dd_scale} 
\end{equation}
is an energy scale associated with the dipole-dipole coupling. In general, $ h_\mathrm{dd}$ is finite also in the thermodynamic limit since $b_{nm}^2{\propto}1/|r_{nm}|^6$ is sufficiently short-ranged in three dimensional space for the sum above to converge. Here, $r_{nm}$ is the internuclear position vector.
However, the precise value of $h_\mathrm{dd}$ depends on microscopic details and cannot be accessed easily.

To summarize, as stated in the main text, it follows that the magnetization in the prethermal plateaus is determined by
\begin{equation}
	\boldsymbol{M}_k = \Tr{\boldsymbol{I} \rho_{F;\, k} }/L = \frac{ - \varepsilon_k }{ h_\mathrm{dd}^2 + |\boldsymbol{w}_k|^2 } \boldsymbol{w}_k \, , \label{eq:magnetization}
\end{equation}
which scales linearly with the AC amplitude, $\boldsymbol{M}_k{\propto} B_\mathrm{AC}$, for small amplitudes compared to the dipole-dipole coupling, ${B_\mathrm{AC}/h_\mathrm{dd}}{\ll 1}$.

In order to characterize the measured prethermal polygon shapes, let us introduce two metrics: the linear size ~$\ell$ and the excursion angle~$\Phi$ away from the spin-lock axis.
To quantify the size of the shapes independent of the number of measured points $N$, we note that all points lie on a circle in $3\mathrm{D}$, and hence we can use the radius of this circle to define the linear size $\ell$ of the polygons:
\begin{equation}
	\ell =  |\boldsymbol{w}_{k;\, \parallel }| = |\boldsymbol{w}_{k} - \hat{\T{n}} \pqty{\hat{\T{n}}\cdot  \boldsymbol{w}_{k}}| \, . \label{eq:linear_dimension}
\end{equation}
As the corners of the polygon, $\boldsymbol{w}_k$, all lie in one plane, we can extract the in-plane components ${\boldsymbol{w}_{k;\, \parallel}{=}\boldsymbol{w}_{k} - \boldsymbol{c}}$ by subtracting the center point $\boldsymbol{c}{=}\hat{\T{n}} \pqty{\hat{\T{n}}\cdot  \boldsymbol{w}_{k}}$ of the polygon, where $\hat{\T{n}}{=}\cos(\alpha) \xhat + \sin(\alpha) \zhat$ is the unit-vector perpendicular to the plane.

In addition, let us define the excursion angle 
\begin{equation}
	\Phi = 2\arctan\lb \frac{\ell}{{| \boldsymbol{c} |}} \rb\, . \label{eq:excursion_angle}    
\end{equation}
The excursion angle is a measure for the size of the deviation of the polygon points from the axis $\xhat$ which is conserved in the absence of the AC drive.
We note that, for regular shapes, these two measures are single valued, i.e., $\Phi$ and $\ell$ should be independent of the index $k$.

Using Eq.~\eqref{eq:magnetization} it is straightforward to show that
\begin{equation}
	\ell \propto \frac{\varepsilon_k B_\mathrm{AC}}{h_\mathrm{dd}^2 + {\left| \boldsymbol{w}_k \right|^2}}
\end{equation}
and
\begin{equation}
	\tan\lb \fr{\Phi}{2}\rb = \varepsilon_k \frac{ {\left| \boldsymbol{u}_k \right|}}{ {\left| \boldsymbol{v}(\alpha) \right|}} \propto B_\mathrm{AC} \, .
\end{equation}

\section{\label{app:theory_quenches}Theory of the quench dynamics}

The transient dynamics when turning on and off the AC field are highly model dependent.
Nevertheless, we can make qualitative predictions about the behavior of the system by considering the thermal properties of the density matrices before and after the quench.
We will, in particular, consider three scenarios, 
(i) initial opening of the shapes, 
(ii) closing of the shapes, and 
(iii) the combination of closing and reopening, corresponding to the scenarios depicted in \zfr{emergence} and \zfr{collisions}. 
We will restrict the following analysis to "square" shapes, for simplicity.

In the following analysis we will analyze the states $\rho_f$ the system is expected to thermalize to after each quench, given the properties of the state $\rho_i$ before each quench.
Following the discussion in the main text, we will further denote states for which the AC field is on (off) by a $\mathrm{on}$ (off) subscript.

The initial opening $\tilde{\rho}_\mathrm{init}\to \tilde{\rho}_{\mathrm{on},f}$ has already been described in detail in App.~\ref{app:prethermal_props}. 
For the closing of the shapes, $\tilde{\rho}_{\mathrm{on},i} \to \tilde{\rho}_{\mathrm{off},f}$, we point out the special role of the $\xhat$-magnetization conservation, such that the thermal state is characterized by both a temperature and a magnetization. To account for the conservation law we introduce an additional Lagrange multiplier $\gamma$ in the prethermal ensemble
\begin{equation}
	\tilde{\rho}_{\mathrm{off},f}= \exp\left( -\beta_{\mathrm{off},f} \lsb \tilde{\mH}_F^{(0)}\pqty{B_\mathrm{AC}{=}0} - \gamma I^x  \rsb\right) / Z \, . \label{eq:thermal_conservation}
\end{equation}
The inverse temperature $\beta_{\mathrm{off},f}$ and the new parameter $\gamma$ are determined by the initial energy $\Tr{\tilde{\rho}_{\mathrm{on},i} \tilde{\mH}_{F}^{(0)}} {\overset{!}{=}}  \Tr{\tilde{\rho}_{\mathrm{off},f} \tilde{\mH}_{F}^{(0)} }$ and  magnetization $\Tr{I_x \tilde{\rho_0}}{\overset{!}{=}} \Tr{I_x \tilde{\rho}_F\pqty{B_\mathrm{AC}{=}0}}$, respectively.

The initial state $\tilde{\rho}_{\mathrm{on},i}$ is given by the thermal state before the quench, i.e., $\tilde{\rho}_{\mathrm{on},i}{=}\tilde{\rho}_{F;\, 0}\pqty{\phi_c} {=} \exp(-\beta \tilde{\mH}_{F;\, 0}^{(0)}\pqty{\phi_c} )/Z$, introduced in App.~\ref{app:prethermal_props}, where the phase at closing is determined by the time $t_\mathrm{c}$ at which the closing happens via $\phi_c {=} \phi_\mathrm{AB} + \pqty{\omega_\mathrm{AC}t_\mathrm{c}}\mathrm{mod}\pqty{2\pi}$.

Combining the high temperature approximation ${\tilde{\rho}_{\mathrm{on},i} {\approx} \pqty{ \mathbf{1} - \beta_{\mathrm{on},i} \tilde{\mH}_{F;\, 0}^{(0)}\pqty{\phi_c} )}/Z}$ and $\tilde{\rho}_{\mathrm{off},f}{\approx} \pqty{ \mathbf{1}  -\beta_{\mathrm{off},f} \lsb \tilde{\mH}_F^{(0)}\pqty{B_\mathrm{AC}=0} - \gamma I_x  \rsb}/Z$, the temperature and magnetization after the quench are immediately given by
\begin{equation}
	\begin{aligned}
		\beta_{\mathrm{off},f} &= \beta_{\mathrm{on},i}&  &\mathrm{and} & \gamma &= 0 \, . \label{eq:temperature_closing}
	\end{aligned}
\end{equation}
We emphasize that this does not imply the equality of the two states, since $\tilde{\mH}_F^{(0)}\pqty{B_\mathrm{AC}{=}0}{\neq} \tilde{\mH}_F^{(0)}\pqty{B_\mathrm{AC}{\neq}0}$. However, our result suggests that the closing quench merely leads to a decay of all components orthogonal to the preserved axis $\xhat$.
In this respect, note also that the final state, after closing the shape, has a finite dipolar order. In contrast, the initial state $\tilde{\rho}_\mathrm{init} {\propto} I_x$ has vanishing dipolar order.

For these reasons, reopening the shapes, $\rho_{\mathrm{on}, i}{\to} \rho_{\mathrm{on}, f}$ is different from the initial opening of the shapes. 
In particular, the temperature after reopening is given by,
\begin{equation}
	\beta_{\mathrm{on}, f} = \beta_{\mathrm{on}, i} \frac{h_\mathrm{dd}^2 + {\boldsymbol{v}\pqty{\alpha}}\cdot {\boldsymbol{w}_k}}{h_\mathrm{dd}^2 + {|\boldsymbol{w}}_k|^2} \, ,\label{eq:temperature_reopening}
\end{equation}
where $\beta_{\mathrm{on}, i}$ is the temperature before the quench. The fraction on the right hand side of Eq.~\eqref{eq:temperature_reopening} is always smaller than $1$, as ${|\boldsymbol{w}}_k| \geq {|\boldsymbol{v}\pqty{\alpha}|}$, with equality only at ${B_\mathrm{AC}{=}0}$. 

Therefore, combining Eqs.~\eqref{eq:magnetization}, \eqref{eq:temperature_closing}, and \eqref{eq:temperature_reopening}, we find that the total magnetization $|\boldsymbol{M}_{\mathrm{on}, f}|$, after closing and reopening, is given by
\begin{equation}
	|\boldsymbol{M}_{\mathrm{on}, f}|/|\boldsymbol{M}_{\mathrm{on}, i}| = \beta_{\mathrm{on}, f}/\beta_{\mathrm{on}, i} = 1 - \sigma.
\end{equation}
Here $|\boldsymbol{M}_{\mathrm{on}, i}|$ is the total magnetization prior to closing, and ${\sigma {=}\pqty{h_\mathrm{dd}^2 + {\boldsymbol{v}\pqty{\alpha}}\cdot {\boldsymbol{w}_k}}/\pqty{h_\mathrm{dd}^2 + {|\boldsymbol{w}}_k|^2}}$. 
Hence, with each closing and reopening, the amplitude of the magnetization is reduced by a factor $\pqty{1-\sigma}$.
However, each closing and reopening process results only in a tiny loss, since ${\sigma {\ll} 1}$ in the experimentally relevant regime ${B_\mathrm{AC}{\ll} h_\mathrm{dd}}$.

\end{appendix}
\bibliography{main.bbl}
\pagebreak

\clearpage
\onecolumngrid
\begin{center}
\textbf{\large{\textit{Supplementary Information} \\ \smallskip
Continuously tracked, stable, large excursion trajectories of dipolar coupled nuclear spins}}\\
\hfill \break
\smallskip
Ozgur Sahin$^{1,\ast}$, Hawraa Al Asadi$^{1,\ast}$, Paul M. Schindler$^{2,\ast}$, Arjun Pillai$^{1}$, Erica Sanchez$^{1}$, Matthew Markham$^{3}$,\\
Mark Elo$^{4}$, Maxwell McAllister$^{1}$, Emanuel Druga$^{1}$, Christoph Fleckenstein$^{5}$, Marin Bukov$^{2,6}$, and Ashok Ajoy$^{1,7}$\\
${}^{1}$\I{{\small Department of Chemistry, University of California, Berkeley, Berkeley, CA 94720, USA.}}\\
${}^{2}$\I{{\small Max Planck Institute for the Physics of Complex Systems, N\"othnitzer Str.~38, 01187 Dresden, Germany.}}\\
${}^{3}$\I{{\small Element Six Innovation, Fermi Avenue, Harwell Oxford, Didcot, Oxfordshire OX11 0QR, UK.}}\\
${}^{4}$\I{{\small Tabor Electronics, Inc., Hatasia 9, Nesher 3660301, Israel.}}\\
${}^{5}$\I{{\small Department of Physics, KTH Royal Institute of Technology, SE-106 91 Stockholm, Sweden.}}\\
${}^{6}$\I{{\small Department of Physics, St. Kliment Ohridski University of Sofia, 5 James Bourchier Blvd, 1164 Sofia, Bulgaria.}}\\
${}^{7}$\I{{\small Chemical Sciences Division Lawrence Berkeley National Laboratory,  Berkeley, CA 94720, USA.}}\\
\end{center}

\twocolumngrid
\beginsupplement

\setcounter{section}{0}


\section{Sample and hyperpolarization strategy}
The sample used in this work is a CVD fabricated single crystal diamond sample that contains $\sim$1ppm NV centers, and natural abundance of $\Cs$ concentration. This corresponds to the diamond $\Cs$ nuclei being distributed with a spin density $\sim$0.92/nm$^3$ in the lattice. The sample is placed flat, i.e. with its [100] face pointing parallel to the hyperpolarization and interrogation magnetic fields (38mT and 7T respectively). 

Hyperpolarization is carried out at $\Bp{=}38$mT through a method previously described~\cite{Ajoy17,Ajoy18}. It employs continuous optical pumping and swept microwave irradiation for ${\sim}$40-60s, and  ratchet-like polarization transfer develops through the traversal of rotating frame Landau-Zener anticrossings~\cite{Zangara18}. Spin diffusion serves to transfer polarization to bulk $\Cs$ nuclei in the diamond lattice.

\section{Experimental setup}
\zsl{setup}

We now provide more details of the apparatus employed for $\Cs$ spin control and interrogation in our experiments. In reality, the $\Cs$ nuclei are hyperpolarized at room temperature at low-fields ($\sim$38mT), and subsequently “shuttled” to high-field (7T) where they are controlled and interrogated. We refer the reader to Ref.~\cite{Sarkar21} for details on the hyperpolarization apparatus and mechanism, and Ref.~\cite{Ajoyinstrument18} for details on the shuttling apparatus.  

Specifically, the detection of the spins proceeds via inductive coupling to an RF cavity and measurement of corresponding $\Cs$ Nuclear Magnetic Resonance (NMR) signals at high-field (7T), where the $\Cs$ resonance frequency is 75.38MHz. This is accomplished via a homemade NMR probe that was detailed in Ref.~\cite{Sahin21}. It consists of a high RF homogeneity saddle coil that allows $>$1M pulses to be applied to the nuclear spins at high power ($\sim$30W) and high duty cycle (${\sim}$50\%). This RF coil is in a saddle geometry and is laser-cut out of OFHC copper with 3 turns and a coil height of 1cm. We measure a Q-factor of 50 at 75MHz and a sample filling factor of $\sim$0.15. The probe also has a z-coil that is employed to apply the time-varying AC field to the sample. This  is a loop of a 2-turn coaxial cable that is wound outside the RF coil (see Ref.~\cite{Sahin21}).

The AC field itself is produced with a Tabor TE5201 arbitrary waveform generator, and is amplified by an AE Techron 7224 amplifier into the z-coil. The spin-lock RF pulses on the other hand are produced by a Varian NMR spectrometer. High signal-to-noise detection of the $\Cs$ nuclear precession is carried out with a quarter wave line and bandpass filter combination following a transmit/receive switch, and the signal is digitized by a high-speed arbitrary waveform generator (Tabor Proteus). The high data acquisition rate of the device (1GS/s) allows for high-fidelity sampling of the $\Cs$ induction signal between the pulses~\cite{Sahin21}. The entire experiment is synchronized via a PulseBlaster digital pulse generator (100MHz).

\begin{figure}[t]
 \centering
 {\includegraphics[width=0.5\textwidth]{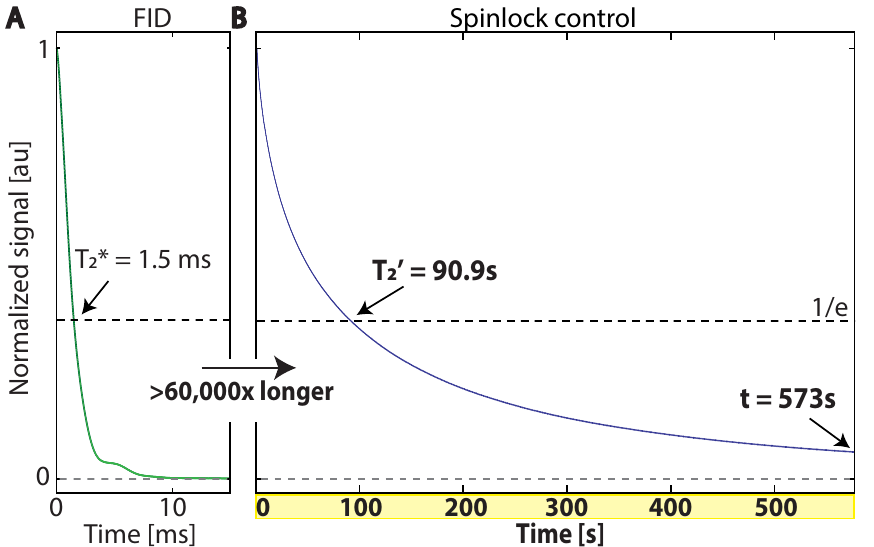}}
 \caption{\T{Effect of spin-locking.} (A) Typical $\Cs$ FID occurs in ${\app}$1.5 ms. (B) Pulsed spin-locking  leads to prethermalization, extend the lifetime by $>$60,000-fold to $T_2^\prime{\app}91$s. Spins are continuously interrogated here for 573s. Data here is reproduced from Ref. ~\cite{Beatrez21} for completeness.}
\zfl{lifetime_ext}
 \end{figure}

\section{Spin-lock pulse sequence and data processing}
\zsl{processing}

For clarity, we describe in \zfr{lifetime_ext} the effect of the spin lock sequence, reproducing data in Ref.\cite{Beatrez21}. The $\Cs$ free induction decay in diamond is $T_2^{\ast}{\app}1.5$ms on account of inter-spin interactions (\zfr{lifetime_ext}A). However, as demonstrated in \zfr{lifetime_ext}B, under pulsed spin-lock it is possible to prolong this coherence to $T_2^\prime{>}90$s. In typical experiments, in this paper, pulse duty cycle is maintained high (19-54\%), and interpulse spacing $\qt{\sim}100\mu$s. Flip angle $\xt$ can be arbitrarily chosen, except for $\xt{=}\pi$~\cite{Ajoy20DD}. 

Let us now provide more detail about the data processing. For each $\tacq$ window between the SL pulses, we directly digitize the heterodyned Larmor precession of the nuclear spins every 1ns using a fast AWT (Tabor Proteus). The amplitude and phase of the Fourier transform of the obtained precession (\zfr{fig2}A-B), sampled at $\fhet{=}20$MHz (the heterodyned Larmor frequency), then directly provide information of the transverse components of the spin vector (as shown in \zfr{fig2}C of the main paper). This homodyne readout is very effective because of the slow Larmor precession and rapid sampling allows the ability to completely capture the signal with ${>}$100 points per precessing period. This oversampling enables high SNR reconstruction due to the ability to suppress noise at any component except at $\fhet$. Ultimately, each readout window provides one such amplitude and phase point, and in typical experiments we apply $N{\gtrsim}$200k pulses. 

\section{Stretched exponential decays}
\zsl{stretched_exp}

\begin{figure}[t]
 \centering
 {\includegraphics[width=0.5\textwidth]{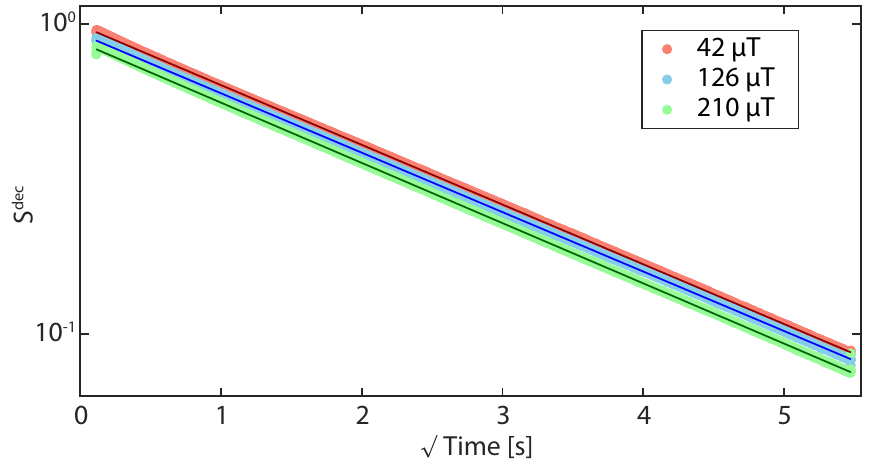}}
 \caption{\T{Stretched exponential nature of $S^{\R{dec}}$.} As a complement to the data in \zfr{amplitude}E-F of the main paper, we plot the decay profile of signal $S$ in a graph considering $S^{\R{dec}}$ vs. $\sqrt{t}$ in a logarithmic scale for three representative values of $\Bac$.  Solid lines are linear fits. Extracted $T_2'$ values for full dataset are plotted in \zfr{amplitude}F.}
\zfl{stretched_exp}
 \end{figure}
 
 \begin{figure}
    \centering
    {\includegraphics[width=0.5\textwidth]{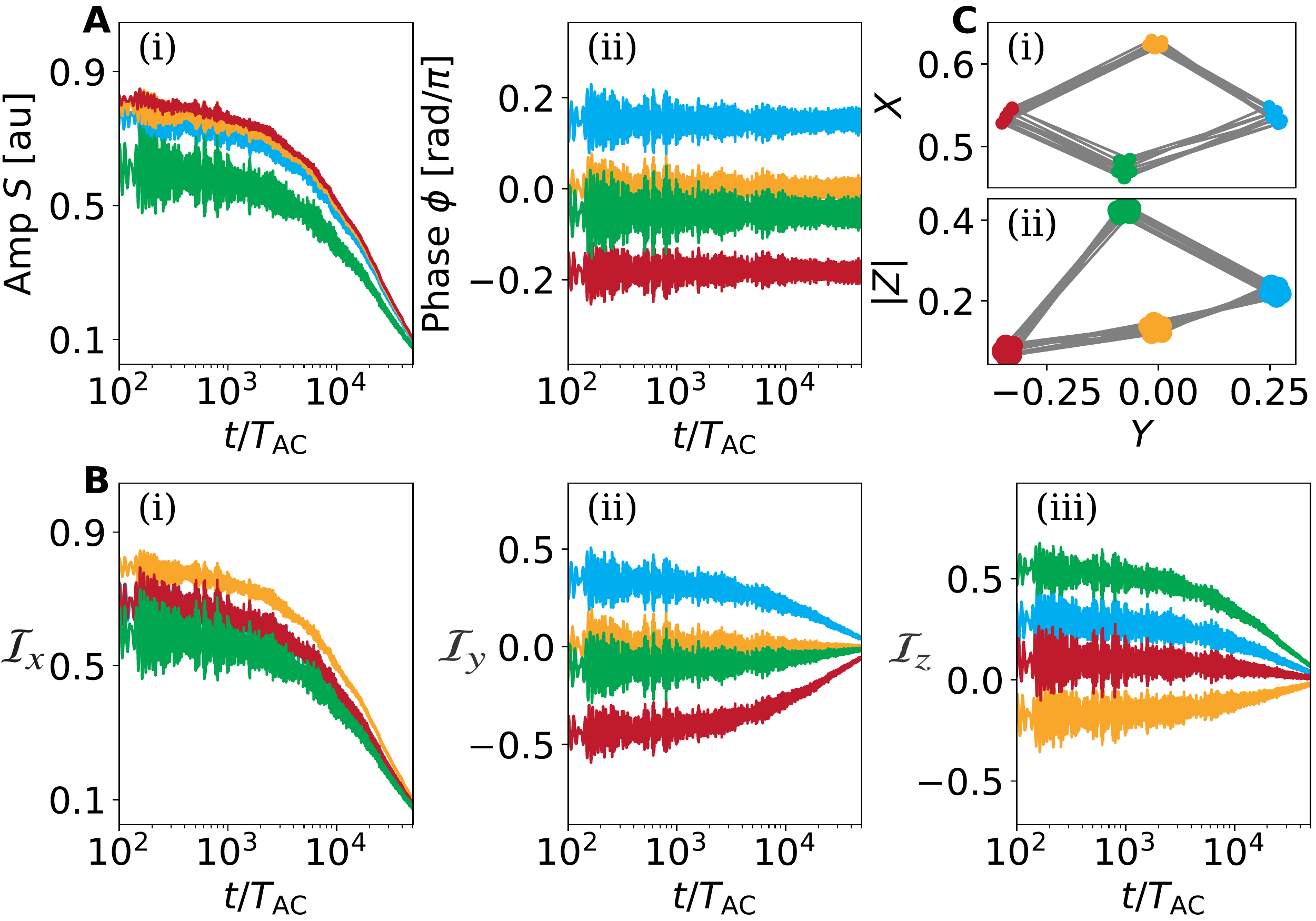}}
    \caption{\textbf{Simulation of square-like spin orbit} for $N{=}4$ (sequence in \zfr{polygons}C(i)) employing Eq.\eqref{eq:Hamiltonian} of the main paper. (A)(i-ii) Plotted are signal $S$ and phase $\varphi$ respectively (analogous to \zfr{longtime}), assuming a SL flip-angle $\xt{=}0.45\pi$,  AC field amplitude $\Bac/J{=}0.5$, and detuning $\delta/J{=}1$, $J\tau {=}0.2$, $t_{\mathrm{pulse}}/\tau {=} 2/3$, $t_{\mathrm{acq}}/\tau {=} 1/3$. (Pseudo-)Random graphs of interacting spins are designed according to the protocol described in \cite{SOM} with the parameters $r_{\mathrm{min}}{=}0.9$ and $r_{\mathrm{max}}{=}1.1$ (in units of $\sqrt[3]{\mu_0\hbar\gamma_n^2}$ ). 
    (B) Extracted magnetization components $\mI_x~(i),\mI_y$~(ii) and $\mI_z$~(iii). (C) Corresponding spin orbit  plotted as projections on the $\xy$~(i) and $\yz$~(ii) planes.
    Data here is shown for $60$ AC cycles starting at $t{=}1000T_\mathrm{AC}$, and we average over $5$ consecutive cycles to increase the SNR.
    We carry out full numerical simulations with $L{=}14$ spins and average over $5$ random graph  realizations.
    }
    \zfl{longtime_simulation}
\end{figure}

\begin{figure}
    \centering
    {\includegraphics[width=0.45\textwidth]{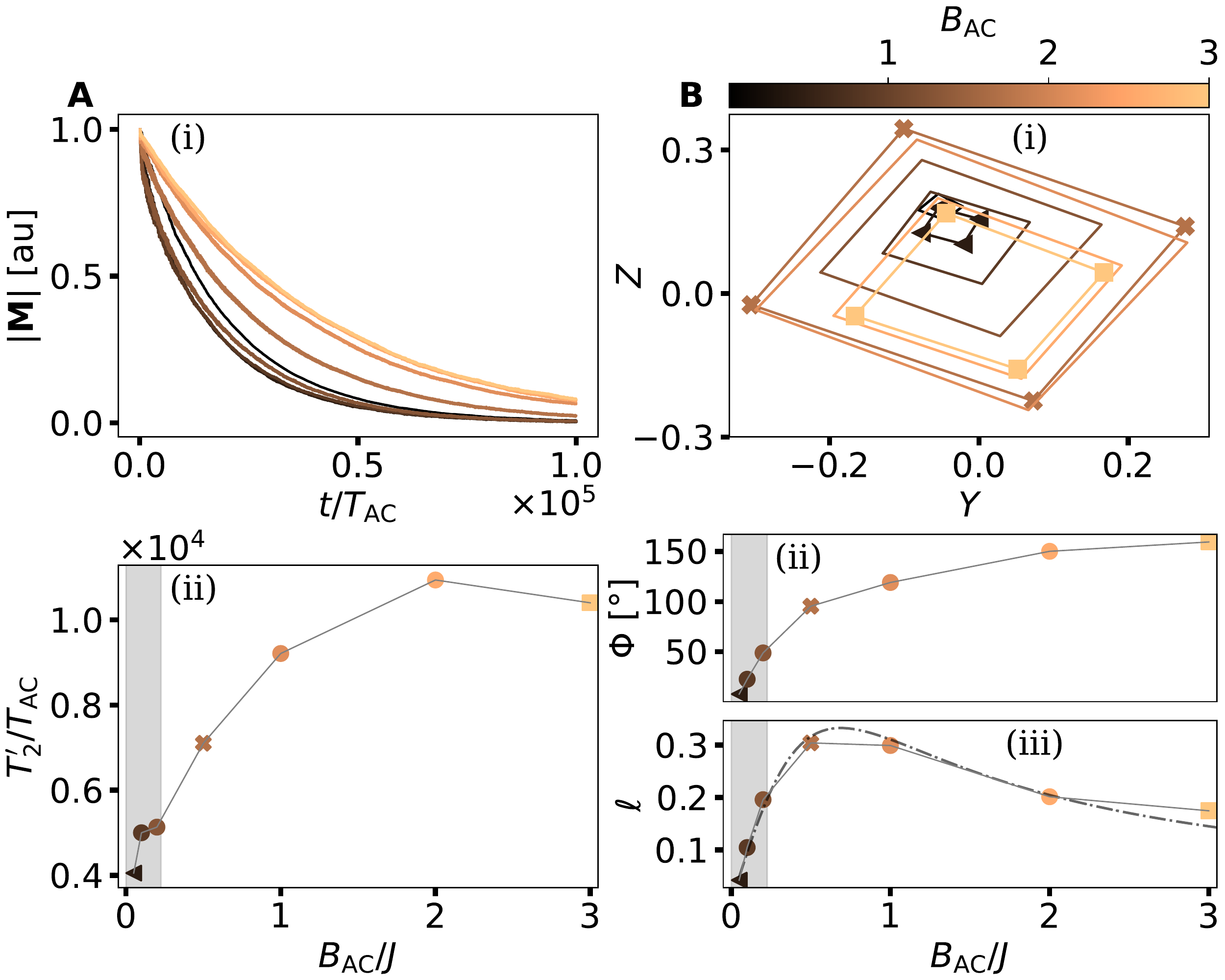}}
    \caption{\textbf{Simulations of orbit scaling with $\Bac$}. In a complement to data in \zfr{amplitude}, we carry out numerical simulations similar to \zfr{longtime_simulation} to simulate square-orbit profiles with increasing amplitude $\Bac/J{\in}[0,3]$ (see colorbar in (B)). 
    (A)(i) Magnetization decay profiles similar to \zfr{amplitude}E. 
    (B)(i) Extracted top-view ($\yz$) profiles of spin orbits for different values of $B_{\mathrm{AC}}$ (to be compared with \zfr{amplitude}B).
    $T_2^\prime$ lifetimes ((A)(ii)), excursion angles $\xPh$ ((B)(ii))  and square radius $\ell$ ((B)(iii)) extracted from (A)(i) and (B)(i) as a function of $\Bac/J$.  Dashed line in (B)(iii) is a fit of the analytical relation, $\ell{\propto} B_\mathrm{AC}/(1+ (B_\mathrm{AC}/h_\mathrm{dd})^2 )$, with $h_\mathrm{dd}/J{\approx} 0.8$.
    The shaded regions in (B)(ii-iii) indicate the experimentally accessible regime. Different markers~(triangle, cross and square) correspond to special magnetic field values~($B_\mathrm{AC}/J{=}0.01,\, 1,\, 3$, respectively).
    Simulation parameters: $L{=}14$, detuning $\delta/J{=}1$, SL flip-angle $\xt{=}0.5\pi$, and comprises an average over $5$ random graph realizations. Further parameters are as in \zfr{longtime_simulation}.
 }
    \zfl{amplitude_simulation}
\end{figure}

\begin{figure}[t]
 \centering
 {\includegraphics[width=0.49\textwidth]{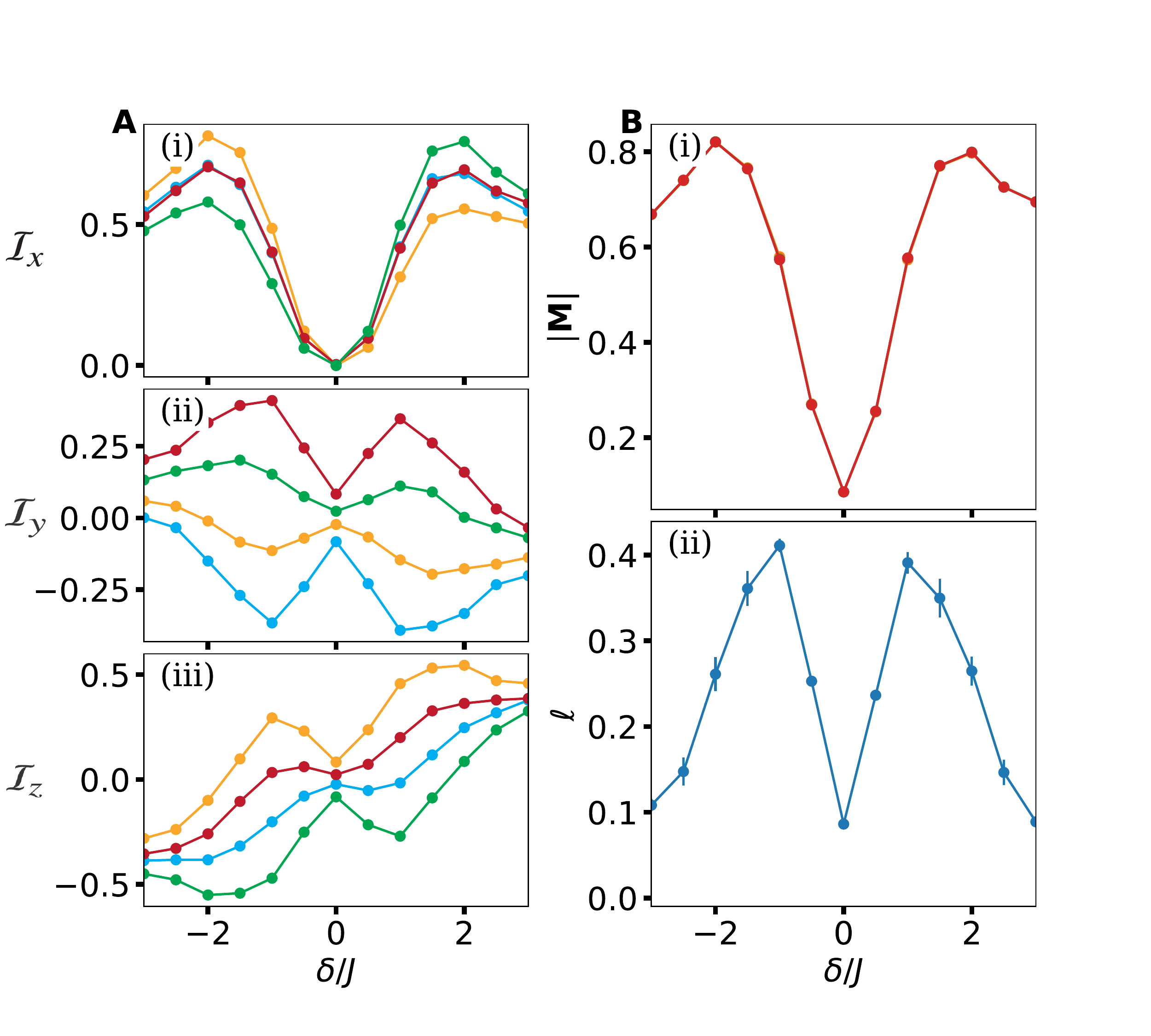}}
 \caption{
 \T{Simulation of effect of detuning $\delta$.}
    (A) Magnetization $\boldsymbol{M}$ in (i) $\xhat$, (ii) $\yhat$ and (iii) $\zhat$ direction. (B) Total magnetization (i) $|\boldsymbol{M}|$ and (ii) linear scale $\ell$ as a function of the detuning $\delta$.
    The different colors indicate the $N{=}4$ different plateaus. 
    Data is extracted after $1000$ pulses and averaged over 5 disorder realizations for a system of $L{=}14$ spins with fixed AC amplitude $B_\mathrm{AC}/J{=}0.5$ and perfect pulse angle $\delta \xt {=} 2\pi/4$. Further parameters are as in \zfr{longtime_simulation}.
 }
\zfl{detuning}
 \end{figure}
 
 \begin{figure}[t]
 \centering
 {\includegraphics[width=0.49\textwidth]{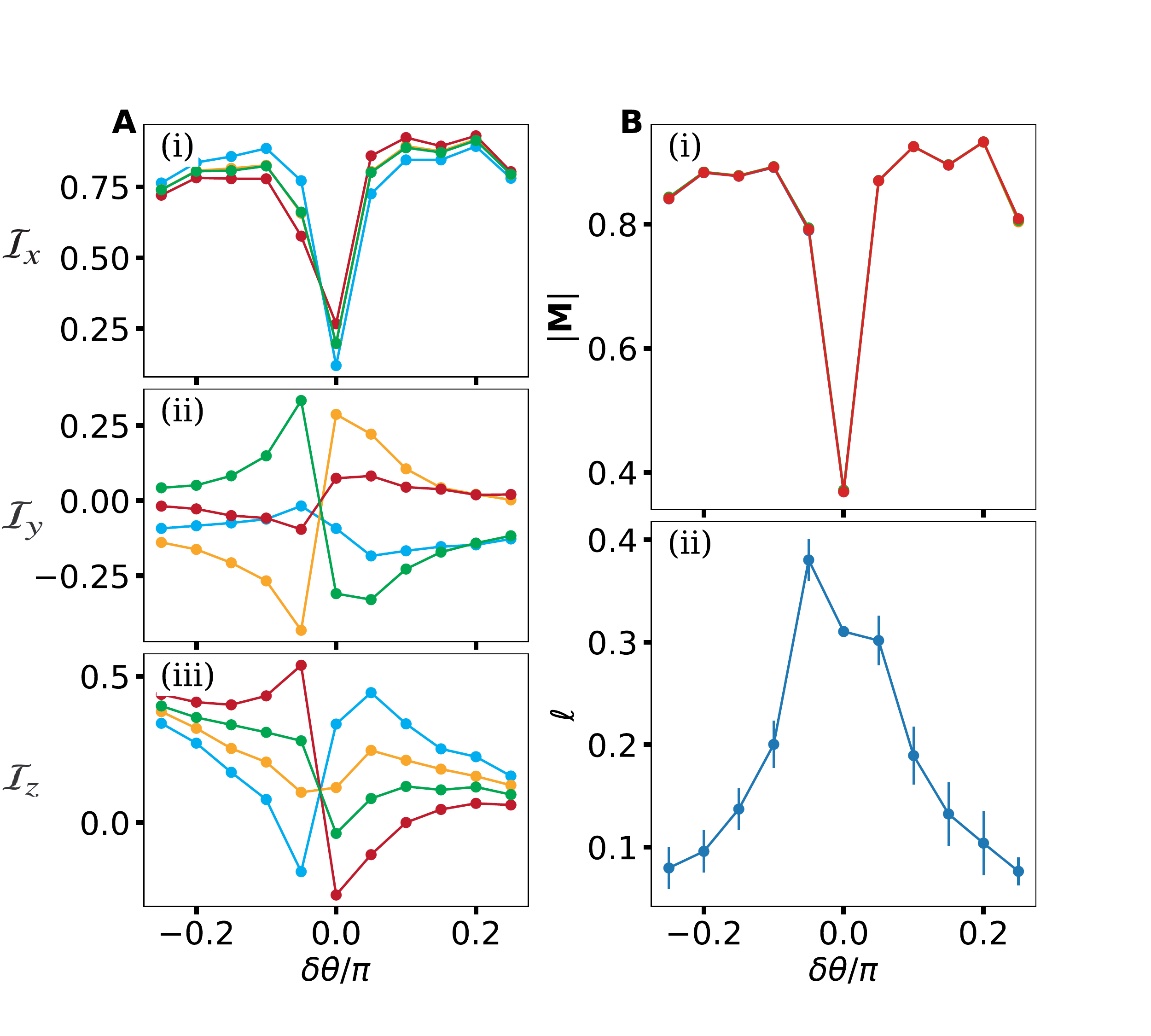}}
 \caption{
 \T{Simulations of effect of imperfect pulse duration.} (A) Magnetization components of $M$ in (i) $\xhat$, (ii) $\yhat$ and (iii) $\zhat$ direction. (B) Total magnetization (i) $|\boldsymbol{M}|$ and linear scale (ii) $\ell$ as a function of the imperfectness $\delta\xt$ in the angle $\xt {=} 2\pi/4 + \delta \xt$. Different colors indicate the $N{=}4$ different plateaus. 
 Data is extracted after $1000$ pulses and averaged over 5 disorder realizations for a system of $L{=}14$ spins with fixed AC amplitude $B_\mathrm{AC}/J{=}0.5$ and detuning $\delta/J{=}1$, and variable accumulated pulse angle $\xt {=} \pi/2 +\delta \xt$. Further parameters are as in \zfr{longtime_simulation}.
 }
\zfl{theta_detuning}
\end{figure}
 
\begin{figure}[t]
 \centering 
 {\includegraphics[width=0.45\textwidth]{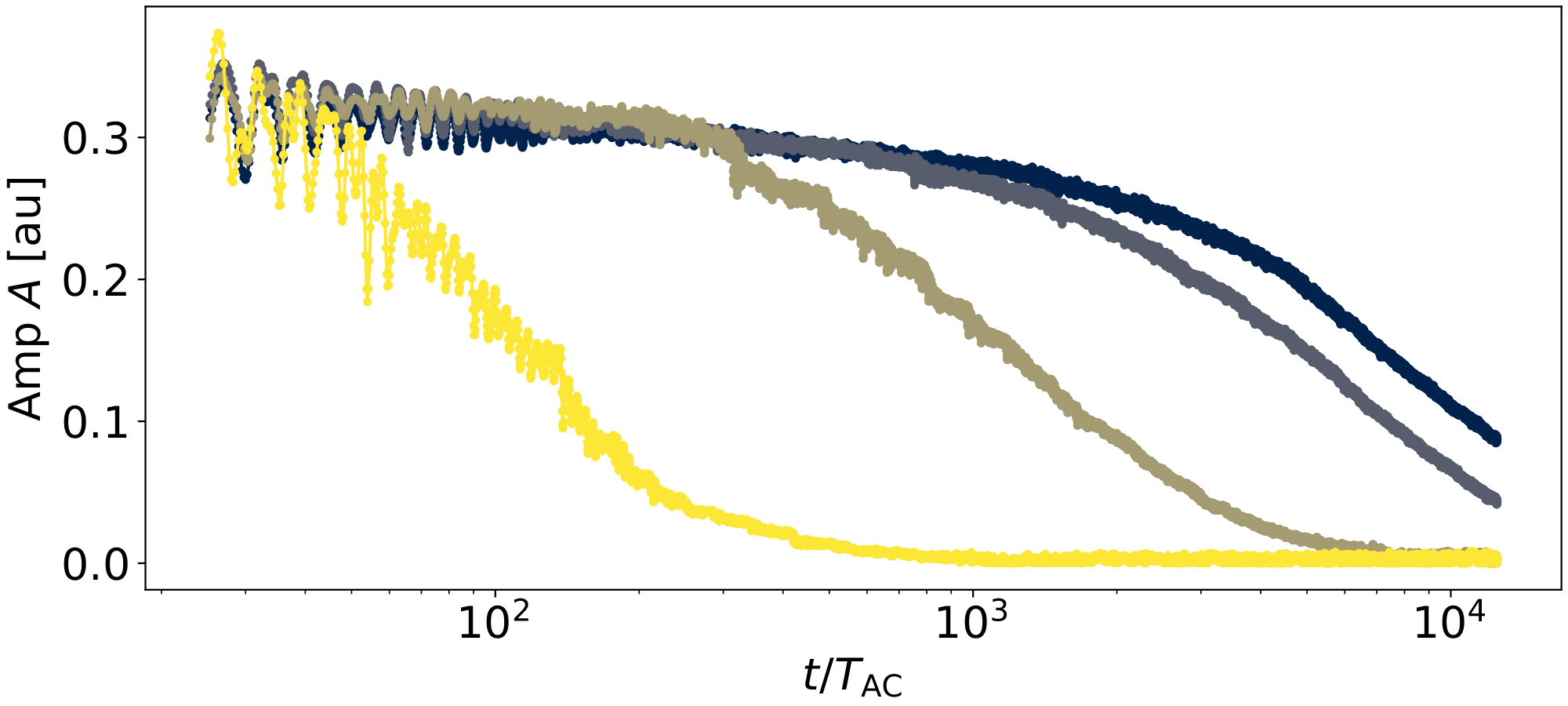}}
 \caption{
    \T{Simulation: Noisy Pulses.} 
    Total magnetization $|\boldsymbol{M}| = \sqrt{ M_x^2 + M_y^2 + M_z^2 }$ as a function of time for disordered kick sequences with varying strength $\eta$. Decrease in the lifetime of the prethermal plateaus with increasing disorder is observed. Data is obtained for $L{=}14$ spins at fixed detuning $\delta/J{=}1.0$, AC amplitude $B_\mathrm{AC}/J{=}0.5$ and fluctuating accumulated phase $\xt{=} \pi/2(1+\zeta)$ with varying uniform fluctuations $\zeta \in \bqty{-\eta, \eta}$, $\eta{=}0,\, \dots ,\, 3.16\%$~(dark to light). Further parameters are as in \zfr{longtime_simulation}. 
 }
\zfl{theta_noise}
 \end{figure} 

In this supplementary note, we consider the decay profiles of the obtained signal in \zfr{amplitude}E in the main paper and demonstrate that they follow stretched exponential profiles ${\propto}\exp[(-t/T_2^\prime)^{1/2}]$ to a good approximation. To see this, we plot in \zfr{stretched_exp} the signal $S^{\R{dec}}$ for three representative $\Bac$ values in a logarithmic scale with respect to $\sqrt{t}$. As in \zfr{amplitude}E, the four manifolds of the raw signal $S$ (see \zfr{longtime}B(i)) are smoothed over to produce respective decay profiles $S^{dec}$. Solid lines are linear fits, and are seen to be linear to a good approximation over the $t{=}35$s period considered. This shows that the decays follow very closely a stretched exponential ${\propto}\exp[(-t/T_2^\prime)^{1/2}]$. We also have previously observed stretched exponential behavior in the prethermalization decay profiles extending to $t{>}500$s~\cite{Beatrez21}. In \zfr{amplitude}F, we extract the respective time constants and calculate error bars from the corresponding data in \zfr{amplitude}E. This same method is employed to report decay constants in \zfr{decay_profile}C.

\section{\label{SOM:theory_numerics} Numerical Simulations}

Our experimental and analytical results are complemented with exact numerical simulations performed on a system of $L$ spins interacting $\Cs$ nuclear quantum spins positioned on a pseudo-random graph. For the design of pseudo-random graphs we respect the following condition~\cite{Beatrez22}: each spin of the graph is required to have at least one nearest neighbor within a distance $r_{\mathrm{max}}$ while no other spin is closer than $r_{\mathrm{min}}$. 
This ensures a certain average spin density and, importantly, avoids weakly coupled outliers. While such weakly coupled spins are in principle expected to emerge to some extent in the experimental system, their influence on the many-body dynamics is negligible. In particular, from a numerical perspective where only a relatively small number of interacting spins can be simulated, outliers are wastefully problematic as each of them reduces the effective interacting Hilbert space by a factor of $2$. 

The pseudo-random graphs are generated iteratively: first, we randomly propose a new spin position $j$ and check if (i) the inter-spin vectors with all other spins $k$ on the graph satisfy $\vert \vec{r}_{jk}\vert>r_{\mathrm{min}}$, while (ii) at least one inter-spin vector satisfies $\vert \vec{r}_{jk} \vert < r_{\mathrm{max}}$. If (i) and (ii) are satisfied, we accept the proposal and update the random graph with the spins at position $j$. Otherwise, we discard the proposed spin. We repeat this procedure until a graph of $L$ spins is constructed. The simulations are performed with $L{=}14$, $r_{\mathrm{min}}{=}0.9$ and $r_{\mathrm{max}}{=}1.1$; here $r_{\mathrm{min}}$ and $r_{\mathrm{max}}$ are given in units of $\sqrt[3]{\mu_0\hbar\gamma_n^2}$.

 For numerical simulations, unless stated explicitly otherwise, we initialize the system in a $\xhat$-polarized pure product state $\vert \psi_0\rangle \!=\!\bigotimes_{j=1}^L  \frac{1}{\sqrt{2}}\left(\vert \! \uparrow_j \rangle \!+\! \vert \downarrow_j \rangle\right)$ and perform numerically exact time evolution using the protocol of Eq.~\eqref{eq:Hamiltonian}. Due to the critical long-range interaction strength, the characteristic energy scales of the model are not immediately obvious (the energy density integral diverges in three spacial dimensions, which indicates the necessity to impose a lattice cutoff scale). However, they can be estimated from the dynamics of a free-induction decay: to this end, the initial state $\vert \psi_0\rangle$ is evolved under $\mathcal{H}_{\mathrm{dd}}$ which induces a decay of the initial $\xhat$-magnetization. Eventually, this provides an energy scale $J=1/\tau_d$ where $\tau_d$ is the timescale on which the $\xhat$-magnetization decays to $1/e$ of its initial value as the system approaches equilibrium. We note that the value of $J$ obtained in this way can only serve as a rough estimate of energy scales, as different initial states might lead to different values of $J$.

To minimize detrimental finite size effects and increase the ergodic properties of the drive we add a small uniformly distributed random noise $\delta \tacq \in [-0.05\tacq,0.05\tacq]$ to the time duration $\tacq$, i.e.~$\tacq \rightarrow \tacq +\delta\tacq$~\cite{Fleckenstein2021a,Fleckenstein2021b} which results in a slightly different unitary for each driving cycle. Such a "noisy" driving protocol can remove left-over synchronization effects emerging from the discreteness of the many-body spectrum. However, since the noise term $\delta\tacq$ breaks the periodicity of the drive, one needs to carefully examine that no additional physics, such as an extra contribution to heating, is induced due to finite $\delta\tacq$. For the system under investigation, this has been demonstrated in Ref.~\cite{Beatrez22}.

Finally, let us present some additions to the data presented in the main text, i.e.~in \zfr{longtime} and \zfr{amplitude}. In \zfr{longtime_simulation}, we present the full numerical data for the simulated time evolution partially depicted in~\zfr{longtime}B. In particular, the simulated data also leads to the square shapes when projected to the $\xy$ or $\yz$ plane.
In addition to the excursion angle $\Phi$ and linear scale $\ell$ shown in \zfr{amplitude}C-D, we show the corresponding raw data and decay times as functions of the AC amplitude in \zfr{amplitude_simulation}.
Let us in particular point out the increase in the lifetime with increasing AC Amplitude by a factor of up to $2$, suggested by the numerical simulation.
This is not observed in the experiment and obtaining analytical expressions for the decay time are beyond reach for our non-integrable long-range interacting systems. 
Notice however that the lifetime is only affected by a factor of $2$ over a large range of AC amplitudes and that the impact of finite size effects on the various quantities is not well-understood.

\section{\label{SOM:xy_phase} Role of the Phase in the Transformation to the Rotating Frame}


In App.~\ref{app:theory_model} we introduced the rotation to the dressed rotating frame, $W(t, \chi){=}\exp\lsb-i \lb \omega_\mathrm{SL} t + \chi \rb I_z \rsb$ which features the phase $\chi$.
A finite phase, $\chi {\neq} 0$, corresponds to a constant shift in the experimentally observed dressed frame phase $\varphi {=} \arctan\lb I_x / I_y  \rb - \chi$. 
While the derivations in App.~\ref{app:theory_model} and following are done for $\chi{=}0$, one can simply account for the additional phase $\chi$ by performing a subsequent static unitary transformation: $W_\mathrm{static}=\exp(-i \chi I_z )$ on the final results.

Note that the phase $\chi$ naturally enters the expression for the experimentally measured phase $\varphi$.
However, due to the subtraction of the ramp-like profile of the phase $\varphi$ from the raw experimental data, see \zfr{fig2} and App.~\ref{app:spin_reconstruction}, the actual observable becomes $\varphi_0 {=} \varphi(B_\mathrm{AC})-\varphi(B_\mathrm{AC}\!=\! 0)$.
Therefore, this subtraction procedure rather can be naturally viewed as fixing the gauge freedom associated with the phase $\chi$, so that $\varphi{=}0$ at $B_\mathrm{AC}{=}0$.

\section{\label{SOM:detuning}Role of detuning and imperfect pulses on observed shapes}

In the main text we discussed two imperfections of the SL pulse drive: the detuning $\delta$ from the Lamor frequency $\omega_L$, and the deviation in the pulse angle $\delta \xt {=} \xt - 2\pi k/N$. Here, we show that they can lead to an enhancement of the signal. In fact, they are essential for reproducing the experimental observations. Eventually, we also investigate the related question of having noise in the kick sequence.

We start with the detuning $\delta$. Let us stress again that, due to the large scale separation $\omega_L/J \leq 10^{-5}$, even a small deviation~($0.01\%$) can lead to a comparably large detuning $\delta\leq J$. 
Note that the detuning affects the direction of the kicks $\xhat\to \hat{\T{n}}=\cos(\alpha) \xhat + \sin(\alpha) \zhat$, cf.~Eq.~\eqref{eq:sl_unitary}. Therefore, the obtained shapes lie on circles which are centered around the $\hat{\T{n}}$ axis.
Note that $\hat{\T{n}}$ has a finite $\zhat$ contribution, $\hat{\T{n}}_z\propto \delta$, which leads to a finite $\zhat$-tilt, as can also be observed in the simulations in \zfr{detuning}.
Therefore, the detuning is crucial for explaining the experimentally observed $\zhat$-tilt.

Moreover, the detuning, together with the  imperfection in the pulse strength $\delta \xt$, leads to a finite $\xhat$-magnetization, $M_x\propto \sin(\alpha) + \delta \xt/\tau$, in the leading-order Floquet Hamiltonian $\mH_{F;\,k}^{(0)}$, cf.~Eq.~\eqref{eq:H_F}.
This leads to an increase in the $\xhat$-polarization strength, as can also be observed in \zfr{detuning}A(i) and \zfr{theta_detuning}A(i).
In addition, as mentioned in App.~\ref{app:prethermal_props} is it vital that $\delta$ or $\delta\vartheta$ are finite, in order for the initial state to have a finite energy density $\epsilon$, and thus also a finite temperature $|\beta| \propto |\sin(\alpha) + \delta \xt/\tau|>0$.
Such an increase in temperature is correlated with the increase in $\yhat$- and $\zhat$-magnetizations in \zfr{detuning} and \zfr{theta_detuning}, cf.~Eq.~\eqref{eq:magnetization}.

Note that the robustness of the shapes with respect to imperfection in the kicks, i.e. for $\delta \xt \neq 0$, also demonstrates that the shapes are not simply a single-particle effect but are stabilized by the many-body interactions.

Finally, let us emphasize that we numerically do not observe a significant change in the lifetime when varying the detuning $\delta$ or the angle $\delta \theta$.
 
To explore this behavior, we simulate noisy pulses $\xt {=} {(1 +\zeta)\times 2\pi/4}$ for the special case of $N{=}4$, where $\zeta$ are random numbers that differ for every pulse uniformly distributed in the interval $[-\eta,\,\eta]$ (with $\eta$ being the noise strength). We observe a strong decrease in the lifetime of the prethermal plateau with increasing $\eta$, see \zfr{theta_noise}. 
This result is not surprising as the periodic nature of the drive is lost with increasing randomness.

\end{document}